\documentclass[12pt]{article}
\usepackage[margin=0.95 in]{geometry}
\usepackage{amsmath} 
\usepackage{floatrow}
\usepackage{slashed}
\usepackage{dsfont}
\usepackage{amssymb,amsfonts}
\allowdisplaybreaks
\usepackage{mathtools}
\usepackage[all]{xy}
\usepackage{graphicx}
\usepackage{amsmath}
\usepackage{amssymb}
\usepackage{float}
\usepackage{array}
\usepackage{tikz}
\usetikzlibrary{calc}
\usepackage{mathtools}
\usepackage{mathrsfs} 
\usepackage[hidelinks]{hyperref}
\usepackage{cite}
\usepackage{tikz}
\usepackage{array, multirow}  
\numberwithin{equation}{section}
\newcommand{\be}{\begin{equation}}
\newcommand{\ee}{\end{equation}}

\def\bea{\begin{eqnarray}}
\def\eea{\end{eqnarray}}
\numberwithin{equation}{section}
\numberwithin{table}{section}

\begin{document}

\hypersetup{pageanchor=false}
\begin{titlepage}
\vbox{\halign{#\hfil    \cr}}  
\vspace*{15mm}
\begin{center}
% {\large \bf Notes on }

% \vspace*{5mm}

{\large \bf The Open/Closed Gromov-Witten/Hurwitz Correspondence }

\vspace*{5mm}

{\large \bf and }

\vspace*{5mm}

{\large \bf Localized World Sheets for Completed Cycles
}

\vspace*{10mm} 

{\large
Jan Troost}
\vspace*{8mm}

% $^c$
Laboratoire de Physique de l'\'Ecole Normale Sup\'erieure \\ 
 \hskip -.05cm
 ENS, CNRS,  Universit\'e PSL,  Sorbonne Universit\'e, \\
 Universit\'e  Paris Cit\'e,
 \hskip -.05cm F-75005 Paris, France	 

\vspace*{0.8cm}
\end{center}

%\noindent
\begin{abstract} {We discuss the open/closed version of the Gromov-Witten/Hurwitz correspondence. The duality equates the relative Gromov-Witten invariants and the count of covers of the target space with prescribed  holonomies at boundaries. We clarify the projective large N limit as well as the role of the completed versus the ordinary cycles associated to the bulk and the boundary vertex operators respectively. We provide an example check of both the correspondence and the fact that  cycles dual to closed strings need to be completed. Moreover, we identify  the connected world sheets that contribute to an equivariantly localized amplitude in the bulk that is solely due to a completion term. We also propose a picture for the completed cycle combinatorics that involves a localization diagram glued to a cut-and-join string interaction. 
  }  
\end{abstract}

\end{titlepage}

\hypersetup{pageanchor=true}

\setcounter{tocdepth}{2}
\tableofcontents
\section{Introduction}
Holography has played a prominent role in our  attempts to understand quantum gravity \cite{tHooft:1993dmi,Susskind:1994vu}. Convincing realizations of holography arise  either from string theories on anti-de Sitter backgrounds \cite{Maldacena:1997re} or from space-times of dimension  three or less. In this paper, we further study an example that lies at the crossroads of these two classes. It is a mathematically proven duality between Gromov-Witten theory on the sphere and the gauge theory that counts its covers \cite{OkounkovPandharipande1,OkounkovPandharipande2,OkounkovPandharipande3}. The first is a two-dimensional theory of gravity coupled to matter while the latter is a pure field theory. We  discussed  physical aspects of the bulk duality and its possible generalizations in \cite{Benizri:2024mpx}. We identified both the topological gravity model as well as the symmetric orbifold gauge theory that constitute the two members of the duality. We showed that summing over all degrees of covering surfaces  leads one to take a grand canonical perspective on the symmetric orbifold field theory. The symmetry of the grand canonical theory is an inverse monoid ${\cal P}_n$ which consists of partial permutations.\footnote{A partial permutation consists of a choice of subset of the set $P_n=\{1,2,\dots,n\}$ and a permutation of that subset.}

In this paper,  we take further steps in our exegesis of the seminal papers \cite{OkounkovPandharipande1,OkounkovPandharipande2,OkounkovPandharipande3}.  In section \ref{FiniteN} we discuss the open/closed grand canonical orbifold at finite $n$. We connect the open/closed Hurwitz theory \cite{OkounkovPandharipande1} with the  state-sum  topological field theory of the inverse monoid of partial permutations \cite{Troost:2025eqm}. The large $n$ limit of the model is described in section \ref{LargeN}. In our simple setting, we will understand hands-on that the limit is well-defined and projective \cite{IvanovKerov}. 
We show how the correspondence can be extended to include boundary (or open string)
sectors and what that perspective implies for the natural boundary versus bulk observables. We give a simple degree counting argument for the identification of the bulk vertex operators as completed cycles in section \ref{BulkDuality}. 
In a second part of our paper, we study the world sheet origins of the completed cycles dual to bulk vertex operators. In section \ref{LocalizationTwoPoint}, we study a two-point function that only obtains contributions from the completion of a  cycle and identify the world sheets that are responsible for the contribution. In order to do so, we make the encompassing results of \cite{OkounkovPandharipande1,OkounkovPandharipande2,OkounkovPandharipande3} very explicit, in both a diagrammatic and operator language. To localize the calculation requires the introduction of the equivariant theory with respect to the rotation of $\mathbb{P}^1$ \cite{OkounkovPandharipande2}. Moreover, in section \ref{RelativeLocalization} we analyze the relative one-point functions that govern the  operator map. Once more, we backtrack the determination of the relative one-point functions to their equivariant localization \cite{OkounkovPandharipande3}. We develop hybrid diagrams for the relative correlators, consisting of localized maps glued to a cut-and-join string theory. Section \ref{Conclusions} contains conclusions and  avenues for further research. 

\section{The Open/Closed Grand Canonical Orbifold}
\label{FiniteN}
In this section, we build upon the works \cite{OkounkovPandharipande1} as well as  \cite{Benizri:2024mpx,Troost:2025eqm}. We use the inverse monoid open/closed topological field theory \cite{Troost:2025eqm} to describe the Hurwitz theory  of branched covers of Riemann surfaces  \cite{OkounkovPandharipande1} as an open/closed theory both at finite and at infinite $n$. In the papers \cite{Benizri:2024mpx,Troost:2025eqm,Benizri:2025bok}, we used  the partial permutation algebra in order to describe both closed  and open grand canonical orbifold theories.
The partial permutation algebra ${\cal B}_n=\mathbb{C}[{\cal P}_n]$ at finite $n$ is made of linear combinations of partial permutations. The set of partial permutations ${\cal P}_n$ contains elements $(d,w)$ that consist of a choice of subset $d \subset P_n$ and a permutation $w$ thereof.\footnote{See \cite{Benizri:2024mpx,Troost:2025eqm} for longer, pedagogical reviews.}  The set $d$ can be thought of as labeling sheets of a cover and the permutation $w$ captures how the sheets are permuted around a point of ramification. Crucially, the cardinality of the set $d$, which is the degree of the cover, lies between $0$ and $n$. 
In other words, the grand canonical perspective  captures covers of multiple degrees  at once.\footnote{This is a natural requirement for a string theory. In particular, it is standard to sum over world sheet instantons of arbitrary degree when computing a string theory path integral.}
In this section, we recall the open/closed topological quantum field theory of the inverse monoid ${\cal P}_n$ of partial permutations at finite $n$ \cite{Troost:2025eqm}.
We consider its permutation invariant bulk operator algebra and its $S_n$ invariant spectrum of boundary conditions in more detail. Our analysis clarifies how to take the large $n$ limit.\footnote{Our example is a simple example of large N limits in gauge theories \cite{tHooft:1973alw}. It is satisfying that in our context, we can be very explicit about the large N limit and that it is well-defined and projective.} 
\subsection{The Open/Closed Inverse Monoid Theory }
\label{Labeled}
The closed grand canonical topological field theory is determined in terms of the center of the semisimple partial permutation algebra ${\cal B}_n=\mathbb{C}[{\cal P}_n]$ and a linear form thereon \cite{Troost:2025eqm}, thus providing a commutative Frobenius algebra. The algebra ${\cal B}_n$ is isomorphic to the direct sum ${\cal B}_n \cong \oplus_{d \subset P_n} \mathbb{C}[S_d]$ of group algebras of symmetric groups permuting the elements of the subsets $d$ of the set $P_n=\{1,2,\dots,n \}$ \cite{IvanovKerov,Troost:2025eqm}. The dimension of the center is $d_{Z(\mathbb{C}[{\cal P}_n])}=\sum_{k=0}^n \binom{n}{k} p(k)$. The center Frobenius algebra is therefore $\mathbb{C}^{d_{Z(\mathbb{C}[{\cal P}_n])}}$ and it is equipped with a linear form specified by parameters $\alpha_{\lambda}$ indexed by the irreducible representations $\lambda$ of the algebra. Their number equals the number of  conjugacy classes which in turn equals the dimension of the center of the algebra \cite{Steinberg}. The open/closed topological quantum field theory associates a  representation to each boundary, thus providing  multiplicities $n^i_\lambda$ for each boundary $i \in \{ 1,2,\dots,b \}$ and for each irreducible representation $\lambda$ \cite{Troost:2025eqm}.

\subsection{The Invariant Orbifold Theory}
\label{Invariant}
There is a natural action of the permutation group $S_n$ on the set $P_n$.
The group $S_n$ also acts on the subset $d$ and on the permutation $w$, by conjugation. The action is inherited by the full grand canonical open/closed ${\cal B}_n$ theory. We can define an $S_n$ invariant subsector of the theory.\footnote{Intuitively, while in subsection \ref{Labeled} we  labeled sheets, in this subsection we develop a label invariant description of covers.} 
%The bulk algebra is a commutative Frobenius algebra which is the  quotient of a Frobenius algebra by a finite group.
The dimension of the commutative invariant algebra ${ A}_n={\cal B}_n^{S_n}$ equals $d_{{ A}_n}=\sum_{k=0}^n p(k)$. There is a linear form inherited from the original algebra.  The set of boundary conditions which we take to be the set of $S_n$ invariant representations, is again compatible with the structure of an open/closed topological quantum field theory \cite{Troost:2025eqm}. %The matrix realization ensures that we have a consistent theory.
We recall that the invariant bulk operators are labeled by partitions $\rho$ where $|\rho|=k$ and $0 \le k \le n$. They are the operators $A_{\rho;n}$  \cite{IvanovKerov} which are equal to the sum of the partial permutations in the orbit of permutations of type $\rho$ under the action of the symmetric group $S_n$. For example, we have:
\begin{equation}
A_{(2,1);4} = (12)(3)+(12)(4)+(13)(2)+(13)(4) + \text{eight other terms} \, .
\end{equation}
Here we use the shorter notation $(12)(3)$ for the partial permutation $(d,w)=(\{1,2,3 \},(12))$. 
In the invariant theory, we have an embedding of the bulk operators into the homomorphisms between $S_n$ invariant representations. Such homomorphisms span the algebra of boundary operators. 

\subsection{The Holonomy Perspective}

In order to make contact between the bulk and the field theory description of the open theory, we want to repackage the partition functions with irreducible boundary conditions for the grand canonical Hurwitz theory in terms of holonomy dependent functions.

\subsubsection{The Group Theory}
To attain the holonomy perspective, we recall how it is obtained in the case of the group topological quantum field theory. 
The partition function for irreducible boundary representations $\lambda_i$ associated to the boundaries $i \in \{1,2,\dots,b \}$ equals \cite{Dijkgraaf:1989pz,Witten:1991we,Blau:1991mp}:
\begin{equation}
Z = \sum_{\lambda \, \text{irrep}} \left( \frac{d_\lambda}{|G|} \right)^{2-2g-b} \delta_{\lambda,\lambda_1,\dots,\lambda_b} \, .
\label{PartitionFunctionDelta}
\end{equation}
where $d_\lambda$ is the dimension of the irreducible representation $\lambda$ of the finite group $G$. 
All boundary irreducible representations $\lambda_i$ need to be identical to obtain a non-zero answer.
There is a more general partition function in which we add 
the multiplicities of the irreducible representations, which translates into a factor of $n_\lambda^i$ associated to each representation $\lambda$ and each boundary $i$: 
\begin{equation}
Z = \sum_\lambda \left( \frac{d_\lambda}{|G|} \right)^{2-2g-b}  \left( \prod_{i=1}^b n_\lambda^i \right) % \delta_{\lambda,\lambda_1,\dots,\lambda_b} 
\, .
\end{equation}
For $n^i_\lambda=\delta_{\lambda,\lambda_i}$, we recover the  expression (\ref{PartitionFunctionDelta}). 
For a more general choice of linear form on the conjugacy classes of group elements, the partition function becomes \cite{Lauda:2006mn,Troost:2025eqm}:
\begin{equation}
Z = \sum_{\lambda \, \text{irrep}} \alpha_\lambda^{2-2g-b}  \left( \prod_{i=1}^b n_\lambda^i \right) %\delta_{\lambda,\lambda_1,\dots,\lambda_b}  
\, ,
\end{equation}
where $\alpha_\lambda$ are the parameters of the linear form.
This is a slight generalization of \cite{Lauda:2006mn}. 
For each $\lambda$, we have a separate term in the partition sum and we multiply in the multiplicities of reducible boundary states associated to each boundary $i$. It becomes natural to define the partition term $Z_\lambda$:
\begin{equation}
Z_\lambda =  \alpha_\lambda^{2-2g-b} 
%\prod_{i=1}^k n_\lambda^i \, ,
\end{equation}
which is the partition function when all boundaries have the boundary condition corresponding to the same irreducible representation $\lambda$.  In the following, we again revert to the choice of linear form $\alpha_\lambda=d_\lambda/|G|$.

\subsubsection*{The Holonomy Dependent Partition Functions}
An alternative manner to specify boundary conditions on the partition function of the $G$  gauge theory is to specify the holonomy of the gauge field around  the boundary circles \cite{Dijkgraaf:1989pz,Witten:1991we,Blau:1991mp}. These holonomy dependent partition functions are discrete Fourier transforms of the functions $Z_\lambda$.  We  include factors of the characters $\chi_{\lambda}(g_i)$ for each boundary $i$ of holonomy $g_i$  and for each irreducible representation $\lambda$
\cite{Dijkgraaf:1989pz,Witten:1991we,Blau:1991mp}. 
%For each irreducible representation and given boundary, we take the same group element. 
We can associate extra factors $N(i,g_i)$   depending on  the boundary $i$ and the group element $g_i$. 
The partition sum then becomes a function of the boundary holonomies $g_i$. We have:
\begin{equation}
Z (g_i) = \sum_{\lambda} (\prod_{i=1}^b  N({i,g_i})\chi_\lambda (g_i)) Z_\lambda \, . 
\end{equation}
For the group theory, there is a direct interpretation of the partition function in terms of counting the principal bundles on surfaces with boundary if we pick the normalizations \cite{Mednyh}:
\begin{equation}
Z %= \frac{|C|}{|G|}% = \sum_\lambda \left( \frac{d_\lambda}{|G|} \right)^{2-2g-b} \prod_i \frac{|k_i|}{|G|} \chi_\lambda(k_i) 
=\sum_\lambda \left( \frac{d_\lambda}{|G|} \right)^{2-2g} \prod_{i=1}^b |C_{[g_i]}| \frac{\chi_\lambda(g_i)}{d_\lambda} 
\, ,
\end{equation}
where $|C_{[g_i]}|$ is the number of elements in the conjugacy class $[g_i]$ of the holonomy $g_i$. 

\subsubsection{The Holonomy Perspective in Inverse Monoid Theories}

For an arbitrary inverse monoid theory, one can also develop a holonomy perspective. This is based on the following  algebraic facts. Firstly, the semisimple monoid algebra is isomorphic to a sum of group algebras \cite{Steinberg,Troost:2025eqm}. 
Secondly, the set of irreducible representations of the direct sum of algebras $A_{1}$ and $A_2$ is the union of irreducibles:
\begin{equation}
\text{Irr} (A_1 \oplus A_2) = \text{Irr}(A_1) \cup \text{Irr}(A_2) \, .
\end{equation}
The dimension of these irreducible representations is the original dimension. 
Suppose that the algebras $A_i$ are group algebras $\mathbb{C}[G_i]$. Then the 
character of an irreducible representation $A_1$  is extended as {zero} on $A_2$. The representation of the identity  in one of the groups becomes a central idempotent in a direct summand. %
Similarly, one can consider the union $\cup_i \, \text{Conj}(G_i)$ of the conjugacy classes of the corresponding groups. There is then again a Fourier transform between the irreducible representations of the direct sum group algebras and the union of the conjugacy classes of the groups. These facts are sufficient to attain the holonomy perspective on a general inverse monoid open/closed topological quantum field theory, by generalizing the group technology reviewed above.

\subsubsection{The Holonomy Perspective in the  Grand Canonical Hurwitz Theory}

For concreteness, we concentrate on the grand canonical Hurwitz theory corresponding to the algebra ${\cal B}_n=\mathbb{C}[{\cal P}_n] \cong \oplus_{d \subset P_n} \mathbb{C} [S_d]$. 
The relevant character formula for the direct sum of algebras over the groups $S_d$ is  constructed as follows. One defines characters that are zero on all group (algebra) elements except those of a given $S_d$. Thus, the sum over all group elements in $\cup_{d \subset P_n}  S_d$ reduces to a sum over group elements in $S_d$ and we have orthogonality of characters there.   Moreover, all irreducibles are in $\cup_{d \subset P_n} \text{Irr} (S_d)$. A sum over irreducibles %
becomes a sum over this set. 
There is a one-to-one correspondence between irreducible representations of the finite inverse monoid ${\cal P}_n$ and the conjugacy classes of the groups $S_d$ where $d$ runs over the subsets of $P_n$. 

After dividing by $S_n$ as in subsection \ref{Invariant}, the subsets become identified at equal cardinality and we have labels corresponding to the conjugacy classes of $S_{|d|}$. 
When we restrict to $S_n$ invariant quantities, we sum over irreducibles of $\oplus_{d=0}^n \mathbb{C} [S_{d}]$. This clarifies the characters and their orthogonality properties in this particular inverse monoid. 
We have that the holonomies or conjugacy class elements take values in  $\cup_{0 \le d \le n} S_{d}$.
The finite $n$ grand canonical path integral as a function of holonomies becomes:
\begin{equation}
Z_n(g_i,q) =\sum_{d=0}^n q^d \sum_{|\lambda|=d} \left( \frac{d_\lambda}{|S_{d}|} \right)^{2-2g} \prod_{i=1}^b \frac{|C_{[g_i]}|}{d_\lambda} \chi_\lambda (g_i) \, ,
\end{equation}
where the parameter $q$ is part of the choice of linear form on the monoid and is akin here to a world sheet instanton weight for a world sheet cover of degree $d$ of the curve of genus $g$. Only when all the holonomies $g_i$ are an element of the same symmetric group $S_d$ does the path integral give a non-zero result. Thus, maximally one value of $d$ will contribute to a partition sum.
Manifestly, another way to arrive at this formula is as  the result for the theory associated to $\oplus \, \mathbb{C} [S_d]$. The micro canonical partition function associated to the group $S_d$ is:
\begin{equation}
Z_d(g_i) =\sum_{|\lambda|=d} \left( \frac{d_\lambda}{|S_{d}|} \right)^{2-2g} \prod_{i=1}^b \frac{|C_{[g_i]}|}{d_\lambda} \chi_\lambda (g_i) \, ,
\label{Microcanonical}
\end{equation}
where the holonomies $g_i$ again all take values in $S_d$. 

\subsubsection{Extended Partition Functions}
It turns out to be helpful to extend the definition of the micro canonical function $Z_d$ to an extended micro canonical Hurwitz partition function. Firstly, we map every group element $g_i \in S_{k_i}$ into an element $g_i^{ext} \in S_{d \ge k_i}$ by trivially extending the permutation. 
%We similarly extend the conjugacy classes. 
We then define the extended micro canonical function:
\begin{equation}
Z_d^{ext}(g_i)= \sum_{|\lambda|=d} \left( \frac{d_\lambda}{|S_{d}|} \right)^{2-2g} \prod_{i=1}^b 
\binom{d}{k_i}
\frac{|C_{[g_i]}|}{d_\lambda} \chi_\lambda (g_i^{ext}) \, , \label{ExtendedPartitionFunctions}
\end{equation}
where $g_i \in S_{k_i}$. 
The function is defined to be zero for any degree $d$ such that  $k_i > d$ for some $i$. The extra combinatorial factor can be understood as arising from choosing $k_i$ among $d$ sheets to be associated to the original group element $g_i$.\footnote{A more detailed justification of the prefactor will be given in  subsection \ref{Isomorphism}.} 
Furthermore, we introduce an extended grand canonical partition function:
\begin{equation}
Z_n^{ext}(g_i,q) = \sum_{d=0}^n q^d Z_d^{ext}(g_i) \, .
\end{equation}
While the information in the extended partition functions is still the same as in the original ones, it  has been combinatorially repackaged. For instance, the holonomies $g_i$ can take values in different symmetric groups and still give rise to a non-zero answer as long as the degree $d$ is high enough. 
%Still, there is neither more nor less information than in the original formulation. 
% 

\subsubsection{The Relation to Hurwitz Numbers and Extended Characters}
Given all these definitions, we can link up the standard topological quantum field theory perspective  \cite{Lauda:2006mn,Troost:2025eqm} with the Hurwitz theory \cite{OkounkovPandharipande1}. We introduce  additional notations in the process. Firstly, we define the insertion functions \cite{OkounkovPandharipande1}:
\begin{equation}
f_{\eta}(\lambda) =\binom{|\lambda|}{|\eta|} |C_\eta| \frac{\chi_\lambda(\eta)}{d_\lambda} \, .
\label{DefinitionEigenvalues}
\end{equation}
These are functions of partitions $\eta$ of an integer $k$ which can be used to  label a conjugacy class of a group element $g$ of $S_k$. 
For a given $|\lambda|$, we can extend the conjugacy class $\eta$ as we extended group elements before. 
Clearly, when $|\eta|=d=|\lambda|$ the insertions are as in the micro canonical theory (\ref{Microcanonical}) while when $|\eta| \neq d$ they correspond to the extended insertions we  introduced in equation (\ref{ExtendedPartitionFunctions}). 

We recall that  the standard Hurwitz counting of covers of the sphere gives the Hurwitz numbers
\cite{CavalieriMiles}:
\begin{equation}
\text{Hu}_d (\boldsymbol{\eta}_i) = \sum_{|\lambda|=d} \left( \frac{d_\lambda}{d!} \right)^2 \prod_{i} f_{\boldsymbol{\eta}_i}(\lambda) = Z_d([g_i] \cong \boldsymbol{\eta}_i)\, .
\end{equation}
where $|\boldsymbol{\eta}_i|=d$ are partitions of the number $d$. 
The extended Hurwitz numbers $\text{Hu}_d(\eta_i)$ were defined in \cite{OkounkovPandharipande1} and are equal to:
\begin{equation}
\text{Hu}_d(\eta_i)= \prod_i \binom{m_1(\boldsymbol{\eta}_i}{m_1(\eta_i)} H_d^X (\boldsymbol{\eta}_i)
\label{ExtendedHurwitz}
\end{equation}
where the bold $\boldsymbol{\eta}_i=\eta_i^{ext}$ are partitions of $d$ which are the extensions of $\eta_i$. The numbers $m_1(\eta)$ are the number of one entries in the partition $\eta$. 
In subsection \ref{Isomorphism} we will establish that the extended partition functions with insertions (\ref{ExtendedPartitionFunctions}) equal the extended Hurwitz numbers (\ref{ExtendedHurwitz}).\footnote{One can follow the combinatorial factors in the derivation (\ref{Combinatorics}) at this stage, if one prefers, and establish equality.} We thus establish that the Hurwitz side of the Gromov-Witten/Hurwitz correspondence \cite{OkounkovPandharipande1} has an interpretation as a two-dimensional open/closed topological quantum field theory  \cite{Lauda:2006mn,Troost:2025eqm}, including at finite $n$.

\section{The Large N Limit and the Correspondence}
\label{LargeN}
\label{LargeNLimit}
In this section, we study the projective limit  $n \rightarrow \infty$. We  recall how the limit takes us from shifted symmetric polynomials to shifted symmetric functions.  Firstly, we review part of the mathematical literature \cite{KerovOlshanski,OkounkovOlshanski,IvanovKerov}. We then connect it to the topological quantum field theory partition functions just constructed and their projective large $n$ limit. 
In the second part of the section, we recall how the large $n$ grand canonical Hurwitz theory is dual to Gromov-Witten theory \cite{OkounkovPandharipande1} and explain how the correspondence extends to the open/closed theories.

\subsection{The Projective Limit of Partial Permutation Algebras}
For each $n$ there is a homomorphism $\psi$ that maps partial permutations to permutations:
\begin{equation}
\psi: {\cal P}_n \rightarrow S_n : (d,w) \mapsto \tilde{w}
\end{equation}
where $\tilde{w}$ is the trivial completion of the permutation $w$ in $S_n$. The homomorphism $\psi$ extends to the algebras. The algebra ${ A}_n$ of $S_n$ orbits $A_{\rho;n}$ of partial permutations lies in the center of the ${\cal B}_n$ algebra.
The  orbits $A_{\rho;n}$  are characterized by a choice of partition $\rho$ of length $|d| \le n$.
To define the large $n$ projective limit of these algebras, we use the homomorphisms of algebras with $m \le n$:
\begin{equation}
\theta_m : {\cal B}_n \rightarrow {\cal B}_m : (d,w) \mapsto (d,w) \quad \text{if} \quad  d \subset P_m \quad \text{and zero otherwise.}
\end{equation}
It commutes with the $S_n$ action and therefore equally well links the invariant subalgebras ${ A}_n$ homomorphically. The algebra ${\cal B}_\infty$ is the projective limit of the algebras ${\cal B}_n$ with respect to the morphisms $\theta_n$. This means that the sequences inside the product over $n$ of elements of ${\cal B}_n$ algebras are retained if they are mapped into each other by the homomorphisms $\theta_m$. The definition makes sure the limit algebra is closed under multiplication (because the product in bigger subalgebras is mapped homomorphically along the sequence).  We can write the elements of the limit as formal infinite sums. Given a partition $\rho$ of $r$ we define the element $A_\rho$ of ${ A}_\infty$:
\begin{equation}
A_\rho = \sum_{|d|=|\rho|, [w]\cong \rho} (d,w) \, .
\end{equation}
It consists of a formal infinite sum of permutations of the  type $\rho$ and incorporates the possibility to mark sheets even when they are not permuted. 
When we run over all partitions $\rho$, the orbits $A_\rho$ form a basis of the limit algebra ${ A}_\infty$. All reference to $n$ has disappeared in the limit and the algebra  has $n$-independent structure constants \cite{IvanovKerov}. 
Therefore, there is a well-defined projective large $n$ limit of the algebra ${ A}_n$ of orbits $A_{\rho;n}$ of partial permutations. We show that the character insertions as well allow for a good large $n$ limit. 
%The bulk algebra of the topological partial permutation theory thus allows for an interesting  large $n$ limit.  Next, we show that the open/closed theory does as well. 

\subsection{The Isomorphism with Shifted Symmetric Functions}
\label{Isomorphism}
Firstly, we introduce the notion of shifted symmetric functions \cite{KerovOlshanski,OkounkovOlshanski}. Shifted symmetric polynomials are polynomials symmetric in the shifted variables $\lambda_i-i$ for $i=\{1,\dots,k \}$.  If we have a series of polynomials $f^n$ that are stable under restriction $f^n(\lambda_n=0)=f^{n-1}$ (and of uniformly bounded degree), then the series defines a shifted symmetric function in the projective limit algebra $\Lambda^\ast$. The generalized characters \begin{equation}
f_{\eta}(\lambda) =\binom{|\lambda|}{|\eta|} |C_\eta| \frac{\chi_\lambda(\eta)}{d_\lambda} \, .
\end{equation}
that we defined in (\ref{DefinitionEigenvalues}) {\em are} shifted symmetric functions in the variables $\lambda_i$ \cite{KerovOlshanski,OkounkovOlshanski}. 
Moreover, the formula for the partition function of the partial permutation theory suggests the existence of the isomorphism $F$
\begin{equation}
F: { A}_\infty \rightarrow \Lambda^\ast : F(A_\rho) = 
 \binom{n}{r} |C_\rho| \frac{\chi_\lambda(\tilde{\rho})}{d_\lambda}
%\frac{p_\rho^{\#}}{z_\rho}
\, ,
\end{equation}
where $r=|\rho|$, $\tilde{\rho}$ is the extension of $\rho$ to $S_n$ and we take the limit $n \rightarrow \infty$ on the right hand side. 
The proof that this is an isomorphic map follows from  homomorphisms and a bit of combinatorics -- we follow the presentation in \cite{IvanovKerov}. 
Firstly, we have the homomorphic projection of orbits $A_\rho$ into the algebra $\mathbb{C}[S_n]$:
\begin{align}
(\psi \circ \theta_n) (A_\rho) &= \binom{n-r+m_1(\rho)}{m_1(\rho)} C_{\tilde{\rho};n} \, ,
\end{align}
where 
%$\tilde{\rho}$ is the $\rho$ partition trivially completed (with ones) to become a partition of $n$ and
$C_{\tilde{\rho};n}$ is the sum of all elements in $S_n$ in the conjugacy class $\tilde{\rho}$.  
When we evaluate the character of these central elements in an irreducible representation $\lambda$ we find \cite{IvanovKerov}:
\begin{align}
\chi_\lambda ( (\psi \circ \theta_n) (A_\rho)) &=
\binom{n-r+m_1(\rho)}{m_1(\rho)}   |C_{\tilde{\rho}}| \chi_\lambda(g_{\tilde{\rho}})
\nonumber \\
&= \binom{n-r+m_1(\rho)}{m_1(\rho)}  \frac{n !}{\prod_i i^{m_i(\tilde{\rho})} m_i(\tilde{\rho})!} \chi_\lambda(g_{\tilde{\rho}})
\nonumber \\
&=  \binom{n-r+m_1(\rho)}{m_1(\rho)}  \frac{n !}{m_1(\tilde{\rho})! \prod_i i^{m_i(\bar{\rho})} m_i(\bar{\rho})!} \chi_\lambda(g_{\tilde{\rho}})
\nonumber \\
&=  \binom{n-r+m_1(\rho)}{m_1(\rho)}  n! \frac{m_1(\rho)!}{m_1(\tilde{\rho})!}
\frac{1}{\prod_i i^{m_i({\rho})} m_i({\rho})!}
\chi_\lambda(g_{\tilde{\rho}})
\nonumber \\
&= \frac{n!}{(n-r)!} \frac{1}{\prod_i i^{m_i({\rho})} m_i({\rho})!}
\chi_\lambda(g_{\tilde{\rho}})
\nonumber \\
&= \binom{n}{r} |C_\rho| \chi_\lambda(g_{\tilde{\rho}})
\label{Combinatorics}
\end{align}
where $\bar{\rho}$ is the partition $\rho$ with all ones removed and we used the cardinality of conjugacy classes of the symmetric groups. 
Since normalized characters $\chi_\lambda(\cdot)/d_\lambda$ define a homorphism from the center of a group algebra to the complex numbers, the whole map $F$ that maps:
\begin{equation}
F (A_\rho) (\lambda) = \frac{1}{d_\lambda} \chi_\lambda ( (\psi \circ \theta_n) (A_\rho))
\end{equation}
is a homomorphism in the $n \rightarrow \infty$ limit and it does map:
\begin{equation}
F(A_\rho)(\lambda) =
f_\rho(\lambda) \, .
\end{equation}
Thus we have established the isomorphism between the orbit sums $A_\rho$ and the shifted symmetric functions $f_\rho$ as well as between their projective limit algebras \cite{IvanovKerov}.

\subsubsection*{Summary}
 The partition function at given covering degree $d$ is a sum over contributions labeled by irreducible representations $\lambda$. Boundary conditions can be alternatively associated to holonomies in $S_d$. Partial permutations are  handy in treating multiple covering degrees at once. They allow to generalize the holonomy argument to any element of $S_k$ for all orders $k$ simultaneously. Thus, boundaries become labeled by arbitrary partitions $\rho$. There is still degree $d$ dependence in $\rho$, but one can sum the degree $d$ over all values from $0$ to $n$. We have shown in two manners (which are equivalent by isomorphism) that a large $n$ projective limit of the properly normalized expressions exists. %The $n$-independence lies in the fact that each term  consists of factors that are  symmetric functions. 
The large $n$ limit lies in taking the sum over degrees to be infinite as well as generalizing the polynomials coding characters to shifted symmetric functions. 

\subsection{The Open/Closed Correspondence}
There is a duality between the large $n$ grand canonical Hurwitz theory and Gromov-Witten theory on curves \cite{OkounkovPandharipande1}. It is  proven by passing through the relative Gromov-Witten theory \cite{OkounkovPandharipande2,OkounkovPandharipande3}. We  stress that the correspondence is valid for the full open/closed theories. 

\subsubsection{The Open/Closed Gromov-Witten Theory}
Firstly, Gromov-Witten theory is formulated in terms of integrals over the compactified moduli spaces of Riemann surfaces. If we wish to extend the duality to the open case, we must consider integrals over the moduli spaces of Riemann surfaces with boundary. Instead of doing so in full, we exploit the following idea.  Consider a surface with boundaries that are circles. Consider the covering of this Riemann surface by another surface with circular boundaries in which the boundaries map to each other. Then a given circle of the cover covers a given circle of the target surface an integer number of times. In other words, there is a winding number $w$. The simple idea is that when we close up the two circles, there will inevitably be a point of ramification in the disk we use to close the covered surface and it will have ramification of degree $w$. 
This idea is formalized in for instance  \cite{LiSong} for the quintessential case of a disk covering a disk.
See also e.g. \cite{KatzLiu}. We will take this correspondence between covering surfaces with boundary circles and points of ramification for granted and will therefore consider the open/closed Gromov-Witten theory to be the relative Gromov-Witten theory, i.e. the theory with bulk insertions and prescribed ramification at points. It would certainly be interesting to put this hypothesis on a firmer footing for general moduli spaces of Riemann surfaces with boundary. 

Under the working hypothesis, the relative Gromov-Witten/Hurwitz correspondence \cite{OkounkovPandharipande1,OkounkovPandharipande2,OkounkovPandharipande3}:
\begin{equation}
\langle \prod_i\tau_{k_i}(\omega),\eta_j \rangle_d^{X,\bullet} = \sum_{|\lambda|=d} \left( \frac{d_\lambda}{d!} \right)^{2-2g} \prod_{i} \frac{p_{k_i+1}(\lambda)}{(k_i+1)!} \prod_j f_{\eta_j}(\lambda) \, 
\label{GWHDuality}
\end{equation}
is an open/closed duality between a topological theory of gravity on the left hand side and a grand canonical infinite $n$  gauge theory on the right hand side. The genus of the Riemann surface $X$ equals $g$ and we have projected the duality on  degree $d$ covers on both sides. The left hand side is a  theory of gravity coupled to a topological sigma model on the curve $X$. The bulk insertions $\tau_{k_i}(\omega)$ are descendants of the volume form $\omega$ on the curve. They correspond in the gauge theory to the shifted symmetric power sums $p_{k_i+1}/(k_i+1)!$. The latter are defined as follows. Recall that we have a one-to-one map between the shifted symmetric functions $f_\rho$ and the orbits $A_\rho$. The highest degree  term in $f_\rho$ is the shifted symmetric power sum  function $p_\rho$ (divided by $\prod_i \rho_i$). The bulk insertions $\tau_k(\omega)$ are dual to one factor power sum $p_{k+1}/(k+1)!$. The counterparts of the single factor power sum symmetric functions  in the algebra ${ A}_\infty$ are by  definition the completed cycles \cite{OkounkovPandharipande1}.\footnote{Note that they are completed orbits rather than completed cycles. We use the conventional misnomer.}${}^{,}$\footnote{See also \cite{Lerche:2023wkj,Benizri:2024mpx} for further discussion of the bulk theory.} The relation between the $f$ and $p$ bases is triangular. 

The extension to include the open sectors is now straightforwardly stated. 
The imposition of ramification profiles $\eta_j$ in the relative Gromov-Witten theory  is translated into the insertions of the shifted symmetric functions $f_{\eta_j}$ on the topological field theory side. 
The relative insertions $\eta_j$
correspond to the open sector because the insertions $f_{\eta_j}$ correspond to circular boundaries with holonomies $\eta_j$. We have explained that these are dual to the specification of boundary conditions in terms of representations. Thus, we have shown that boundary conditions are coded by the symmetric functions $f_{\eta_j}$.
%\footnote{There are further multiplicities $n_\lambda$ to consider, as mentioned previously.} 

\subsubsection{Generalizations of the Open/Closed Theory}
A  generalization for the right hand side of the duality (\ref{GWHDuality}) is given by invoking the multiplicities $n_\lambda^j$ that we can associate to boundaries and writing:
\begin{equation}
Z_d =  \sum_{|\lambda|=d} \left( \frac{d_\lambda}{d!} \right)^{2-2g} \prod_{i} \frac{p_{k_i+1}(\lambda)}{(k_i+1)!} \prod_j n^j_\lambda f_{\eta_j}(\lambda)
\, .
\end{equation}
These types of factors were analyzed in the context of the KP hierarchy in \cite{Buryak:2025mij}, restricted to a single integer raised to the power of the   number of boundaries.\footnote{One can set all $n^i_\lambda$ equal to obtain such a special case.} Interestingly, the integer was demonstrated to index a 
 sequence of $\tau$ functions which are non-trivial transforms of one another. In the context of the Toda hierarchy for the Gromov-Witten theory at hand, it would be instructive to formulate the full open/closed hierarchy, including the integer multiplicities. 
Furthermore, we remark in passing that from a string theory point of view, it is also natural to associate open string (Chan-Paton) matrix algebras to  boundaries of the surface that are intervals. 

In short, we  interpreted the relative Gromov-Witten/Hurwitz correspondence as a duality between open/closed theories. By carefully analyzing the map to standard topological theories, we are able to generalize the partition function to include integer multiplicities. The triangular change of basis between bulk and holonomy insertions can be reinterpreted (after exponentiation) as a translation of open string deformations into closed string deformations and vice versa.

\section{The Bulk Operator Duality }
\label{BulkDuality}
In the rest of the paper, a guiding goal is to understand the bulk operator map that is instrumental in the Gromov-Witten/Hurwitz correspondence \cite{OkounkovPandharipande1} better. We have the operator map:
\begin{equation}
\tau_k(\omega) \overset{\text{GW/H}}{\leftrightarrow} \frac{p_{k+1}}{(k+1)!} \overset{\text{F}}{\leftrightarrow} 
%\frac{\overline{A_{k+1}}}{k!} \overset{\text{def}}{=} 
\frac{\overline{(k+1)}}{k!} \, , \label{OperatorMap}
\end{equation}
where the left hand side is a standard descendant vertex operator associated to the volume form $\omega$ of the target space and the right hand side are the completed cycles corresponding to the power sum symmetric functions of order $k+1$. It was discussed in \cite{OkounkovPandharipande1,Benizri:2024mpx} how the operator map comes about through the isolation of individual vertex operators connected to the full surface through a neck that allows for a degeneration or spectral decomposition. We want to study the map (\ref{OperatorMap}) itself more closely.

\subsection{The Degree}
Our first insight in the completed cycles will come from the proof that the  counterpart to bulk operators must be polynomials $p$ of fixed degree. Along with the fact that the observables are shifted symmetric, this singles out the multiplicative basis of shifted symmetric power sums as the dual to the bulk vertex operators $\tau_k(\omega)$. 
To understand the fixed degree of the dual, we study the bulk charge constraints and how these propagate throughout the calculations and ensuing correspondence of \cite{OkounkovPandharipande1}. 
\subsubsection{Ghost Number Conservation}
The Gromov-Witten theory is a topological sigma-model coupled to topological gravity. The corresponding two-dimensional string amplitudes are integrated over moduli spaces of Riemann surfaces. The integral is well-defined and assigns a particular dimension to the compactified moduli space and a differential degree to the vertex operators. A dimensional constraint on the correlators is a consequence -- it can also be thought of as due to ghost number conservation.
The constraint for the Gromov-Witten correlators
\begin{equation}
\langle \prod_{i=1}^s  \tau_{k_i} (\omega) \rangle^X_d
\end{equation}
on a target curve $X$ with insertions $\tau_{k_i}(\omega)$ of descendants of the volume form $\omega$ equals:
\begin{equation}
2g-2+d(2-2g(X)) =  \sum_{i=1}^s k_i  \label{ChargeConstraint}
\end{equation}
where $d$ is the degree of the cover. 
%and $g(X)=0$ for target $X=\mathbb{CP}^1$. 
We can think of the net contribution $k$ for each vertex operator $\tau_k(\omega)$ as indicating that it is a differential form of degree $2k$.\footnote{Moreover, each insertion carries a charge $1$ on the right hand side which is canceled by a contribution of $1$ to the complex dimension on the left.}
%${}^{,}$\footnote{For relative Gromov-Witten theory with respect to a partiton $\nu$ we add $l(\nu)$ to the left hand side because each cycle counts as one marked point. }
%

\subsubsection{The Localization and the Degree}
To track the degree of the bulk operator towards the fixed degree of the dual polynomial, we very briefly review how the correspondence is constructed \cite{OkounkovPandharipande1}. Firstly, it is proven that one can isolate each vertex operator and that it is thus sufficient to calculate the non-equivariant relative one-point functions of $\tau_k(\omega)$ in order to be able to compute all correlators.

To describe the result of the localization calculation for the one-point functions, we use a Fock space of fermions \cite{OkounkovPandharipande1}. 
On the Fock space of complex fermions $\psi_i$ labeled by $i \in \mathbb{Z}+1/2$, we have the bilinear $gl(\infty)$ operators
\begin{equation}
E_{ij} = :\psi_i \psi_j^\ast: \, .
\end{equation}
We also define the operators
\begin{equation}
{\cal E}_r (z) = \sum_{k \in \mathbb{Z}+\frac{1}{2}} 
e^{z(k- \frac{r}{2})} E_{k-r,k} + \frac{\delta_{r,0}}{\zeta(z)} \, . \label{DefinitionE}
\end{equation}
The fermion can be bosonized and the boson has oscillator excitations generated by the operators $\alpha_r = {\cal E}_r(0)$ (for $r \neq 0$). 
We need further  operators $ A(tz,z)$:
\begin{equation}
%A(z)=
A(tz,z) = S(z)^{tz} \sum_{k \in \mathbb{Z}} 
\frac{\zeta(z)^k}{(tz+1)_k} {\cal E}_k(z)
\end{equation}
where we introduced the functions:
\begin{equation}
\zeta(z) = e^{\frac{z}{2}}-e^{-\frac{z}{2}} \, , \qquad S(z) = \frac{\sinh \frac{z}{2}}{\frac{z}{2}} \, .
\end{equation}
We can finally state the generating function for the equivariant one-point functions \cite{OkounkovPandharipande3}:
\begin{equation}
G(z|\nu;t) = \sum_{k} z^{k+1} \langle \tau_k(0) | \nu \rangle = \langle  A(tz,z) e^{\alpha_1} | \nu \rangle
\, ,
\label{EquivariantOnePoint}
\end{equation}
where 
\begin{equation}
|\nu\rangle = \frac{1}{z(\nu)} \prod \alpha_{-\nu_i} | 0 \rangle \, ,
\end{equation}
with $z(\nu)=|\text{Aut}(\nu)| \prod_i {\nu_i}$ i.e. the number of symmetries of the partition $\nu$ multiplied into the products of its members.  
The non-equivariant $t \rightarrow 0$ limit of the one-point function is easily obtained from the commutator:
\begin{equation}
{ A}(0,z) = e^{\alpha_1} {\cal E}_0 (z) e^{-\alpha_1} 
\, ,
\end{equation}
which implies the non-equivariant equality:
\begin{equation}
\prescript{\text{n}}{}G(z|\nu)=\lim_{t \rightarrow 0} G(z|\nu;t) %= \lim \sum_{g,k} z^{k+1} \langle \tau_k(0) | \nu \rangle 
=  \langle e^{\alpha_1} {\cal E}_0(z) | \nu \rangle \, .
\label{NonEquivariantOnePointFunction}
\end{equation}
The operator ${\cal E}_0(z)$ captures the insertion of the operators $\tau_k(\omega)$ with the $k+1$-st order in $z$ corresponding  to the $\tau_k(\omega)$ insertion -- see equation (\ref{EquivariantOnePoint}). However, we also have correspondence between the operators ${\cal E}_0$ and their eigenvalues on a zero charge state $v_\lambda$ in the fermionic Hilbert space \cite{OkounkovPandharipande1}:
\begin{align}
% {\cal E}_0(z) &= \sum_{k \in \mathbb{Z}+\frac{1}{2}} e^{zk} E_{kk} +2/\sinh(2/z)
% \nonumber \\
{\cal E}_0(z) v_\lambda &= e(\lambda,z)v_\lambda \,  ,
\end{align}
where
\begin{align}
e(\lambda,z) &= \sum_{i=0}^\infty e^{z (\lambda_i-i+\frac{1}{2})}
\end{align}
with a direct relation to the fixed degree polynomial $p_{k+1}$:
\begin{align}
p_{k+1}(\lambda) &= (k+1)! [z^{k+1}] e(\lambda,z)
 \, ,
\end{align}
where the symbol $[z^{k+1}]$ instructs to project onto the term of degree $k+1$ in $z$.  
We thus see a direct translation from the descendant order $k$ of the operator $\tau_{k}(\omega)$ to the $k+1$-st order fixed degree shifted symmetric  polynomial $p_{k+1}$ in the representation labels $\lambda_i$.

\subsubsection{The Limit of a  Commutator}

The final result (\ref{NonEquivariantOnePointFunction}) for the   non-equivariant relative one-point functions contains three factors. 
The $\nu$ vacuum codes a particular branching at one point of the sphere and the $e^{\alpha_1}$ factor is an operator that creates any number of trivial sheets at another point. These are straightforwardly interpreted ingredients in the final formula. The main ingredient ${\cal E}_0(u z)$ demands a closer look.\footnote{We introduce a parameter $u$ that keeps track of the genus expansion, i.e. a string coupling. Because for the non-equivariant one-point function, we have the constraint $2g-2+2d=k$, it accompanies the parameter $z$.}
Therefore, we revisit the commutator that gives rise to the operator ${\cal E}_0( u z)$. It originates in the commutation relation \cite{OkounkovPandharipande2}:
\begin{align}
e^{\frac{u}{t} {\cal F}_2} \alpha_{-m} e^{-\frac{u}{t} {\cal F}_2} 
&= \sum_{n=0}^\infty \frac{1}{n!} [\frac{u}{t} {\cal F}_2,\alpha_{-m}]_n
= {\cal E}_{-m}( \frac{u m }{t} ) 
\, ,
\end{align}
where ${\cal F}_2=\sum_{k \in \mathbb{Z}+1/2} \frac{k^2}{2} E_{kk}$ is the operator in the fermionic Fock space that implements transposition. 
%Each extra commutator brings the same factor $m(k+m/2)$ in the sum. With the $1/n!$, the result exponentiates with 
The coupling $u/t$ keeps track of the number of ${\cal F}_2$ insertions we commuted. 
The operator ${\cal E}_0(uz)$ arises from taking the limit  $t \rightarrow 0$ and $m \rightarrow 0$ while keeping $z=m/t$ fixed. 
This also requires taking the limit for the oscillator $\alpha_{-tz}$ with continuous index.\footnote{This is reminiscent of the following physical set-up. One considers a charged complex scalar field in two dimensions in a background magnetic field. A continuous change in the magnetic field allows for a tuning of the oscillator modes of the complex field.
}
As $t \rightarrow 0$, the commutator remains finite because the coupling to the branch point insertion ${\cal F}_2$ is taken to be $\frac{u}{t}$ with $t \rightarrow 0$ and this blowing up compensates the operator commutator going to zero. In this manner, we find the rule that the power of $z$ (or the string coupling $u$) also captures the number of ${\cal F}_2$ insertions.
The operator ${\cal E}_0(u z)$ is  most directly described as:
\begin{equation}
{\cal E}_0(u z) =
\lim_{t \rightarrow 0} e^{\frac{u}{t} {\cal F}_2} \alpha_{-tz} e^{- \frac{u}{t} {\cal F}_2} \, .
\end{equation}
The operator ${\cal E}_0$ does not introduce new sheets (since the corresponding oscillator has zero index) but it {\em is} inserted at a point. The commutator is non-trivial and the power of $z$ keeps track of the number of branch point insertions that were involved in the commutator.\footnote{In the $t \rightarrow 0$ limit of the calculation, a normal ordering ambiguity appears that is resolved as 
in the definition (\ref{DefinitionE}) of the  operator ${\cal E}_0$. }

\subsection{Localized Correlators}
\label{Localization}
Tracking the fixed degree of the bulk  observable gives one answer to the question why bulk operators correspond to completed cycles in the field theory. We also attempted to interpret the final formula for the non-equivariant relative one-point function directly.  These analyses are only partially satisfying because the crucial calculations are done after regularizing through equivariant localization.  Therefore, if we want to get a handle on the world sheet origins of the completed cycle contributions, we are encouraged to identify them in the equivariant calculations themselves.
We remind the reader that the logic of the series of papers \cite{OkounkovPandharipande1,OkounkovPandharipande2,OkounkovPandharipande3} is to prove the non-equivariant Gromov-Witten correspondence \cite{OkounkovPandharipande1} by equivariant localization on $\mathbb{P}^1$ \cite{OkounkovPandharipande2} which is extended to relative invariants in \cite{OkounkovPandharipande3}.

Equivariant localization is achieved on $\mathbb{P}^1$ by exploiting the $\mathbb{C}^\ast$ action on the target, inherited by the compactified moduli space of maps \cite{Kontsevich:1994na}. Through degeneration and localization, correlators become a sum over products of linear Hodge integrals associated to the two fixed points of $\mathbb{P}^1$. The latter can be computed through their relation to Hurwitz numbers which in turn have a Fock space description.  
In the rest of this section, we will illustrate these general theorems \cite{OkounkovPandharipande1,OkounkovPandharipande2,OkounkovPandharipande3} and exploit  examples to pinpoint equivariant localization contributions responsible for the completion of single cycles.

\subsubsection{Connected Low Genus Correlators}
Momentarily, we will pick a few simple correlators to study in detail. Firstly, we identify interesting classes of candidate correlators at low genus. 
%
% 
%
%\subsubsection{The Constraints}
We consider a spherical target space $X$. We then have the constraint equation:
\begin{equation}
2g-2 + 2d = \sum_{i=1}^s k_i \, .
\end{equation}
We note that extra $\tau_0(\omega)$ insertions leave the constraint equation invariant. These insertions add and subtract one to and from the dimension of the moduli space (because there is an extra insertion on the one hand and one fixes a point of the world sheet to be mapped to a particular point of the base -- see \cite{Witten:1989ig}). They can be inserted in any correlator while respecting the constraint. 
Spherical one point functions can be non-zero for even $k=2d-2$. For $d=0$ or $1$, they can be connected, but not for higher degree (since they must correspond to the identity permutation). Non-trivial  contributions arise from:
\begin{equation}
\langle \tau_0(\omega)^s \rangle_{d=1,g=0} \, .
\end{equation}
%These are spherical instanton contributions. 
For two-point functions, there is a natural possibility to obtain a connected world sheet. We can pick $k_1=d-1=k_2$. This corresponds to, among other contributions, the leading single cycle contribution in the completed cycles dual to the operators $\tau_k$: 
\begin{equation}
\langle \tau_k \tau_k \rangle_{d=k+1,g=0} \, .
\end{equation}

At one loop, we have more intriguing possibilities. 
When the covering world sheet is a surface of genus one, the constraints become
\begin{equation}
2d = \sum_{i=1}^s k_i 
%\quad \text{and} \quad d \le k_i+1
\, .
\end{equation}
There are interesting one-point functions like the degree $1$ one-point function of $\tau_2$ (optionally completed with any number of $\tau_0$ operators):
\begin{equation}
\langle \tau_2 %\tau_0^s 
\rangle_{d=1,g=1} \, .
\end{equation}
If we have two insertions, and the surface is connected, the permutations must be inverse,   they must be single cycle and equal to the degree in cycle length. Since there are no $(k)$ cycle contributions in completed cycles $\overline{(k+1)}$,  as a consequence of a parity constraint, the only possibility then is that $k_1=d-1$ and $k_2=d+1$:
\begin{equation}
\langle \tau_{d-1} \tau_{d+1} \rangle_{g=1,d} \, . \label{TwoPointFunctions}
\end{equation}
These are generically non-zero. 
In the following, we will focus on the examples of the $\tau_2$ one-point function at degree one and the two-point function $\langle \tau_1 \tau_3 \rangle$ at degree two.

\subsection{The Bulk Localization}
\label{OscillatorApproach}
We study in detail the equivariant calculation of  example correlators. On the one hand, this will give a pedagogical inroad to the powerful mathematics results in \cite{OkounkovPandharipande1,OkounkovPandharipande2,OkounkovPandharipande3}. On the other hand, it will provide us with concrete examples of localized world sheets that contribute to the completion of cycles. To identify the world sheets, we need to recall a number of results and techniques of the equivariant localization on $\mathbb{P}^1$ \cite{OkounkovPandharipande2}.

\subsubsection{Linear Hodge Integrals}
Since they are the main factors in the localization formula, we digress on the linear Hodge integrals and their relation to Hurwitz numbers as captured in the ELSV formula \cite{Ekedahl:2000fyi}. 
%We moreover find it illuminating to recall the localization derivation of the latter result. 
The relevant integrals over the compactified moduli space ${\overline{{\cal M}}_{g,n}}$ of Riemann surfaces are linear in the Hodge class:
\begin{equation}
H^\circ_g(z) = \prod z_i \int_{\overline{{\cal M}}_{g,n}} \frac{1-\lambda_1+\dots + (-1)^{g} \lambda_g}{\prod_i (1-z_i \psi_i)} \, . \label{LinearHodgeIntegral}
\end{equation}
The integrals will correspond to connected diagrams and are labeled by the genus $g$ of the Riemann surfaces with $n$ punctures whose shape we integrate over.
For the parameters $z_i$ equal to the parts of a partition $\mu$, the ELSV formula states that the linear Hodge integral is given in terms of the Hurwitz numbers $C_g(\mu)$ \cite{Ekedahl:2000fyi}:
\begin{equation}
C_g(\mu) = \frac{b!}{z(\mu)} (\prod \frac{\mu_i^{\mu_i}}{\mu_i!})H_g(\mu)
\, .  \label{ELSV}
\end{equation}
The Hurwitz number $C_g(\mu)$ counts covers of degree $|\mu|$ with ramification profile $\mu$ and simple branch points over $b=2g+|\mu|+l(\mu)-2$ points (where $l(\mu)$ is the number of parts in the partition $\mu$).\footnote{The count includes disconnected covers such that we must exponentiate the connected Hodge numbers $H_g^{\circ}$ in order to get their disconnected counterpart $H_g$ on the right hand side.}${}^{,}$\footnote{A derivation  through equivariant localization of the ELSV equality is reviewed for a mathematical audience in \cite{Liu}.} This connection allows one to establish an oscillator formula for the linear Hodge integrals \cite{OkounkovPandharipande2}.

\subsubsection{The Generating Function}

The generating function of Gromov-Witten correlators can be computed through degeneration and then localization on the fixed points $0$ and $\infty$ of a $\mathbb{C}^\ast$ action on the sphere  \cite{OkounkovPandharipande2}. At degree $d$, the generating function with insertions at $0$ and $\infty$ reads:
\begin{align}
G_d(z_i,w_j,u) &=\frac{1}{z(\mu)} \sum_{|\mu|=d}
(\frac{u}{t})^{l(\mu)} (-\frac{u}{t})^{l(\mu)} 
t^{-d-n} (-t)^{-d-m} (\prod_i \frac{\mu_i^{\mu_i}}{\mu_i!})^2 
\nonumber \\
 & \qquad \qquad \qquad
H(\mu,tz,\frac{u}{t}) \times  H(\mu,-tw,- \frac{u}{t})
\, .
\label{DegenerationFormula}
\end{align}
 There is an intermediate sum over partitions $\mu$ of the degree $d$ and there are two linear Hodge integrals $H$ associated to the two fixed points where the insertions $z_i$ and $w_j$ reside. The string coupling is denoted $u$ and the equivariant parameter $t$. All other factors are appropriate weight and symmetry factors. 

We have two expressions for the linear Hodge integrals $H$. On the one hand, we have the direct definition in terms of an integral over compacified moduli spaces (\ref{LinearHodgeIntegral}). On the other hand we have a  generalisation of ELSV \cite{OkounkovPandharipande2} which allows us to write the Hodge integrals as generalized Hurwitz numbers which in turn have the oscillator expression \cite{OkounkovPandharipande2}: 
\begin{equation}
H(t z,\frac{u}{t}) = (\frac{u}{t})^{-n} \langle
\prod_{i=1}^n { A}(t z_i, u z_i) \rangle \, .
\end{equation}
\subsubsection*{Oscillator Calculations}
The oscillator expression can be computed using basic commutators like:
\begin{equation}
[\alpha_k , {\cal E}_r(z)] = \zeta(kz) {\cal E}_{k+r}(z) 
\, ,
\end{equation}
as well as the commutation of operators ${\cal E}_{a_1} (z_1)$ and ${\cal E}_{a_2} (z_2)$:
\begin{equation}
[ {\cal E}_{a_1} (z_1) , {\cal E}_{a_2} (z_2) ]=
\zeta \left( \det \left( \begin{array}{cc} a_1 & z_1 \\ a_2 & z_2 \end{array} \right) \right) {\cal E}_{a_1+a_2}(z_1+z_2) \, .
\end{equation}
Simple consequences of the definitions of operators can be exploited, like the  vacuum expectation value:
\begin{equation}
\langle {\cal E}_0 (z) \rangle = (\zeta(z))^{-1} \, .
\end{equation}
Strictly negative index ${\cal E}_k$ operators annihilate the left vacuum and strictly positive index operators the right vacuum. In this manner, using a generalized Wick theorem, one can  evaluate all linear Hodge integrals \cite{OkounkovPandharipande2}. 

\subsubsection*{Moduli Space Integral Calculations}
To compute the disconnected Hodge number $H(\mu,tz,u/t)$, it is  useful to transform the calculation into a list of diagrams each of which corresponds to a product of integrals over compactified moduli spaces of Riemann surfaces. The contributing diagrams in the calculation are described in \cite{OkounkovPandharipande2}. One then computes the amplitude associated to each diagram and takes the sum over diagrams. 

The diagrams consist of vertices $v$. Each vertex $v$ has an assigned genus $g_v$. Out of the vertices come edges of degrees that equal the members of the partition $\mu$. The number of edges coming out of vertex $v$ is denoted $e_v$. Moreover, for each $z_i$, we also have a half-edge coming out of a vertex. The number of these at vertex $v$ equals $n_v$. We call these markings. To each vertex, we associate an integral over the compactified moduli space  $\overline{\cal{M}}_{g_v,e_v+n_v}$. The dimension of the moduli space is $3g_v-3+e_v+n_v$ at each vertex.
We multiply the result over all vertices in a diagram.

Note that we have only described the left Hodge number $H$. There is also a diagram for the full Gromov-Witten generating function $G$. We consider a left diagram as just described, with edges pointing towards the right. We consider a right diagram of the same type, with edges pointing to the left. When these two match up because they correspond to the same partition, we have a term in the sum over partitions $\mu$ in the degeneration formula (\ref{DegenerationFormula}).
The sum of all the degrees (on the left and separately on the right) must equal $d$.  The total genus of the diagram equals:
\begin{align}
\sum_v (2g_v-2)+e_v &= 2g-2
\, .
\end{align}
Note that the genus of a disconnected surface can be negative. 

We have briefly reviewed two ways of computing equivariant bulk correlators \cite{OkounkovPandharipande2}. One is through the calculation of diagrams which we can split into left and right, then to compute the corresponding moduli space integrals. The second manner is through computing the Hodge numbers by oscillator manipulations. Let us turn to example calculations and identify the localized world sheets that contribute.
In the rest of the section, we compute a one-point function and in section \ref{TwoPoint} we turn to a two-point function.

\subsection{A One-Point Function}
We compute the $\tau_2$ one-point function 
\begin{equation}
\langle \tau_2 \rangle_{d=1,g=1} = \frac{1}{24} + \frac{7}{5760}
\end{equation}
in two manners that are equivalent by the Gromov-Witten/Hurwitz correspondence. Firstly, we use the map to the completed cycles in Hurwitz theory
\begin{equation}
\tau_k(\omega) \leftrightarrow \frac{1}{k!} \overline{(k+1)} \, ,
\end{equation}
as well as the completed cycles \cite{OkounkovPandharipande1}:
\begin{align}
\overline{(1)} &= (1) - \frac{1}{24} ()
\nonumber \\
\overline{(2)} &= (2)
\nonumber \\
\overline{(3)} &= (3) + (1,1) + \frac{1}{12} (1) + \frac{7}{2880} () 
\nonumber \\
\overline{(4)} &= (4) + 2 (2,1) + \frac{5}{4} (2)  \, ,
\label{CompletedCycles}
\end{align}
expressed in partial permutation orbits summarized by their partition. 
Next, we evaluate the vacuum expectation value in the Hurwitz theory
\begin{equation}
\langle \tau_2 \rangle_{d=1}\overset{\text{GW/H}}{=} \frac{1}{2} 
\langle \frac{1}{12} (1) + \frac{7}{2880} () \rangle_{d=1} = \frac{1}{24} + \frac{7}{5760}
\end{equation}
and we  find the announced result, given the definition of extended Hurwitz numbers (\ref{ExtendedHurwitz}) used for the second term. 

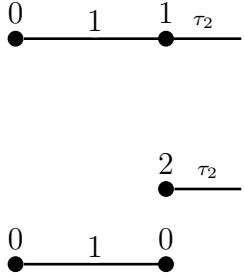
\begin{figure}[H]
\begin{tikzpicture}[
    node distance=1.5cm, % Adjust spacing between nodes
    every node/.style={fill=black, minimum size=6pt, circle, inner sep=1pt} % Style for nodes (small black dots)
]
\useasboundingbox (0,-5) rectangle (4,2);
    % Nodes
    \node[label=0] (A) at (0,0) {}; % Node A (small black dot)
    \node[label=1] (Ap) at (2,0) {};
 %   \node[fill=none,font=\scriptsize] (App) at (0,1) {$1$};
   % \node[label=1](C) at (2,2) {}; % Node C (small black dot, placed above B)

    % Edges
    \draw[-,line width=1pt] (A) -- +(2,0)   node[fill=white,midway,above] {1}; 
    \draw[-,line width=1pt] (2,0) -- (3,0) node[fill=none,midway,above,font=\scriptsize] {$ \tau_2$}; 
    % \draw[dashed,line width=1pt] (E) -- (F);
    % \draw[-,line width=1pt] (B) --  (D) node[fill=white,midway,above] {2}; 
    % \draw[dashed,line width=1pt] (G) -- (H);

 % Nodes
    \node[label=0] (B) at (0,-3) {}; % Node A (small black dot)
%    \node[fill=none,font=\scriptsize] (Bppp) at (0,-1.5) {$2$};
    \node[label=0] (Bp) at (2,-3) {};
    \node[label=2] (Bpp) at (2,-2) {}; % Node B (small black dot)
   % \node[label=1](C) at (2,2) {}; % Node C (small black dot, placed above B)

    % Edges
    \draw[-,line width=1pt] (B) -- +(2,0)   node[fill=white,midway,above] {1}; 
    \draw[-,line width=1pt] (Bpp) -- +(1,0) node[fill=none,midway,above,font=\scriptsize] {$ \tau_2$}; 
    % \draw[dashed,line width=1pt] (E) -- (F);
    % \draw[-,line width=1pt] (B) --  (D) node[fill=white,midway,above] {2}; 
    % \draw[dashed,line width=1pt] (G) -- (H);

\end{tikzpicture}
\caption{The two diagrams contributing to the $\tau_2(\omega)$ one-point function. The edge is of degree $1$ and the vertices are labelled by genera. A right hand vertex is marked by the $\tau_2$ insertion. }
\label{OnePointTau1}
\end{figure}

The second manner to compute the one-point function is in the Gromov-Witten theory directly. At degree one there is a single partition $\mu=(1)$ to take into account in the degeneration formula (\ref{DegenerationFormula}). The relevant localization diagrams are  identified in Figure \ref{OnePointTau1}. 
\noindent
Both diagrams have one edge of degree one. We attach the operator on the right hand side of the diagram.\footnote{The equivariant $\tau_k(\infty)$ class corresponds directly to the $\tau_k(\omega)$ class in the formalism of \cite{OkounkovPandharipande2}.} We then decorate the vertices with the genus assignments that will lead to a non-zero result -- there is only one such assignment. Note that the second diagram is also of total genus $1$ due to the disconnected sphere factor.  

The diagrams are easily evaluated immediately in this simple case, but as
 a preparation for more intricate calculations to come, we proceed in elementary steps. 
The left hand side of the diagrams are degree one vertices. They correspond to the moduli space integral calculation:
\begin{equation}
 H(1,\frac{u}{t}) = \frac{t^2}{u^2} \int_{0,1} \frac{1}{1-\psi}+\dots=\frac{t^2}{u^2}+\dots \, .
\end{equation}
Equivalently, we obtain through oscillator evaluation:
\begin{align}
H((1),\frac{u}{t}) &= (\frac{u}{t})^{-1} \langle { A} (1, \frac{u}{t}) \rangle
= (\frac{u}{t})^{-1} S(\frac{u}{t}) \frac{1}{\zeta(\frac{u}{t})} = \frac{t^2}{u^2} \, .
\end{align}
Thus, for the left part of the $(1)$ edge, we have obtained an exact result -- there are no higher genus corrections. We summarize the result for the left diagram in an intuitive notation (indicating the genus of the vertex and the degree of the edge):
\begin{equation}
L_{0,(1)} = \frac{t^2}{u^2} \, .
\end{equation}
The right parts of the diagram can either be made up of an edge of degree one with a $\tau_2$ mark attached, at genus one, or a disconnected $\tau_2$ mark on a vertex of genus two. The corresponding moduli space integrals are\footnote{The result for these integrals can be found in a number of ways. A systematic analytic method to compute all linear Hodge integrals is described in \cite{LinearHodge} based on much earlier work on moduli space integrals that goes back in the physics literature to the integrable hierarchy of two-dimensional topological gravity \cite{Witten:1989ig}. There is also a handy implementation in Sage \cite{AdmcyclesAuthors,AdmcyclesPaper}. We listed most of the moduli space integrals we use in Appendix \ref{ModuliSpaceIntegrals}.}
\begin{equation}
\int_{2,1} \psi_1^2 \lambda_2 = \frac{7}{5760} 
\, , \qquad \qquad \int_{1,2} \psi_1^2 = \frac{1}{24} \, .
\label{OnePointTauTwo}
\end{equation}
A $\tau_2$ insertion corresponds to the presence of the $\psi_1^2$ factor while the dimension of the moduli space dictates the term in the Hodge class $\Lambda$ that needs to appear in the integrand.
These numbers give rise to the final result for the right hand diagrams, with powers of $u$ and $t$ restored and an intuitive notation for genus, partition and marking:
\begin{align}
R_{1,(1),\tau_2} &= \frac{1}{24} t^3
&R_{{(1)}} &= \frac{t^2}{u^2}
& R_{2,\tau_2} &= \frac{u^2}{t^2} \frac{7}{5760} t^3 \, .
\end{align}
All remaining factors in the degeneration formula (\ref{DegenerationFormula}) are summarized in the symmetry factor for the $(1)$ partition:
\begin{equation}
S_{(1)} = u^2 t^{-5} \, .
\end{equation}
Combining the contributions of the symmetry factors and left and right half of the two full diagrams in Figure \ref{OnePointTau1}, we find:
\begin{align}
\langle \tau_2(\omega) \rangle_{g=1,d=1} &= u^2 t^{-5}  \times  \frac{t^2}{u^2} (\frac{1}{24} t^3+\frac{t^2}{u^2} \times \frac{u^2}{t^2} \frac{7}{5760} t^3)
\nonumber \\
&= \frac{1}{24} + \frac{7}{5760} \, .
\end{align}
The final result indeed has the trivial $u$ dependence that we expect from a genus one result.  The equivariant parameter $t$ dropped out.

 We can explicitly identify the surfaces that contribute to the localized sum. We have a connected surface of genus one that covers the sphere once. There is also a disconnected surface of genera two and zero respectively. The genus zero surface covers the sphere once while the genus two surface is shrunk to the marked point which resides at one of the fixed points of the torus action on $\mathbb{P}^1$.

\section{Localized World Sheets and the Two-Point Function}
\label{LocalizationTwoPoint}
\label{TwoPoint}
In this section, we study a second purely closed example of the powerful theorems of  \cite{OkounkovPandharipande1,OkounkovPandharipande2,OkounkovPandharipande3}. 
Previously, we argued that interesting two-point functions (\ref{TwoPointFunctions}) arise at genus one.  We will study the simplest non-trivial member of this set:
\begin{equation}
\langle \tau_1(\omega) \tau_3(\omega) \rangle_{d=2,g=1} \, .
\label{TwoPointTau1Tau3}
\end{equation}
It is non-zero because the Hurwitz operator  (\ref{CompletedCycles}) dual to $\tau_3(\omega)$ is completed. By working out the diagrams for the correlator, we can determine which world sheets are responsible for the completion.
\subsection{Initial Remarks on a  Two-Point Function}
Before delving into the calculation of the Gromov-Witten correlator proper, we prepare the grounds.
\subsubsection{The Field Theory Correlator}
Firstly, we evaluate the correlator on the Hurwitz field theory side.
We again use the duality map to  the completed cycles (\ref{CompletedCycles})
% \begin{align}
% \overline{(2)} &= (2)
% \nonumber \\
% \overline{(4)} &= (4) + 2 (2,1) + \frac{5}{4} (2) \, 
% \end{align}
which implies that the field theory correlator at degree $d=2$ equals:
\begin{equation}
\langle \tau_1(\omega) \tau_3(\omega) \rangle_{d=2} \overset{\text{GW/H}}{=} \frac{1}{1!} \frac{1}{3!} \langle (2) \frac{5}{4} (2) \rangle_{d=2} =  \frac{1}{2!} \frac{5}{24}  = \frac{5}{48} \, .
\end{equation}
The first equality is the Gromov-Witten/Hurwitz correspondence. 
In the resulting Hurwitz correlator, we used that in degree two, we can only accommodate the transposition terms in the completed cycles. We also  normalized the result of the Hurwitz correlator with a $1/d!$ factor.

\subsubsection{Motivation}
One reason to concentrate on this correlator is that the $\overline{(4)}$ completed cycle is the first one to have a non-trivial cycle as a completion contribution -- see equations (\ref{CompletedCycles}). At degree two and higher, the completion can  contribute to $\tau_3$ correlators. The two-point function therefore provides a first simple bulk correlator in which we see the effect of a non-trivial completion of the single cycle operators in the boundary theory.
We turn to the calculation of the same correlator (\ref{TwoPointTau1Tau3}) on the Gromov-Witten  side.
\subsubsection{Disconnected and Connected Correlators}
We make a preliminary structural remark on the Gromov-Witten calculation. We will be calculating a fully disconnected Gromov-Witten correlator $G$ at a given total degree. It is defined as the exponential of connected Gromov-Witten correlators $G^\circ$. The relation is provided by:
\begin{align}
 G (z) % &= \exp G^{\circ} (z_i)
% \nonumber \\
%  &= \exp \sum_{d=0}^\infty G_d^\circ
%  \nonumber \\
  &= \exp \sum_{d=0}^\infty \sum_{n=0}^\infty q^d G_d^{\circ} (z_1,\dots,z_n)
%  \nonumber \\
%  &= \sum_{d_{\text{tot}}} \sum_{n_{tot}} G_{d_\text{tot}} (z_i) 
\, ,
 \end{align}
 and one uses the vanishing of all zero-point functions except at degree one. 
For example, the one-point disconnected correlator $G_2(z_1)$ at total degree $d=2$ equals \cite{OkounkovPandharipande2}:
\begin{equation}
G_2(z_1) = \frac{1}{2} G^\circ_1()^2 G^\circ_0(z_1) + G^\circ_1() G^\circ_1(z_1) + G^\circ_2(z_1)  \, .
\end{equation}
For two-point functions at degree $1$, we find:
\begin{equation}
G_1(z_1,z_2) = G_1^\circ(z_1,z_2) + G^{\circ}_1(z_1) G^{\circ}_0(z_2) + G^{\circ}_1(z_2) G^{\circ}_0(z_1)
 + 
G_1^{\circ}() G_0^{\circ}(z_1) G_0^{\circ}(z_2) 
+G_0^\circ(z_1,z_2) G_1^\circ() \, . \nonumber
\end{equation}
Finally, the most relevant to the calculation at hand is the two-point disconnected correlator at degree $2$ in terms of its connected counterparts:
 \begin{align}
 G_2(z_1,z_2) &=
 \frac{1}{2}  { {G^\circ}_1}^2  { G^\circ}_0 (z_1)  { G^\circ}_0 (z_2)+\frac{1}{2}  { {G^\circ}_1}^2  { G^\circ}_0( {z_1}, {z_2})
  + { G^\circ}_1  { G^\circ}_1( {z_1})  { G^\circ}_0 (z_2)
 \nonumber \\
  & + { G^\circ}_1  { G^\circ}_0 (z_1)  { G^\circ}_1( {z_2})
  + { G^\circ}_1  { G^\circ}_1( {z_1}, {z_2})
  + { G^\circ}_1( {z_1})  { G^\circ}_1( {z_2})
 \nonumber \\
  & + { G^\circ}_2( {z_1})  { G^\circ}_0 (z_2)
  + { G^\circ}_0 (z_1)  { G^\circ}_2( {z_2})
  + { G^\circ}_2( {z_1}, {z_2}) \, .
  \end{align}
  Thus, we can organize the calculation of the disconnected two-point functions in terms of connected zero, one and two-point functions  at degrees zero, one and two. 

  Importantly, for the calculation at hand, we have that the non-equivariant one-point functions of $\tau_k(\omega)$ operators at  odd $k$ are automatically zero because of the charge constraint (\ref{ChargeConstraint}). 
  As a consequence, for the non-equivariant correlators (indicated with an upper index $\text{n}$), one has:
 \begin{align}
 \prescript{{\text{n}}}{} G(z_1,z_2) &=\frac{1}{2} 
 (\prescript{{\text{n}}}{}
 {G^\circ_1})^2
 \,
 \prescript{{\text{n}}}{} {G^\circ_0}(z_1,z_2)
  +\prescript{{\text{n}}}{}G^\circ_1
  \, \prescript{{\text{n}}}{}G^\circ_1(z_1,z_2)
  +\prescript{{\text{n}}}{}G^\circ_2(z_1,z_2) \, .
  \end{align}
  We will not exploit this structure in the following calculation, but will use it as an a posteriori check. 
  Moreover, if we glimpse over the fence of the correspondence, from the Hurwitz completed cycle formulas,
  we surmise that the non-equivariant connected two-point functions of degree one and zero will also be zero such that only the last term will be important. That observation too, we shall check independently.
\label{Predictions}

\subsubsection{The Diagrams and the Techniques}
\begin{figure}

\begin{tikzpicture}[
  scale=1,
  vertex/.style={circle, fill=black, inner sep=1.5pt}
]

% --- Diagram 1 ---

\node[vertex] (d1) at (0,0) {};
\node[vertex] (d1p) at ($(d1) + (2,0)$) {};

\draw (d1) -- +(2,0);

\node[vertex] (e) at ($(d1) + (2,1)$) {};
\draw (e) -- +(1,0);

\node[vertex] (f) at ($(d1) + (2,2)$) {};
\draw (f) -- +(1,0);

% Diagram 2

\node[vertex] (d2) at ($(d1) + (5,0)$) {};
\node[vertex] (d2p) at ($(d2) + (2,0)$) {};
\node[vertex] (d2pp) at ($(d2) + (2,1)$) {};
\draw (d2) -- +(2,0);
\draw (d2p) -- +(1,0) node[fill=none,midway,above] { ${ \scriptstyle \tau_1}$};
\draw (d2pp) -- +(1,0) node[fill=none,midway,above] { ${ \scriptstyle \tau_3}$};

% Diagram 3 

\node[vertex] (d3) at ($(d1) + (10,0)$) {};
\node[vertex] (d3p) at ($(d3) + (2,0)$) {};
\node[vertex] (d3pp) at ($(d3) + (2,1)$) {};
\draw (d3) -- +(2,0);
\draw (d3p) -- +(1,0) node[fill=none,midway,above] { ${ \scriptstyle \tau_3}$};
\draw (d3pp) -- +(1,0) node[fill=none,midway,above] { ${ \scriptstyle \tau_1}$};

% Diagram 4

\node[vertex] (d4) at ($(d1) + (0,-3)$) {};
\node[vertex] (d4p) at ($(d4) + (2,0)$) {};

\draw (d4) -- +(2,0);
\draw (d4p) -- +(1,1);
\draw (d4p) -- +(1,-1);

% --- Diagram 5 ---

\node[vertex] (d5) at ($(d1) + (5,-4)$) {};
\node[vertex] (d5p) at ($(d5) + (2,0)$) {};

\draw (d5) -- +(2,0);

\node[vertex] (d5pp) at ($(d5) + (2,1.5)$) {};
\draw (d5pp) -- +(1,0);

\node[vertex] (d5ppp) at ($(d5) + (2,2)$) {};
\draw (d5ppp) -- +(1,0);

\node[vertex] (d5pppp) at ($(d5) + (0,1)$) {};
\node[vertex] (d5ppppp) at ($(d5) + (2,1)$) {};
\draw (d5pppp) -- +(2,0);

% --- Diagram 6 ---

\node[vertex] (d6) at ($(d1) + (10,-4)$) {};
%\node[fill=none] at ($(d6)+(0,1.7)$)  {$D_{6}$};
\node[vertex] (d6p) at ($(d6) + (2,0)$) {};

\draw (d6) -- +(2,0);

\node[vertex] (d6pp) at ($(d6) + (2,1)$) {};
\draw (d6pp) -- +(1,0) node[fill=none,midway,above] { ${ \scriptstyle \tau_1}$};

\node[vertex] (d6ppp) at ($(d6) + (2,2)$) {};
\draw (d6ppp) -- +(1,0) node[fill=none,midway,above] { ${ \scriptstyle \tau_3}$};

\node[vertex] (d6pppp) at ($(d6) + (0,1)$) {};
\node[vertex] (d6ppppp) at ($(d6) + (2,1)$) {};
\draw (d6pppp) -- +(2,0);

% Diagram 7 

\node[vertex] (d7) at ($(d1) + (0,-8)$) {};
\node[vertex] (d7p) at ($(d7) + (2,0)$) {};

\draw (d7) -- +(2,0) ;

\node[vertex] (d7pp) at ($(d7) + (2,1)$) {};
\draw (d7pp) -- +(1,0) node[fill=none,midway,above] { ${ \scriptstyle \tau_3}$};

\node[vertex] (d7ppp) at ($(d7) + (2,2)$) {};
\draw (d7ppp) -- +(1,0) node[fill=none,midway,above] { ${ \scriptstyle \tau_1}$};

\node[vertex] (d7pppp) at ($(d7) + (0,1)$) {};
\node[vertex] (d7ppppp) at ($(d7) + (2,1)$) {};
\draw (d7pppp) -- +(2,0);

% Diagram 8

\node[vertex] (d8) at ($(d1) + (5,-7)$) {};
\node[vertex] (d8p) at ($(d8) + (2,0)$) {};
\node[vertex] (d8pp) at ($(d8) + (0,-1)$) {};
\node[vertex] (d8ppp) at ($(d8) + (2,-1)$) {};

\draw (d8) -- +(2,0);
\draw (d8p) -- +(1,1);
\draw (d8p) -- +(1,-1);
\draw (d8pp) -- +(2,0);

% --- Diagram 9 ---

\node[vertex] (d9) at ($(d1) + (10,-8)$) {};
\node[vertex] (d9p) at ($(d9) + (2,0)$) {};

\draw (d9) -- +(2,0);

\node[vertex] (d9pp) at ($(d9) + (2,1)$) {};
\draw (d9pp) -- +(1,0);

\node[vertex] (d9pppp) at ($(d9) + (0,1)$) {};
\draw (d9pppp) -- +(2,0);
\draw (d9p) -- +(1,0);

% --- Diagram 10 ---

\node[vertex] (d10) at ($(d1) + (0,-12)$) {};
\node[vertex] (d10p) at ($(d10) + (2,0.5)$) {};
\node[vertex] (d10pp) at ($(d10) + (2,-0.5)$) {};
\node[vertex] (d10ppp) at ($(d10) + (2,1)$) {};
\node[vertex] (d10pppp) at ($(d10) + (2,1.5)$) {};

\draw (d10) -- (d10p);
\draw (d10) -- (d10pp);
\draw (d10ppp) -- +(1,0);
\draw (d10pppp) -- +(1,0);

% --- Diagram 11 ---

\node[vertex] (d11) at ($(d1) + (5,-12)$) {};
%\node[fill=none] at ($(d11)+(0,1.7)$)  {$D_{11}$};
\node[vertex] (d11p) at ($(d11) + (2,0.5)$) {};
\node[vertex] (d11pp) at ($(d11) + (2,-0.5)$) {};
\node[vertex] (d11ppp) at ($(d11) + (2,0.5)$) {};
\node[vertex] (d11pppp) at ($(d11) + (2,1)$) {};

\draw (d11) -- (d11p);
\draw (d11) -- (d11pp);
\draw (d11ppp) -- +(1,0) node[fill=none,midway,above] { ${ \scriptstyle \tau_1}$};
\draw (d11pppp) -- +(1,0) node[fill=none,midway,above] { ${ \scriptstyle \tau_3}$};

% --- Diagram 12 ---

\node[vertex] (d12) at ($(d1) + (9,-12)$) {};
\node[vertex] (d12p) at ($(d12) + (2,0.5)$) {};
\node[vertex] (d12pp) at ($(d12) + (2,-0.5)$) {};
\node[vertex] (d12ppp) at ($(d12) + (2,0.5)$) {};
\node[vertex] (d12pppp) at ($(d12) + (2,1)$) {};

\draw (d12) -- (d12p);
\draw (d12) -- (d12pp);
\draw (d12ppp) -- +(1,0) node[fill=none,midway,above] { ${ \scriptstyle \tau_3}$};
\draw (d12pppp) -- +(1,0) node[fill=none,midway,above] { ${ \scriptstyle \tau_1}$};

% --- Diagram 13 ---

\node[vertex] (d13) at ($(d1) + (0,-16)$) {};
\node[vertex] (d13p) at ($(d13) + (2,0.5)$) {};
\node[vertex] (d13pp) at ($(d13) + (2,-0.5)$) {};

\draw (d13) -- (d13p);
\draw (d13) -- (d13pp);
\draw (d13p) -- +(1,0.5);
\draw (d13p) -- +(1,-0.5);

% --- Diagram 14 ---

\node[vertex] (d14) at ($(d1) + (5,-16)$) {};
\node[vertex] (d14p) at ($(d14) + (2,0.5)$) {};
\node[vertex] (d14pp) at ($(d14) + (2,-0.5)$) {};

\draw (d14) -- (d14p);
\draw (d14) -- (d14pp);
\draw (d14p) -- +(1,0);
\draw (d14pp) -- +(1,0);

% --- Diagram 15 ---

\node[vertex] (d15) at ($(d1) + (10,-16)$) {};
\node[vertex] (d15p) at ($(d15) + (2,-0.5)$) {};
\node[vertex] (d15pp) at ($(d15) + (0,-1)$) {};
\node[vertex] (d15ppp) at ($(d15)+(2,0.5)$) {};
\node[vertex] (d15pppp) at ($(d15)+(2,1)$) {};

\draw (d15) -- (d15p);
\draw (d15p) -- (d15pp);
\draw (d15ppp) -- +(1,0);
\draw (d15pppp) -- +(1,0);

\node[fill=none,font=\scriptsize] at ($(d1)+(0,1.7)$)  {${1}$};
\node[fill=none,font=\scriptsize] at ($(d2)+(0,1.7)$)  {${ 2}$};
\node[fill=none,font=\scriptsize] at ($(d3)+(0,1.7)$)  {${ 3}$};
\node[fill=none,font=\scriptsize] at ($(d4)+(0,1.2)$)  {${ 4}$};
\node[fill=none,font=\scriptsize] at ($(d5)+(0,2)$)  {${5}$};
\node[fill=none,font=\scriptsize] at ($(d6)+(0,2)$)  {${ 6}$};
\node[fill=none,font=\scriptsize] at ($(d7)+(0,2.1)$)  {${ 7}$};
\node[fill=none,font=\scriptsize] at ($(d8)+(0,1.0)$)  {${ 8}$};
\node[fill=none,font=\scriptsize] at ($(d9)+(0,2)$)  {${9}$};
\node[fill=none,font=\scriptsize] at ($(d10)+(0,1.7)$)  {${ 10}$};
\node[fill=none,font=\scriptsize] at ($(d11)+(0,1.7)$)  {${ 11}$};
\node[fill=none,font=\scriptsize] at ($(d12)+(0,1.7)$)  {${ 12}$};
\node[fill=none,font=\scriptsize] at ($(d13)+(0,1.7)$)  {${13}$};
\node[fill=none,font=\scriptsize] at ($(d14)+(0,1.7)$)  {${ 14}$};
\node[fill=none,font=\scriptsize] at ($(d15)+(0,1.7)$)  {${ 15}$};

\end{tikzpicture}

\caption{The first set of 15 diagrams contributing to the two-point function.}
\label{Diagrams1}

\end{figure}
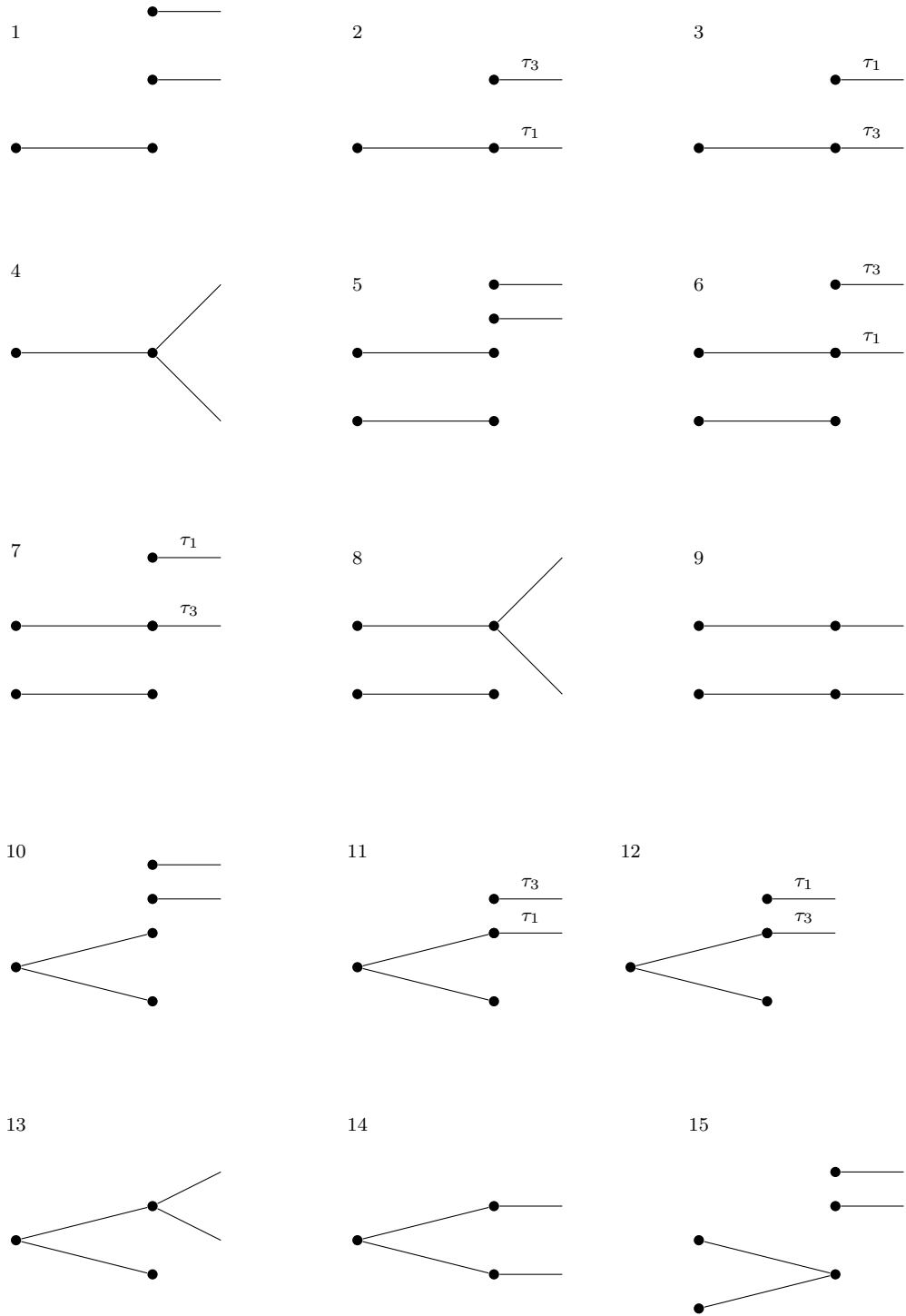

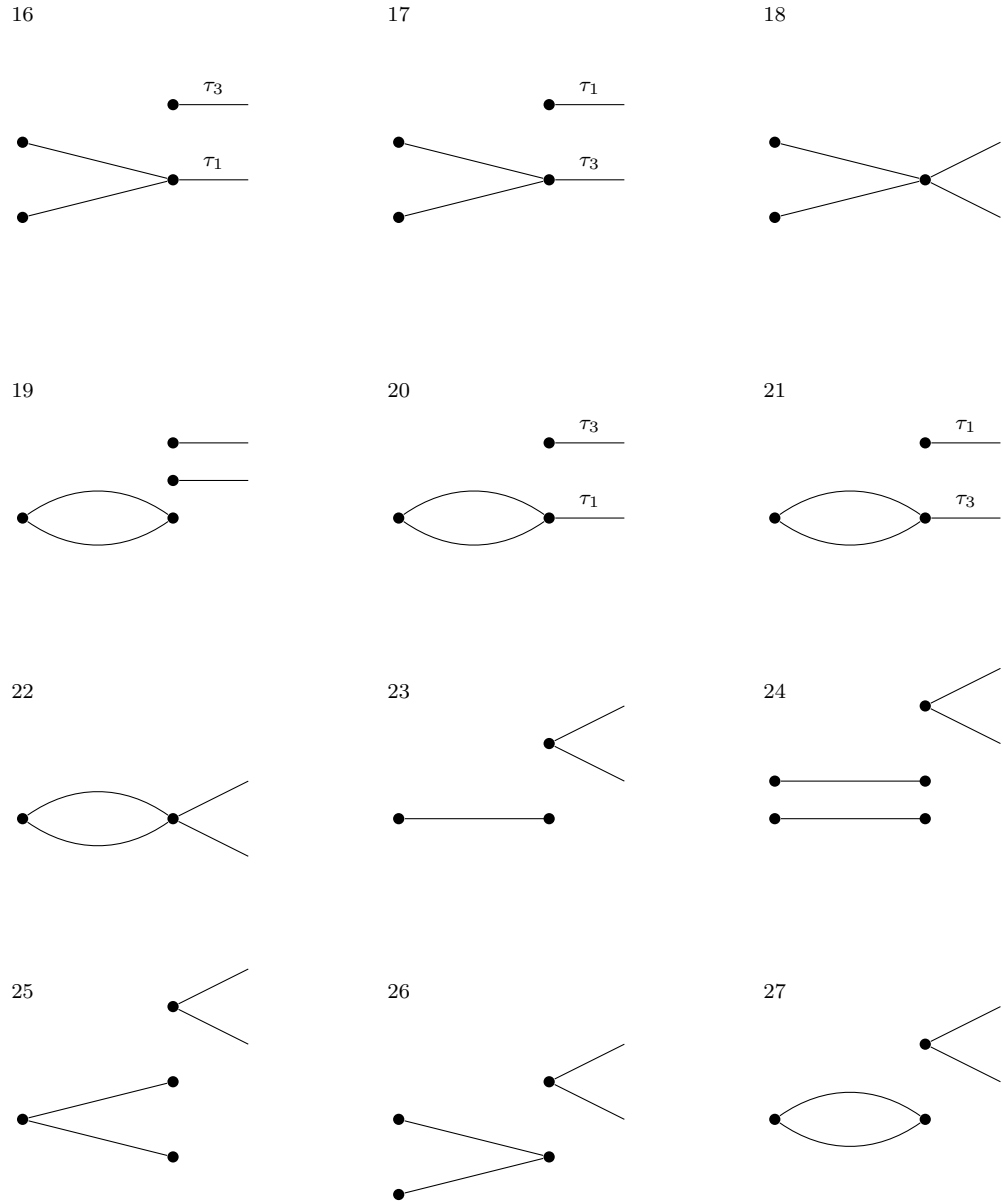
\begin{figure}

\begin{tikzpicture}[
  scale=1,
  vertex/.style={circle, fill=black, inner sep=1.5pt}
]

\node[fill=none] (d1) at (0,0) {};

% --- Diagram 16 ---

\node[vertex] (d16) at (0,0) {};
\node[fill=none,font=\scriptsize] at ($(d16)+(0,1.7)$)  {${ 16}$};
\node[vertex] (d16p) at ($(d16) + (2,-0.5)$) {};
\node[vertex] (d16pp) at ($(d16) + (0,-1)$) {};
\node[vertex] (d16ppp) at ($(d16)+(2,0.5)$) {};
%\node[vertex] (d16pppp) at ($(d16)+(2,1)$) {};

\draw (d16) -- (d16p);
\draw (d16p) -- (d16pp);
\draw (d16ppp) -- +(1,0) node[fill=none,midway,above] { ${ \scriptstyle \tau_3}$};
\draw (d16p) -- +(1,0) node[fill=none,midway,above] { ${ \scriptstyle \tau_1}$};
%\draw (d16pppp) -- +(1,0);

% --- Diagram 17 ---

\node[vertex] (d17) at ($(d16) + (5,0)$) {};
\node[fill=none,font=\scriptsize] at ($(d17)+(0,1.7)$)  {${ 17}$};
\node[vertex] (d17p) at ($(d17) + (2,-0.5)$) {};
\node[vertex] (d17pp) at ($(d17) + (0,-1)$) {};
\node[vertex] (d17ppp) at ($(d17)+(2,0.5)$) {};
%\node[vertex] (d17pppp) at ($(d17)+(2,1)$) {};

\draw (d17) -- (d17p);
\draw (d17p) -- (d17pp);
\draw (d17ppp) -- +(1,0) node[fill=none,midway,above] { ${ \scriptstyle \tau_1}$};
\draw (d17p) -- +(1,0) node[fill=none,midway,above] { ${ \scriptstyle \tau_3}$};
%\draw (d17pppp) -- +(1,0);

% --- Diagram 18 ---

\node[vertex] (d18) at ($(d16) + (10,0)$) {};
\node[fill=none,font=\scriptsize] at ($(d18)+(0,1.7)$)  {${ 18}$};
\node[vertex] (d18p) at ($(d18) + (2,-0.5)$) {};
\node[vertex] (d18pp) at ($(d18) + (0,-1)$) {};
%\node[vertex] (d18ppp) at ($(d18)+(2,0.5)$) {};
%\node[vertex] (d18pppp) at ($(d18)+(2,1)$) {};

\draw (d18) -- (d18p);
\draw (d18p) -- (d18pp);
\draw (d18p) -- +(1,0.5);
\draw (d18p) -- +(1,-0.5);
%\draw (d18pppp) -- +(1,0);

% --- Diagram 19 ---

\node[vertex] (d19) at ($(0,-5)$) {};
\node[vertex] (d19p) at ($(d19) + (2,0)$) {};
%\node[vertex] (d19pp) at ($(d19) + (0,-1)$) {};
\node[vertex] (d19ppp) at ($(d19)+(2,0.5)$) {};
\node[vertex] (d19pppp) at ($(d19)+(2,1)$) {};

\draw (d19) to[out=35, in=145] (d19p);

\draw (d19) to[out=-35, in=-145](d19p);
\draw (d19ppp) -- +(1,0);
\draw (d19pppp) -- +(1,0);

% --- Diagram 20 ---

\node[vertex] (d20) at ($(d19) + (5,0)$) {};
\node[vertex] (d20p) at ($(d20) + (2,0)$) {};
%\node[vertex] (d20pp) at ($(d20) + (0,-1)$) {};
%\node[vertex] (d20ppp) at ($(d20)+(2,0.5)$) {};
\node[vertex] (d20pppp) at ($(d20)+(2,1)$) {};

\draw (d20) to[out=35, in=145] (d20p);

\draw (d20) to[out=-35, in=-145](d20p);
\draw (d20p) -- +(1,0) node[fill=none,midway,above] { ${ \scriptstyle \tau_1}$};
\draw (d20pppp) -- +(1,0) node[fill=none,midway,above] { ${ \scriptstyle \tau_3}$};

% --- Diagram 21 ---

\node[vertex] (d21) at ($(d19)+(10,0)$) {};
\node[vertex] (d21p) at ($(d21) + (2,0)$) {};
%\node[vertex] (d21pp) at ($(d21) + (0,-1)$) {};
%\node[vertex] (d21ppp) at ($(d21)+(2,0.5)$) {};
\node[vertex] (d21pppp) at ($(d21)+(2,1)$) {};

\draw (d21) to[out=35, in=145] (d21p);

\draw (d21) to[out=-35, in=-145](d21p);
\draw (d21p) -- +(1,0) node[fill=none,midway,above] { ${ \scriptstyle \tau_3}$};
\draw (d21pppp) -- +(1,0) node[fill=none,midway,above] { ${ \scriptstyle \tau_1}$};

% --- Diagram 22 ---

\node[vertex] (d22) at ($(d19) + (0,-4)$) {};
\node[vertex] (d22p) at ($(d22) + (2,0)$) {};
%\node[vertex] (d22pp) at ($(d22) + (0,-1)$) {};
%\node[vertex] (d22ppp) at ($(d22)+(2,0.5)$) {};
%\node[vertex] (d22pppp) at ($(d22)+(2,1)$) {};

\draw (d22) to[out=35, in=145] (d22p);

\draw (d22) to[out=-35, in=-145](d22p);
\draw (d22p) -- +(1,0.5);
\draw (d22p) -- +(1,-0.5);

% --- Diagram 23 ---

\node[vertex] (d23) at ($(d19) + (5,-4)$) {};
\node[vertex] (d23p) at ($(d23) + (2,0)$) {};

\draw (d23) -- +(2,0);

\node[vertex] (e) at ($(d23) + (2,1)$) {};
\draw (e) -- +(1,0.5);
\draw (e) -- +(1,-0.5);

% --- Diagram 24 ---

\node[vertex] (d24) at ($(d19) + (10,-4)$) {};
%\node[fill=none] at ($(d24)+(0,1.7)$)  {$D_{24}$};
\node[vertex] (d24p) at ($(d24) + (2,0)$) {};

\draw (d24) -- +(2,0);

\node[vertex] (d24pp) at ($(d24) + (2,1.5)$) {};
\draw (d24pp) -- +(1,0.5);
\draw (d24pp) -- +(1,-0.5);

%\node[vertex] (d24ppp) at ($(d24) + (2,2)$) {};
%\draw (d24ppp) -- +(1,0);

\node[vertex] (d24pppp) at ($(d24) + (0,0.5)$) {};
\node[vertex] (d24ppppp) at ($(d24) + (2,0.5)$) {};
\draw (d24pppp) -- +(2,0);

% --- Diagram 25 ---

\node[vertex] (d25) at ($(d19) + (0,-8)$) {};
%\node[fill=none] at ($(d25)+(0,1.7)$)  {$D_{25}$};
\node[vertex] (d25p) at ($(d25) + (2,0.5)$) {};
\node[vertex] (d25pp) at ($(d25) + (2,-0.5)$) {};
\node[vertex] (d25ppp) at ($(d25) + (2,1.5)$) {};
%\node[vertex] (d25pppp) at ($(d25) + (2,1.5)$) {};

\draw (d25) -- (d25p);
\draw (d25) -- (d25pp);
\draw (d25ppp) -- +(1,0.5) %node[fill=none,midway,above] { ${ \scriptstyle \tau_1}$}
;
\draw (d25ppp) -- +(1,-0.5) 
%node[fill=none,midway,below] { ${ \scriptstyle \tau_3}$}
;

% --- Diagram 26 ---

\node[vertex] (d26) at ($(d19) + (5,-8)$) {};
%\node[fill=none] at ($(d26)+(0,1.7)$)  {$D_{26}$};
\node[vertex] (d26p) at ($(d26) + (2,-0.5)$) {};
\node[vertex] (d26pp) at ($(d26) + (0,-1)$) {};
\node[vertex] (d26ppp) at ($(d26)+(2,0.5)$) {};
%\node[vertex] (d26pppp) at ($(d26)+(2,1)$) {};

\draw (d26) -- (d26p);
\draw (d26p) -- (d26pp);
\draw (d26ppp) -- +(1,0.5) %node[fill=none,midway,above] { ${ \scriptstyle \tau_1}$}
;
\draw (d26ppp) -- +(1,-0.5) %node[fill=none,midway,below] { ${ \scriptstyle \tau_3}$}
;

%\draw (d26p) -- +(1,0);
%\draw (d26pppp) -- +(1,0);

% --- Diagram 27 ---

\node[vertex] (d27) at ($(d19) + (10,-8)$) {};

%\node[fill=none] at ($(d27)+(0,1.7)$)  {$D_{27}$};

\node[vertex] (d27p) at ($(d27) + (2,0)$) {};
%\node[vertex] (d27pp) at ($(d27) + (0,-1)$) {};
%\node[vertex] (d27ppp) at ($(d27)+(2,0.5)$) {};
\node[vertex] (d27pppp) at ($(d27)+(2,1)$) {};

\draw (d27) to[out=35, in=145] (d27p);

\draw (d27) to[out=-35, in=-145](d27p);
\draw (d27pppp) -- +(1,0.5) %node[fill=none,midway,above] { ${ \scriptstyle \tau_1}$}
;
\draw (d27pppp) -- +(1,-0.5) %node[fill=none,midway,below] { ${ \scriptstyle \tau_3}$}
;
\node[fill=none,font=\scriptsize] at ($(d19)+(0,1.7)$)  {${19}$};
\node[fill=none,font=\scriptsize] at ($(d20)+(0,1.7)$)  {${20}$};
\node[fill=none,font=\scriptsize] at ($(d21)+(0,1.7)$)  {${ 21}$};
\node[fill=none,font=\scriptsize] at ($(d22)+(0,1.7)$)  {${ 22}$};
\node[fill=none,font=\scriptsize] at ($(d23)+(0,1.7)$)  {${ 23}$};
\node[fill=none,font=\scriptsize] at ($(d24)+(0,1.7)$)  {${24}$};
\node[fill=none,font=\scriptsize] at ($(d25)+(0,1.7)$)  {${ 25}$};
\node[fill=none,font=\scriptsize] at ($(d26)+(0,1.7)$)  {${ 26}$};
\node[fill=none,font=\scriptsize] at ($(d27)+(0,1.7)$)  {${ 27}$};

\end{tikzpicture}

\caption{The remaining set of 12 diagrams contributing to the two-point function.}
\label{Diagrams2}

\end{figure}

We have a first look at the relevant diagrams and the techniques necessary to compute them.
At degree two, there are two partitions $\mu$ that contribute in the degeneration formula (\ref{DegenerationFormula}), namely $\mu=(2)$ and $\mu=(1,1)$. The two markings are situated on the right of the diagram once more. 
Thus, on the left side of the diagram, we either have a degree $2$ edge or two degree $1$ edges. The latter   are either emanating from the same vertex on the left or two different vertices. 
%We have necessarily one or two vertices on the left for $\mu=(1,1)$ and one for $\mu=2$. 
No other vertices will arise on the left because there are no marks.
%The $(1,1)$ edges can be attached to the same vertex or not. 
On the right hand side, the same description for the edges hold. Full diagrams are obtained by matching edges in the middle and will therefore combine left partition  $\mu$ diagrams with right partition $\mu$ diagrams. On the right, we have to decorate vertices of edges or new vertices with the two distinct insertions $\tau_1$ and $\tau_3$.  All possible 27 diagrams are shown in Figures \ref{Diagrams1} and \ref{Diagrams2}. 
Various superfluous labels have been left unspecified. For instance, if there is a single edge (necessarily between two vertices indicated as nodes), it is of degree $2$ while if there are two, then they are of degree $1$. If there is a symmetry between the external markings, then we have not labeled them. If there is not, then we have drawn the two diagrams with alternative labelings $(\tau_{1},\tau_3)$ or $(\tau_{3},\tau_1)$. 

We still need to specify the genus of every vertex in order to complete our description of the diagrams. 
Most vertices come with only one possibility for the ascribed genera. For instance, an isolated vertex with a $\tau_1$ marking will necessarily have genus one (as we will confirm shortly) while a vertex with only a $\tau_3$ marking will necessarily have genus two. Further reasonings on the total genus of the diagram and various vanishing results will determine all genus assignments for all vertices.   %This is because at lower genus, there is no contribution proportional to $z_1^2$ and higher genus contributions will be zero. We have the integral 
For example, diagram $D_1$ has genera $1$ and $2$ on the marked vertices and must therefore have genera $0$ on the other vertices. After all, the constraint equation imposes that the total genus is one. Let us continue to study diagram $D_1$ in a little more detail. The $\tau_1$ marking will have  a factor equal to the modular integral:
\begin{align}
\int_{1,1} \psi_1 &= \frac{1}{24}
% \nonumber \\
% \int_{1,1} \psi_1^3 &= 0 
\, , \label{OnePointTauOne}
\end{align}
while the isolated $\tau_3$ mark contributes a factor
\begin{align}
-\int_{2,1} \psi_1^3 \lambda_1 &= -\frac{1}{480}.
\label{OnePointTauThree}
\end{align}
Note again that the power of the tangent class $\psi_1$ corresponding to $\tau_3$ is set by the index $k$ of the operator $\tau_k$, by definition. The choice of edge and Hodge classes covers all possibilities given the dimension of the compactified moduli space $\overline{\cal M}_{2,1}$ at hand.  We thus supplemented the $\psi_1^3$ integrand with the appropriate Hodge class $-\lambda_1$ (since no  edge classes are available). 
 We are left with a left and a right vertex at genus zero with an incoming degree two edge. These will each contribute a factor:
\begin{equation}
 \int_{0,1} \frac{2}{1-2 \psi_e}= \frac{1}{2}  \, ,
\end{equation}
where we made use of a regularization of unstable moduli space integrals \cite{OkounkovPandharipande1}.\footnote{See  Appendix \ref{ModuliSpaceIntegrals} once more.} 
After taking into account an overall symmetry factor of $-2$ (up to powers of $(u,t)$ which all cancel in the end), we find:
\begin{equation}
D_1 = -2 \times \frac{1}{2} \times \left( \frac{1}{2} \times \frac{1}{24} \times (- \frac{1}{480}) \right)
=  \frac{1}{23040}
\end{equation}
where the first factor is a symmetry factor, the second the left edge, the third the right edge, and the following factors give the contribution of the isolated markings. The symmetry factor and the power counting will be done in more detail shortly. We have determined the value of the  diagram $D_1$.   

While most techniques we use to evaluate the diagrams are standard, we also further use regularized unstable moduli space integrals that may be less familiar. Firstly, a degree one single edge at genus zero contributes:
\begin{equation}
\int_{0,1} \frac{1}{1-\psi}  = 1
\end{equation}
while two joint degree one edges at genus zero give rise to:
\begin{equation}
\int_{0,2} \frac{1}{1-\psi_1} \frac{1}{1-\psi_2} = \frac{1}{2} \, .
\end{equation}
Other unstable contributions to keep in mind arise when we attach a $\tau_1$ or $\tau_3$ marking to a genus zero vertex with an edge. For instance, if we have a  degree $1$ edge  and a  $\tau_{1}$ or $\tau_3$  marking at genus zero, we must evaluate the integral:
\begin{equation}
t z_1 \int_{0,2} \frac{1}{1-t z_1 \psi_1} \frac{1}{1-\psi_2} = \frac{t z_1}{1+tz_1} \approx t z_1 - t^2 z_1^2 + t^3 z_1^3 - t^4 z_1^4 + \dots
\label{SphereOneEdgeOneMarking}
\end{equation}
This will give rise to a contribution proportional to $-t^2$ for the operator $\tau_1$ corresponding to the coefficient of $z_1^2$ and $-t^4$ for $\tau_3$. For a degree $2$ edge, a similar calculation must be performed. 

Moreover, we note that if the total 
genus of a diagram is lower than the expected non-equivariant genus,\footnote{The  equivariant theory  can absorb the charge in the equivariant parameter and can therefore have contributions at other genera.} then the result will carry extra strictly positive powers of $t^2 u^{-2}$ (since $u$ is the genus counting parameter and it enters in this combination). The result then vanishes in the non-equivariant limit and we disregard these diagrams.\footnote{An example will be provided by diagram $D_5$.}  When the genus of a contribution is higher, the partial result must conspire with other diagrams to obtain a vanishing  total coefficient. Indeed, the limit $t \rightarrow 0$ is well-defined. 
%(Diagram $D_4$ can be used to illustrate this case.)
These initial remarks equip the reader with  basic techniques. 

Finally, while the  diagrams are straightforwardly calculable through moduli space integrals alone,  we found it very useful in practice to have a check on the calculation through the oscillator approach recalled in subsection \ref{OscillatorApproach}. While the oscillator approach allows for a more rapid calculation, the localization diagrams are closer to a geometric world sheet intuition. These complementary characteristics motivate us to work out both these perspectives on the correlator at hand.\footnote{At this junction it is possible to skip the pedagogical buildup and jump ahead to either subsection \ref{ValuesHalfDiagrams} or subsection \ref{TheDiagrams} to perform the final calculations right away.}

\subsection{The Symmetry Factors and the Left Hand Diagrams}
As we have seen, the calculation of the correlator on the Gromov-Witten side is based on the degeneration  (\ref{DegenerationFormula}) which splits the computation into the evaluation of left and right relative (generalized) Hurwitz number calculations combined with a symmetry gluing factor. At degree $d=2$ we sum over intermediate partitions $\mu=(1,1)$ or $(2)$. 
\subsubsection{The Symmetry Factors}
We shall gather all the factors that are not in the left and right Hurwitz numbers $H$  in terms of a symmetry factor $S_\mu$ associated to the partition $\mu$.   From the degeneration formula (\ref{DegenerationFormula}) we deduce that the symmetry factor for the partition $\mu=(2)$ equals:
 \begin{equation}
 S_{(2)} = -2 \frac{u^2}{t^{8}} \, ,
 \label{SymmetryFactor2}
 \end{equation} 
 while for the partition $\mu=(1,1)$ we have:
 \begin{equation}
 S_{(1,1)} = \frac{1}{2} \frac{u^4}{t^{10}}
 \, . \label{SymmetryFactor11}
 \end{equation}
 The first will enter all diagrams with a single edge and the second all diagrams with two edges. 
This takes care of all symmetry factors. The rest of the computation is a calculation of (disconnected)  numbers $H$.

\subsubsection{The Equivariant Zero-point Functions - the Left Hand Diagrams}
The linear Hodge integrals $H$  are half-diagrams. Imagine a vertical (dotted) line cutting through all the edges. We concentrate on either the left or the right hand side of the diagram separately.
For starters, we consider the left hand side of our two-point function diagrams, which have zero external insertions. Then, up to genus one,  the left hand diagrams are of the types drawn in Figures \ref{LeftHandDegreeTwo} and \ref{LeftHandDegreesOneOne}.
\begin{figure}[H]

\begin{tikzpicture}[
    node distance=1.5cm, % Adjust spacing between nodes
    every node/.style={fill=black, minimum size=6pt, circle, inner sep=1pt} % Style for nodes (small black dots)
]
\useasboundingbox (0,-3) rectangle (4,2);
    % Nodes
    \node[label=0] (A) at (0,0) {}; % Node A (small black dot)
    \node[label=1] (B) at (0,-2) {}; % Node B (small black dot)
   % \node[label=1](C) at (2,2) {}; % Node C (small black dot, placed above B)
   \node[fill=none] (C) at (2,0) {};
   \node[fill=none] (D) at (2,-2) {};

    \node[fill=none] (E) at (2,0.5) {};
    \node[fill=none] (F) at (2,-0.5) {};
     \node[fill=none] (G) at (2,-1.5) {};
    \node[fill=none] (H) at (2,-2.5) {};

    % Edges
    \draw[-,line width=1pt] (A) --  (C) node[fill=white,midway,above] {2}; 
    \draw[dashed,line width=1pt] (E) -- (F);
    \draw[-,line width=1pt] (B) --  (D) node[fill=white,midway,above] {2}; 
    \draw[dashed,line width=1pt] (G) -- (H);

\end{tikzpicture}
\caption{The left hand diagrams for degree two contributions up to genus one}
\label{LeftHandDegreeTwo}
\end{figure}
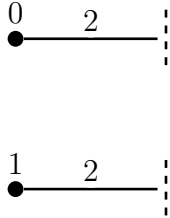
\begin{figure}[H]

\begin{tikzpicture}[
    node distance=1.5cm, % Adjust spacing between nodes
    every node/.style={fill=black, minimum size=6pt, circle, inner sep=1pt} % Style for nodes (small black dots)
]
\useasboundingbox (0,-7) rectangle (4,2);
    % Nodes
    \node[label=0] (A) at (0,0) {}; % Node A (small black dot)
    \node[label=0] (B) at (0,-1) {}; % Node B (small black dot)
   % \node[label=1](C) at (2,2) {}; % Node C (small black dot, placed above B)
   \node[fill=none] (C) at (2,0) {};
   \node[fill=none] (D) at (2,-1) {};

    \node[fill=none] (E) at (2,0.5) {};
    \node[fill=none] (F) at (2,-2.5) {};

     \node[label=0] (G) at (0,-3.5) {};
    \node[label=1] (H) at (0,-5.5) {};
    \node[fill=none] (I) at (2,-3) {};
    \node[fill=none] (J) at (2,-4) {};
    \node[fill=none] (K) at (2,-5) {};
    \node[fill=none] (L) at (2,-6) {};

    % Edges
    \draw[-,line width=1pt] (A) --  (C) node[fill=white,midway,above] {1}; 
    \draw[dashed,line width=1pt] ($(C)+(0,0.5)$) -- +(0,-2.0);
    \draw[dashed,line width=1pt] (F) -- +(0,-1.75);
    % \draw[dashed,line width=1pt] ($(D)+(0,0.5)$) -- +(0,-1);
    \draw[dashed,line width=1pt] ($(K)+(0,0.25)$) -- +(0,-1.5);
    
    \draw[-,line width=1pt] (B) --  (D) node[fill=white,midway,above] {1}; 
    \draw[-,line width=1pt] (G) -- (I)
    node[fill=white,midway,above] {1};
    \draw[-,line width=1pt] (G) -- (J)
    node[fill=white,midway,above] {1};
    \draw[-,line width=1pt] (H) -- (K)
    node[fill=white,midway,above] {1};
    \draw[-,line width=1pt] (H) -- (L)
    node[fill=white,midway,above] {1};

\end{tikzpicture}
\caption{The left hand diagrams for degree $(1,1)$ contributions up to genus one.}
\label{LeftHandDegreesOneOne}
\end{figure}
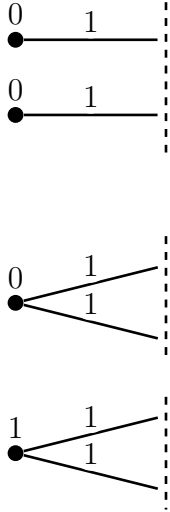
\noindent
These are the diagrams with vertex genera $0$ and $1$ -- they suffice for our calculation. We  compute the corresponding Hodge numbers $H$ on the one hand through moduli space integrals and on the other using the  Fock space realization.
The relevant moduli space integrals are:
\begin{align}
H^\circ_0 (z_1=1,z_2=1) &= \int_{0,2} \frac{1}{(1-\psi_1)(1-\psi_2)} = \frac{1}{2}
\nonumber \\
H^\circ_0 (z_1=1) &= \int_{0,1} \frac{1}{1-\psi}=1
\nonumber \\
H^\circ_0 (z_1=2) &=  \int_{0,1} \frac{2}{1-2 \psi} = \frac{1}{2}
\nonumber \\
H^\circ_1(z_1=2) &=  \int_{1,1} \frac{2 \Lambda}{1-2 \psi}  = 2 \int_{1,1} ((-\lambda_1) + 2 \psi)=\frac{1}{12}
\nonumber \\
H^\circ_1(z_1=1,z_2=1) &= \int_{1,2} \frac{\Lambda}{(1-\psi_1)(1-\psi_2)} = \frac{1}{24} \, .
\label{LeftHodgeIntegrals}
\end{align}
The index $(g,n)$ on the integral is an abbreviation for the integration over the compactified moduli spaces $\overline{{\cal M}}_{g,n}$. These are all the connected contributions which we multiply and sum over genera to obtain the disconnected Hurwitz numbers $H(1,1,u/t)$ and $H(2,u/t)$. Let us compute the latter in the oscillator representation as well. We start with the easiest case:
\begin{align}
H((2),\frac{u}{t}) &= (\frac{u}{t})^{-1} \langle { A}(2,\frac{2u}{t}) \rangle
\nonumber \\
&= (\frac{u}{t})^{-1} S(\frac{2u}{t})^2 \frac{1}{\zeta(\frac{2u}{t})}
\nonumber \\
&= \frac{1}{4} u^{-3} t^3 \zeta(\frac{2u}{t})
\nonumber \\
& \approx  \frac{1}{2} \frac{t^2}{u^2} + \frac{1}{12}
+ \frac{1}{240} \frac{u^2}{t^2} + \dots
\end{align}
The genus zero and the genus one contribution match those in equation (\ref{LeftHodgeIntegrals}) as they must. There is an infinite set of higher genus corrections.  Next, we compute the disconnected Hurwitz number:
\begin{align}
H((1,1),\frac{u}{t}) &= (\frac{u}{t})^{-2} \langle { A}(1,\frac{u}{t}) { A}(1,\frac{u}{t}) \rangle
\nonumber \\
&= (\frac{u}{t})^{-2} ( S(\frac{u}{t})^2/\zeta(\frac{u}{t})^2 + \frac{1}{2} S(\frac{u}{t})^2 \langle {\cal E}_{-1} (\frac{u}{t}) {\cal E}_{1} (\frac{u}{t}) \rangle
\nonumber \\
&= u^{-4} t^4 + \frac{1}{2} u^{-4} t^4 \zeta(\frac{u}{t})^2 
\nonumber \\
& \approx \frac{t^4}{u^4}+\frac{1}{2} \frac{t^2}{ u^2}+\frac{1}{24}+ \frac{1}{720} \frac{u^2}{t^2} + \dots
\, .
\end{align}
Again, this is exact. The disconnected part equals:
\begin{equation}
L_{(1,1)_d} = u^{-4} t^4 \, .
\end{equation}
We therefore have the exact connected part $L_{(1,1)_c}$ as well and it contains an infinite set of higher loop corrections. The leading terms match those computed in (\ref{LeftHodgeIntegrals}) using the moduli space integrals.

This provides us with exact expressions for the left hand diagrams (though we will only need them up to genus one). The  symmetry factors are also  exact. We can then expand the right hand expressions to the necessary order (which depends on the diagram at hand) and be sure that we have taken into account all relevant contributions. 

\subsection{The Right Hand Diagrams}
The right hand side diagrams are multiple. They have edges of the types we discussed, decorated by two distinct markings. All right hand side diagrams occur in our full diagrams in Figures \ref{Diagrams1} and \ref{Diagrams2}. For instance, there are five right hand side diagrams with one edge of degree $2$. They correspond to (the right of) diagrams $D_{1,2,3,4,23}$. 
In our calculation of the right hand diagrams, we favor the oscillator approach. 
%. In the oscillator approach, for the two-point function with two edge insertions, we need to evaluate a four-point function of operators ${ A}$.
In particular, we need to evaluate the two oscillator expressions:
\begin{equation}
H(2,t z_1,t z_2,\frac{u}{t}) = (\frac{u}{t})^{-3}
\frac{1}{2} \langle { A}(2, \frac{2u}{t})
{ A}(t z_1, u z_1) { A}(tz_2, u z_2) \rangle 
\,  \label{DegreeTwoTwoPoint}
\end{equation}
and
\begin{equation}
H(1,1,t z_1,t z_2,\frac{u}{t}) = (\frac{u}{t})^{-4} \langle { A}(1,  \frac{u}{t}) { A}(1,  \frac{u}{t})
{ A}(t z_1, u z_1) { A}(tz_2, u z_2) \rangle 
\,   \label{DegreeOneOneTwoPoint}
\end{equation}
corresponding to the $\mu=(2)$ and $\mu=(1,1)$ partitions respectively. 
Each operator ${ A}$ contains an infinite sum over operators ${\cal E}_k$. 
There will be many connected and disconnected contributions and they can correspond to multiple diagrams at once. 
%

% \subsection{Raising Insertion Number and Degree}
Previously, we treated the equivariant zero-point function at degree two. Here, we pedagogically build up to the oscillator calculation of the expressions (\ref{DegreeTwoTwoPoint}) and (\ref{DegreeOneOneTwoPoint}) by gradually increasing the number of insertions as well as the degree of the correlator. Thus, the following subsections gradually increase in complexity while decreasing in the description of calculational details. We will regularly perform checks using the diagrams and moduli space integrals, but will soon see that the oscillator calculation is more powerful.
%\footnote{In the following, we determine many other correlators implicitly in the following.}    
\subsubsection{The Equivariant One-point at Degree Zero}
For starters, we evaluate the right hand diagram for the equivariant $\tau_1$ one-point function at degree zero using the Fock space representation:
\begin{align}
H(t z_1,\frac{u}{t}) &= (\frac{u}{t})^{-1} \langle { A} (t z_1,u z_1) \rangle
\nonumber \\
          &= (\frac{u}{t})^{-1} \langle
S(u z_1)^{tz_1} \sum_k \frac{\zeta(uz_1)^k}{(t z_1 +1)_k} {\cal E}_k (uz_1) \rangle
\nonumber \\
% &=  (\frac{u}{t})^{-1} 
% S(u z_1)^{tz_1} \frac{1}{\zeta(uz_1)}
% \nonumber \\
&= (\frac{u}{t})^{-1}     (\frac{\zeta(u z_1) }{u z_1})^{t z_1}  \frac{1}{\zeta(uz_1)}
\nonumber \\
& \approx \frac{t^2}{u^{2}} \frac{1}{t z_1}   -\frac{1}{24} t z_1 + \frac{1}{24} t^2 z_1^2  + \frac{7}{5760} {u^2 t z_1^3}  - \frac{1}{480} t^2 u^2 z_1^4 +\dots
\end{align}
We  used that the vacuum expectation values of ${\cal E}_k$ is zero for non-zero index $k$. We find, as the coefficient of $z_1^2$, the genus one contribution proportional to $t^2$ and the diagram  has the value $1/24$. We found this result previously in equation (\ref{OnePointTauOne}).
 We repeat the exercise for $\tau_3$, which is the same computation to higher order in $z_1$. It agrees with the genus $2$ moduli space integral (\ref{OnePointTauThree}).
The $z_1^3$ coefficient matches the first integral in (\ref{OnePointTauTwo}). We have confirmed the degree $0$ isolated one-point functions of $\tau_1$ and $\tau_3$ that decorate many diagrams. In passing we have demonstrated that they receive a contribution from a unique genus (namely genus $1$ for $\tau_1$ and genus $2$ for $\tau_3$) as claimed previously. 

\subsubsection{The Degree Zero Two-Point Functions}
\begin{figure}[H]
\begin{tikzpicture}[
    node distance=1.5cm, % Adjust spacing between nodes
    every node/.style={fill=black, minimum size=6pt, circle, inner sep=1pt} % Style for nodes (small black dots)
]
\useasboundingbox (0,-3) rectangle (4,2);
    % Nodes
    \node[fill=none] (A) at (0,0) {}; % Node A (small black dot)
    \node[fill=none] (B) at (0,-2) {}; % Node B (small black dot)
   % \node[label=1](C) at (2,2) {}; % Node C (small black dot, placed above B)
   \node[label={1}] (C) at (2,-0.25) {};
   \node[label=2] (D) at (2,-2) {};
   \node[fill=none] (Cp) at (3,-0.25) {};
   \node[fill=none] (Dp) at (3,-2.5) {};
   \node[label=2] (K) at (2,0.5) {};
   \node[fill=none] (Kp) at (3,0.5) {};
   \node[fill=none] (Lp) at (3,-1.5) {};

    \node[fill=none] (E) at (0,0.5) {};
    \node[fill=none] (F) at (0,-0.5) {};
     \node[fill=none] (G) at (0,-1.5) {};
    \node[fill=none] (H) at (0,-2.5) {};

    % Edges
    \draw[-,line width=1pt] (Cp) --  (C) node[fill=none,midway,above] {$\tau_3$}; 
     \draw[-,line width=1pt] (Kp) --  (K) node[fill=none,midway,above] {$\tau_1$}; 
    \draw[dashed,line width=1pt] (E) -- (F);
    \draw[-,line width=1pt] (Dp) --  (D) node[fill=none,midway,below] {$\tau_3$}; 
     \draw[-,line width=1pt] (Lp) --  (D) node[fill=none,midway,above] {$\tau_1$}; 
    \draw[dashed,line width=1pt] (G) -- (H);

\end{tikzpicture}
\caption{The right hand diagrams for degree zero two-point functions}
\label{DegreeZeroTwoPoint}
\end{figure}
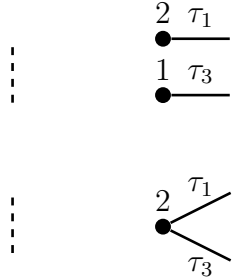

Next, 
we  compute the degree zero two-point functions in the oscillator language:
\begin{equation}
H(tz_1,t z_2,\frac{u}{t})
= (\frac{u}{t})^{-2} \langle { A}(tz_1,u t_1)
{ A}(tz_2,u t_2) \rangle \, .
\end{equation}
%It is the first case where we may have contributions from all ${\cal E}_k$. 
We have:
\begin{align}
H(tz_1,t z_2,\frac{u}{t})
& = (\frac{u}{t})^{-2} S(uz_1)^{tz_1} S(u z_2)^{t z_2} \sum_{k \ge 0}
\zeta(uz_1)^k \zeta(uz_2)^{-k}  \frac{1}{(t z_1+1)_k (tz_2+1)_{-k}} \nonumber \\
& \langle {\cal E}_{k}(u z_1) {\cal E}_{-k}(u z_2) \rangle
\nonumber \\
 &= (\frac{u}{t})^{-2}S(uz_1)^{tz_1} S(u z_2)^{t z_2} (\sum_{k \ge 1}
\zeta(uz_1)^k \zeta(uz_2)^{-k}  \frac{1}{(t z_1+1)_k (tz_2+1)_{-k}} 
\nonumber \\
& \frac{\zeta(k(z_1+z_2)}{\zeta(z_1+z_2)}
+ \frac{1}{\zeta(z_1) \zeta(z_2)}) \, .
\end{align}
We  concentrate on terms that are quadratic in $z_1$. This implies that  only the values $k \le 2$  contribute. We perform a Taylor expansion in the $z_i$ and concentrate on the term:
\begin{align}
H(tz_1,t z_2,\frac{u}{t})
& = \dots %+\frac{25}{576} t^4 z_1^2 z_2^2
- \frac{73}{11520}  t^4 u^2 z_1^2 z_2^4 +\dots
\end{align}
The coefficient is the total (right hand) two-point function at degree zero of $\tau_1$ and $\tau_3$. We disentangle the moduli space integrals that contribute.  See the diagrams in Figure \ref{DegreeZeroTwoPoint}.
\noindent
We have a disconnected contribution that corresponds to the product of
a $\tau_1$ operator at genus $1$ disconnected from a $\tau_3$ marking at genus $2$. The disconnected contribution is the product:
\begin{align}
H^{d}(tz_1,t z_2,\frac{u}{t})
& = \dots %+\frac{1}{576} t^4 z_1^2 z_2^2
- \frac{1}{11520}  t^4 u^2 z_1^2 z_2^4 +\dots
\end{align}
where $-1/11520=1/24 \times (-1/480)$.
Thus, the connected contribution is:
\begin{align}
H^\circ(tz_1,t z_2,\frac{u}{t})
& = \dots 
%+\frac{1}{24} t^4 z_1^2 z_2^2
- \frac{1}{160}  t^4 u^2 z_1^2 z_2^4 +\dots
\end{align}
This checks out with the moduli space integral:
% modular integrals that we have not yet computed. Let us do so immediately:
% \begin{equation}
% \int_{1,2} \psi_1 \psi_2 = \frac{1}{24} = 24/576
% \end{equation}
% and 
\begin{equation}
-\int_{2,2} \psi_1 \psi_2^3 \lambda_1 =-\frac{1}{160}=-\frac{72}{11520} \, .
\end{equation}
Finally, we remark that we only needed to go up to the summand with  $k=2$  to get the $z_1^2 z_2^4$ term. However, if we want to exchange the role of $z_1$ and $z_2$, we  need to compute the summands up to $k=4$ in order to get the correct (i.e. the same) coefficient for $z_1^4 z_2^2$.  Thus, only at the contribution $k=4$ do we restore the symmetry between the two insertions (at this order in $(z_1,z_2)$).

\subsubsection{The Degree One One-Point Functions}
We have understood the two point function contributions to $\langle \tau_1(\omega) \tau_3(\omega) \rangle$ at degree $d=0$.\footnote{The reader who wonders at this stage how this matches coefficients in the completed cycle will find solace in the fact that the contributions we found are at genus two while any non-zero contribution to the non-equivariant two-point function at degree zero would have genus three. Indeed, the product of completed cycles has no degree zero term. The non-equivariant limit of our calculations will be zero. This does not imply that the non-equivariant limit of higher degree diagrams that contain these subdiagrams is necessarily zero, as we will witness.} Our goal is to reach degree $2$. In the service of pedagogy, we first tackle the intermediate degree $1$.  We compute the Hurwitz number corresponding to a degree one edge with one marking:
\begin{align}
H(1,tz_1,\frac{u}{t}) &= (\frac{u}{t})^{-2} \langle { A}(tz_1,u z_1) { A}(1,\frac{u}{t}) \rangle
% \nonumber \\
% &= (\frac{u}{t})^{-2} \langle { A}(tz_1,u z_1) 
% e^{\alpha_1} e^{\frac{u}{t} {\cal F}_2} \alpha_{-1} \rangle 
\, .
\end{align}
We  pick terms with ${\cal E}_k {\cal E}_{-k}$ and $k>0$ or the disconnected $k=0$ term:
% Let us check:
% \begin{align}
% { A}(1,\frac{u}{t})_{proj. 0} &=  S(\frac{u}{t}) \frac{}{\zeta (\frac{u}{t})} = (\frac{u}{t})^{-1} \, .
% \end{align}
% 
\begin{align}
H(1,tz_1,\frac{u}{t}) &= (\frac{u}{t})^{-2} \langle { A}(tz_1,u z_1) { A}(1,\frac{u}{t}) \rangle 
% \nonumber \\
%  &= (\frac{u}{t})^{-2} \langle
% S(u z_1)^{tz_1} \sum_k \frac{\zeta(uz_1)^k}{(t z_1 +1)_k} {\cal E}_k (uz_1)
% S(\frac{u}{t})^{1} \sum_k \frac{\zeta(\frac{u}{t})^k}{(1 +1)_k} {\cal E}_k (\frac{u}{t}) \rangle
\nonumber \\
&= (\frac{u}{t})^{-2}S(u z_1)^{tz_1}  \frac{1}{\zeta(uz_1)} \frac{ S(\frac{u}{t})}{\zeta(\frac{u}{t})}
\nonumber \\
& +  (\frac{u}{t})^{-2} 
S(u z_1)^{tz_1} \sum_{k > 0} \frac{\zeta(uz_1)^k}{(t z_1 +1)_k} {\cal E}_k (uz_1)
S(\frac{u}{t})^{1} 
\frac{\zeta(\frac{u}{t})^{-k}}{(1 +1)_{-k}} {\cal E}_{-k} (\frac{u}{t}) \rangle
% \nonumber \\
% & =  (\frac{u}{t})^{-2}S(u z_1)^{tz_1}  1/\zeta(uz_1) S(\frac{u}{t})/\zeta(\frac{u}{t})
% \nonumber \\
% & +  (\frac{u}{t})^{-2} 
% S(u z_1)^{tz_1} \sum_{k > 0} \frac{\zeta(uz_1)^k}{(t z_1 +1)_k} \langle {\cal E}_k (uz_1)
% S(\frac{u}{t})^{1} 
% \frac{\zeta(\frac{u}{t})^{-k}}{(1 +1)_{-k}} {\cal E}_{-k} (\frac{u}{t}) \rangle
\, .
\end{align}
The values of relevant Pochhammer symbols are $(1+1)_0=1, (1+1)_{-1}=1$ and $1/(1+1)_{-2}=0$. Thus, we only pick up one further term which corresponds to the operators $\tau_{1}$ or $\tau_3$ connected to the edge of degree one:
\begin{align}
H(1,tz_1,\frac{u}{t}) &
= (\frac{u}{t})^{-2}S(u z_1)^{tz_1}   \frac{1}{\zeta(uz_1)} \frac{ S(\frac{u}{t})}{\zeta(\frac{u}{t})}
\nonumber \\
& +  (\frac{u}{t})^{-2} 
S(u z_1)^{tz_1}  \frac{\zeta(uz_1)}{(t z_1 +1)_1} \langle {\cal E}_1 (uz_1)
S(\frac{u}{t})
\frac{\zeta(\frac{u}{t})^{-1}}{(1 +1)_{-1}} {\cal E}_{-1} (\frac{u}{t}) \rangle
\nonumber \\
% &= (\frac{u}{t})^{-2}S(u z_1)^{tz_1}  \frac{1}{\zeta(uz_1)} \frac{ S(\frac{u}{t})}{\zeta(\frac{u}{t})}
% \nonumber \\
% & +  (\frac{u}{t})^{-2} 
% S(u z_1)^{tz_1}  \frac{\zeta(uz_1)}{(t z_1 +1)_1} \langle {\cal E}_1 (uz_1)
% S(\frac{u}{t})^{1} 
% \frac{\zeta(\frac{u}{t})^{-1}}{(1 +1)_{-1}} {\cal E}_{-1} (\frac{u}{t}) \rangle
% \nonumber \\
&= (\frac{u}{t})^{-2}S(u z_1)^{tz_1}   \frac{1}{\zeta(uz_1)} \frac{ S(\frac{u}{t})}{\zeta(\frac{u}{t})}
\nonumber \\
& +  (\frac{u}{t})^{-2} 
S(u z_1)^{tz_1} S(\frac{u}{t}) \zeta(\frac{u}{t})^{-1} \frac{\zeta(uz_1)}{t z_1 +1}
%\langle {\cal E}_1 (uz_1)
% {\cal E}_{-1} (\frac{u}{t}) \rangle
% \nonumber \\
% &= \text{disconnected}
% \nonumber \\
% & +  (\frac{u}{t})^{-2} 
% S(u z_1)^{tz_1} (\frac{u}{t})^{-1} \frac{\zeta(uz_1)}{t z_1 +1} \times 1
% \nonumber \\
% & \approx \text{disc}+ (t^3 z_1)/u^2 - (t^4 z_1^2)/u^2 + (t^3/24 + t^5/u^2) z_1^3 - (
%  t^6 z_1^4)/u^2
 \nonumber \\
 & \approx (\text{$k=0$ term})- \frac{t^4 z_1^2}{u^2}- \frac{
 t^6 z_1^4}{u^2} + \text{other powers of $z_1$}
\, .
\end{align}
%The auxiliary function of \cite{OkounkovPandharipande1} comes into play and we used \cite{OkounkovPandharipande1}, subsection 3.3.2].
There is a disconnected term, but we also  find a connected sphere contribution with $t^4$ power and coefficient $-1$ for $\tau_1$ as well as a $t^6$ power and coefficient $-1$ for $\tau_3$. 
These correspond to the unstable moduli space integrals we calculated in equation (\ref{SphereOneEdgeOneMarking}). 
We confirmed all the degree one one-point functions.

\subsubsection{The Degree One Two-Point Functions}
We move to the degree one  two-point function.
This is the first calculation in which we insert three  ${ A}$ operators:
\begin{align}
H(1,t z_1, tz_2, \frac{u}{t}) &= (\frac{u}{t})^{-3}
S(uz_1)^{tz_1} S(u z_2)^{t z_2} 
S(\frac{u}{t})^{1} 
\sum_{k_i}
\zeta(uz_1)^{k_1} \zeta(uz_2)^{k_2} 
\zeta(\frac{u}{t})^{k_3} 
\nonumber \\
&
\frac{1}{(t z_1+1)_{k_1} (tz_2+1)_{k_2} (1+1)_{k_3}} 
\langle {\cal E}_{k_1}(u z_1) {\cal E}_{k_2}(u z_2) 
{\cal E}_{k_3} (\frac{u}{t}) \rangle \, .
\end{align}
Because of the way we ordered our operators, we have that $k_3 \le 0$. Because of the Pochhammer symbol,  all terms with $k_3 \le -2$ vanish. We are left with $k_3=0$ and $k_3=-1$ terms. The term with $k_3=-1$ equals:
\begin{align}
H^{-1}(1,t z_1, tz_2, \frac{u}{t}) &= (\frac{u}{t})^{-3}
S(uz_1)^{tz_1} S(u z_2)^{t z_2} 
S(\frac{u}{t})^{1} 
\sum_{k}
\zeta(uz_1)^{k} \zeta(uz_2)^{-k+1} 
\zeta(\frac{u}{t})^{-1} 
\nonumber \\
&
\frac{1}{(t z_1+1)_{k} (tz_2+1)_{-k+1} } 
\langle {\cal E}_{k}(u z_1) {\cal E}_{-k+1}(u z_2) 
{\cal E}_{-1} (\frac{u}{t}) \rangle \, .
\end{align}
It is  sufficient to calculate the terms with $k=0,1,2$ in order to obtain the coefficient of  $z_1^2$.
We compute the three relevant vacuum expectation values:
\begin{align}
\langle {\cal E}_{0}(u z_1) {\cal E}_{+1}(u z_2) 
{\cal E}_{-1} (\frac{u}{t}) \rangle
&= \frac{1}{\zeta(u z_1)} 
\nonumber \\
\langle {\cal E}_{1}(u z_1) {\cal E}_{0}(u z_2) 
{\cal E}_{-1} (\frac{u}{t}) \rangle 
&= \zeta (u z_2)  + \frac{1}{\zeta(u z_2)} 
\nonumber \\
\langle {\cal E}_{2}(u z_1) {\cal E}_{-1}(u z_2) 
{\cal E}_{-1} (\frac{u}{t}) \rangle 
&=  \zeta(u (z_1+2 z_2))  \, .
\end{align}
Using these, we find the total:
\begin{align}
H^{-1}(1,t z_1, tz_2, \frac{u}{t}) &= (\frac{u}{t})^{-4} 
S(uz_1)^{tz_1} S(u z_2)^{t z_2} 
(\zeta(uz_2) \frac{1}{(t z_2+1)_{1} } \frac{1}{\zeta(u z_1)}
\nonumber \\
&
+
\zeta(u z_1) \frac{1}{(1+tz_1)_1} (\zeta(u z_2)+1/\zeta(uz_2))
\nonumber \\
 &
+\zeta(u z_1)^2 \zeta(uz_2)^{-1} \frac{1}{(1+tz_1)_2(1+tz_2)_{-1}} \zeta (u (2z_2+z_1))) \, .
% \nonumber \\
% & \approx 
\end{align}
The term with $k_3=0$ equals:
\begin{align}
H^0(1,t z_1, tz_2, \frac{u}{t}) 
% &= (\frac{u}{t})^{-3}
% S(uz_1)^{tz_1} S(u z_2)^{t z_2} 
% \frac{t}{u}
% \sum_{k}
% \zeta(uz_1)^{k} \zeta(uz_2)^{-k} 
% \nonumber \\
% &
% \frac{1}{(t z_1+1)_{k} (tz_2+1)_{-k} } 
% \langle {\cal E}_{k}(u z_1) {\cal E}_{-k}(u z_2) 
%  \rangle 
%  \nonumber \\
 &= (\frac{u}{t})^{-4}
S(uz_1)^{tz_1} S(u z_2)^{t z_2}  \Big(\frac{1}{\zeta(u z_1) \zeta(u z_2)}
 \\
&
+\frac{\zeta(u z_1)\zeta(u z_2)^{-1}}{(t z_1+1)_{1} (tz_2+1)_{-1} } 
+\frac{\zeta(u z_1)^2 \zeta(u z_2)^{-2}}{(t z_1+1)_2 (tz_2+1)_{-2}} \frac{\zeta(2 (u (z_1+z_2)))}{\zeta(u (z_1+z_2))} \Big) \nonumber 
 \, .
\end{align}
We single out the $z_1^2 z_2^4$ term in the total:
\begin{align}
H(1,tz_1,tz_2,\frac{u}{t}) 
& \approx \dots+ \left(- \frac{1}{24} \frac{t^8}{u^2}
+ \frac{1}{480} t^6 - \frac{73}{11520} t^6 \right) z_1^2 z_2^4 + \dots
\end{align}
Let us understand the $z_1^2 z_2^4$ coefficient. There are five half-diagrams. Two of them give the previous $-73/11520$ which is indeed what we get by expanding the  term $H^0$. 
We have three more diagrams, which are the degree one with $\tau_1$ attached and $\tau_3$ loose, degree one with $\tau_3$ attached and $\tau_1$ loose and the degree one fork with $\tau_1$ and $\tau_3$ attached. Let us compute these half-diagrams. The first equals
$-1 \times -1/480=1/480$. The second one equals $-1 \times 1/24=-1/24$. The  genus is one for the first,  and genus $0$ for the second.  We have matched four out of five half-diagrams.
The fifth diagram is the degree one fork. At genus $g=1$ it is zero.

\subsubsection{The Edge Degree Two One-point Functions}
We move to the degree two one-point function. We first consider the edge of degree two.
We do as for the edge of degree one, but have some more contributions:
\begin{align}
H(2,tz_1,\frac{u}{t}) &= (\frac{u}{t})^{-2} \langle { A}(tz_1,u z_1) { A}(2,\frac{2u}{t}) \rangle 
\nonumber \\
 &= (\frac{u}{t})^{-2} \langle
S(u z_1)^{tz_1} \sum_k \frac{\zeta(uz_1)^k}{(t z_1 +1)_k} {\cal E}_k (uz_1)
S(\frac{2u}{t})^{2} \sum_k \frac{\zeta(\frac{2u}{t})^k}{(2 +1)_k} {\cal E}_k (\frac{2u}{t}) \rangle
\nonumber \\
&= (\frac{u}{t})^{-2}S(u z_1)^{tz_1} /\zeta(uz_1) S(\frac{2u}{t})^2/\zeta(\frac{2u}{t})
\nonumber \\
& +  (\frac{u}{t})^{-2} 
S(u z_1)^{tz_1} \sum_{2 \ge k > 0} \frac{\zeta(uz_1)^k}{(t z_1 +1)_k} {\cal E}_k (uz_1)
S(\frac{2u}{t})^{2} 
\frac{\zeta(\frac{2u}{t})^{-k}}{(2 +1)_{-k}} {\cal E}_{-k} (\frac{2u}{t}) \rangle
% \nonumber \\
% & = \text{disconnected combination of previous calculations}
% \nonumber \\
% & +  (\frac{u}{t})^{-2} 
% S(u z_1)^{tz_1} \sum_{k=1,2} \frac{\zeta(uz_1)^k}{(t z_1 +1)_k} 
% S(\frac{2u}{t})^{2} 
% \frac{\zeta(\frac{2u}{t})^{-k}}{(2 +1)_{-k}}
% \langle {\cal E}_k (uz_1) {\cal E}_{-k} (\frac{2u}{t}) \rangle
% \, .
% \end{align}
% We can easily compute the whole thing:
% \begin{align}
%H(2,tz_1\frac{u}{t}) 
\nonumber \\
&= (\frac{u}{t})^{-2} S(uz_1)^{t z_1} S( \frac{2u}{t})^2  \Big( \frac{1}{\zeta(u z_1) \zeta(\frac{2u}{t})}
 \nonumber \\
 &+ \frac{\zeta(uz_1) \zeta(\frac{2u}{t})^{-1}}{(tz_1+1)_1 (2+1)_{-1}}
  + \frac{\zeta(uz_1)^2 \zeta(\frac{2u}{t})^{-2}}{(tz_1+1)_2 (2+1)_{-2}} \frac{\zeta(2u (z_1+2/t))}{\zeta(u (z_1+2/t))} \Big)
 \nonumber \\
  & \approx \dots+( \frac{23}{48} \frac{t^4}{u^2} + \frac{25}{288} t^2 + \frac{73}{5760} u^2 + \dots ) z_1^2 
  \nonumber \\
 & +( - \frac{1}{8} \frac{t^6}{u^2} - \frac{1}{960} t^4 + \frac{101}{5760} t^2 u^2 + \dots) z_1^4 + \dots
 \label{DegreeTwoOnePoint}
\end{align}
We understand the numbers at low genus orders from moduli space integrals. We have 
$-1/8$ for a connected genus zero diagram and $-1/960$ for a genus one disconnected diagram for the $\tau_3$ operator one-point function.

\subsubsection{The Edge Degree Two Two-point Functions}
The edge degree $2$ two-point functions follow the familiar pattern:
\begin{align}
H(2,tz_1,tz_\frac{2u}{t}) &= (\frac{u}{t})^{-3} \langle { A}(tz_1,u z_1) { A}(tz_2,uz_2) { A}(2,\frac{2u}{t}) \rangle 
\nonumber \\
 &= (\frac{u}{t})^{-3} S(u z_1)^{tz_1}
 S(u z_2)^{tz_2}S(\frac{2u}{t})^{2}
 \sum_{k_1} \frac{\zeta(uz_1)^{k_1}}{(t z_1 +1)_{k_1}}
 \sum_{k_2} \frac{\zeta(uz_1)^{k_2}}{(t z_2 +1)_{k_2}}
  \sum_{k_3} \frac{\zeta(\frac{2u}{t})^{k_3}}{(2 +1)_{k_3}}
  \nonumber \\
  &
 \langle
  {\cal E}_{k_1} (uz_1)
 {\cal E}_{k_2} (uz_2)
 {\cal E}_{k_3} (\frac{2u}{t}) \rangle
\, . \label{22}
\end{align}
We need $k_3 \le 0$ on the right hand side, but also $k_3 \ge -2$ in order for the Pochhammer symbol to be non-trivial. Thus, we have $k_3 \in \{ 0,-1,-2 \}$. In order to obtain the $z_1^2$ coefficient, we can restrict $k_1 \le 2$ and it needs to be positive and therefore $k_1 \in \{ 0,1,2 \}$. We therefore have nine terms. These invoke the expectation values:
\begin{align}
\langle
  {\cal E}_{0} (uz_1)
 {\cal E}_{0} (uz_2)
 {\cal E}_{0} (\frac{2u}{t}) \rangle &= (\zeta(u z_1) \zeta(u z_2) \zeta(\frac{2u}{t}))^{-1}
 \nonumber \\
 \langle {\cal E}_{0} (uz_1)
 {\cal E}_{1} (uz_2)
 {\cal E}_{-1} (\frac{2u}{t}) \rangle &= (\zeta(u z_1))^{-1}
 \nonumber \\
 \langle {\cal E}_{0} (uz_1)
 {\cal E}_{2} (uz_2)
 {\cal E}_{-2} (\frac{2u}{t}) \rangle &=(\zeta(u z_1))^{-1}
 \frac{\zeta(2 (u z_2 + \frac{2u}{t})}{\zeta(u z_2+\frac{2u}{t})}
 \nonumber \\
 \langle {\cal E}_{1} (uz_1)
 {\cal E}_{-1} (uz_2)
 {\cal E}_{0} (\frac{2u}{t}) \rangle &=(\zeta(\frac{2u}{t}))^{-1}
 \nonumber \\
 \langle {\cal E}_{1} (uz_1)
 {\cal E}_{0} (uz_2)
 {\cal E}_{-1} (\frac{2u}{t}) \rangle &= \zeta(u z_2)+(\zeta(u z_2))^{-1}
 \nonumber \\
 \langle {\cal E}_{1} (uz_1)
 {\cal E}_{1} (uz_2)
 {\cal E}_{-2} (\frac{2u}{t}) \rangle &= \zeta (2 u z_2+\frac{2u}{t})
 \nonumber \\
 \langle {\cal E}_{2} (uz_1)
 {\cal E}_{-2} (uz_2)
 {\cal E}_{0} (\frac{2u}{t}) \rangle &=(\zeta(\frac{2u}{t}))^{-1}
 \frac{\zeta(2 (u z_1 + u z_2))}{\zeta(u z_1+uz_2)}
 \nonumber \\
\langle {\cal E}_{2} (uz_1)
 {\cal E}_{-1} (uz_2)
 {\cal E}_{-1} (\frac{2u}{t}) \rangle &= \zeta(u z_1+2 u z_2)
 \nonumber \\
 \langle {\cal E}_{2} (uz_1)
 {\cal E}_{0} (uz_2)
 {\cal E}_{-2} (\frac{2u}{t}) \rangle &=
 \zeta(2 u z_2) \frac{\zeta(2(uz_1+u z_2+\frac{2u}{t}))}{\zeta(u z_1 + u z_2 + \frac{2u}{t})} + \zeta(uz_2)^{-1}
 \frac{\zeta(2(u z_1+\frac{2u}{t}))}{\zeta(u z_1 + \frac{2u}{t})} \, .
 \end{align}
We can plug the formulas into a symbolic manipulation program and find the Taylor expansion. We will only match the coefficients to moduli space integrals through the final diagram calculation.

\subsubsection{The One-One Partition One-Point Functions}
We move to the degree two cases which have an intermediate $(1,1)$ partition. We start with the one-point functions. Again, we have three operators ${ A}$:
\begin{align}
H(1,1, tz_1, \frac{u}{t}) &= (\frac{u}{t})^{-3}
S(uz_1)^{tz_1} S(\frac{u}{t}) 
S(\frac{u}{t}) 
\sum_{k_i}
\zeta(uz_1)^{k_1} \zeta(\frac{u}{t})^{k_2} 
\zeta(\frac{u}{t})^{k_3} 
\nonumber \\
&
\frac{1}{(t z_1+1)_{k_1} (1+1)_{k_2} (1+1)_{k_3}} 
\langle {\cal E}_{k_1}(u z_1) {\cal E}_{k_2}(\frac{u}{t}) 
{\cal E}_{k_3} (\frac{u}{t}) \rangle \, .
\end{align}
We have that $k_{3} =0,-1$. This implies that $k_2=1,0,-1$. And then we can compute $k_1$ from the values of $k_{2}$ and $k_3$.
We therefore must compute the vevs:
\begin{align}
\langle {\cal E}_{0}(u z_1) {\cal E}_{0}(\frac{u}{t}) 
{\cal E}_{0} (\frac{u}{t}) \rangle
&= \frac{1}{\zeta(u z_1) \zeta(\frac{u}{t}) \zeta(\frac{u}{t})}
\nonumber \\
\langle {\cal E}_{1}(u z_1) {\cal E}_{-1}(\frac{u}{t}) 
{\cal E}_{0} (\frac{u}{t}) \rangle
&=  \frac{1}{ \zeta(\frac{u}{t})}
\nonumber \\
\langle {\cal E}_{1}(u z_1) {\cal E}_{0}(\frac{u}{t}) 
{\cal E}_{-1} (\frac{u}{t}) \rangle
&= \zeta(\frac{u}{t}) + 1/\zeta(\frac{u}{t})
\nonumber \\
\langle {\cal E}_{2}(u z_1) {\cal E}_{-1}(\frac{u}{t}) 
{\cal E}_{-1} (\frac{u}{t}) \rangle
&= \zeta(\frac{2u}{t}+uz_1)
\nonumber \\
\langle {\cal E}_{0}(u z_1) {\cal E}_{1}(\frac{u}{t}) 
{\cal E}_{-1} (\frac{u}{t}) \rangle
&= \frac{1}{\zeta(uz_1)} \, .
\end{align}
 We identify the $z_1^2$ and $z_1^4$ coefficients:
\begin{align}
H(1,1,tz_1,\frac{u}{t}) & \approx
\dots +
(-\frac{47}{24} \frac{t^6}{u^4} + \frac{1}{48} \frac{t^4}{u^2} + \frac{49}{576} t^2 + \dots) z_1^2
\nonumber \\
& +(- 2 \frac{t^8}{u^4} - \frac{1}{480} \frac{t^6}{u^2} + \frac{13}{320} t^4 + \dots ) z_1^4 + \dots
% \nonumber \\
% &+(-(2 t^8)/u^4 - t^6/(480 u^2) + (13 t^4)/320 + (35 t^2 u^2)/2304+\dots) z_1^4 
\label{DegreeOneOneOnePoint}
\end{align}

\subsubsection{The  One-One Partition Two-Point Function}
Finally, we tackle the  disconnected Hodge number determined by the vacuum expectation value of four ${ A}$ operators:
\begin{align}
H(1,1, tz_1, tz_2, \frac{u}{t}) &= (\frac{u}{t})^{-4}
S(uz_1)^{tz_1} S(uz_2)^{tz_2}S(\frac{u}{t}) 
S(\frac{u}{t}) 
\nonumber \\
&
\sum_{k_i}
\frac{\zeta(uz_1)^{k_1}\zeta(uz_2)^{k_2} \zeta(\frac{u}{t})^{k_3} 
\zeta(\frac{u}{t})^{k_4} }{(t z_1+1)_{k_1} (1+t z_2)_{k_2} (1+1)_{k_3}
 (1+1)_{k_4}} 
 \nonumber \\
 &
\langle {\cal E}_{k_1}(u z_1) {\cal E}_{k_2}(u z_2){\cal E}_{k_3}(\frac{u}{t}) 
{\cal E}_{k_4} (\frac{u}{t}) \rangle \, .
\label{112}
\end{align}
We must contemplate the $k_4=0,-1$ contributions combined with the $k_3=-1,0,1$ contributions appropriately. The other factors can in principle give rise to an infinite number of contributions, but as per usual, if we concentrate on the $z_1^2 z_2^4$ term,  the values $k_1=2,1,0$ are enough. We have five choices for the last two, three for the first, for fifteen possibilities total. We compute these fifteen terms:
\begin{align}
\langle {\cal E}_{0}(u z_1) {\cal E}_{0}(u z_2){\cal E}_{0}(\frac{u}{t}) 
{\cal E}_{0} (\frac{u}{t}) \rangle
&= (\zeta(uz_1) \zeta(uz_2) \zeta(\frac{u}{t})^2)^{-1}
\nonumber \\
\langle {\cal E}_{1}(u z_1) {\cal E}_{-1}(u z_2){\cal E}_{0}(\frac{u}{t}) 
{\cal E}_{0} (\frac{u}{t}) \rangle
&= \zeta(\frac{u}{t})^{-2}
\nonumber \\
\langle {\cal E}_{2}(u z_1) {\cal E}_{-2}(u z_2){\cal E}_{0}(\frac{u}{t}) 
{\cal E}_{0} (\frac{u}{t}) \rangle
&= \frac{\zeta(2u (z_1+z_2)}{\zeta(u(z_1+z_2) \zeta(\frac{u}{t})^2}
\nonumber \\
\langle {\cal E}_{0}(u z_1) {\cal E}_{1}(u z_2){\cal E}_{-1}(\frac{u}{t}) 
{\cal E}_{0} (\frac{u}{t}) \rangle
&= (\zeta(u z_1) \zeta(\frac{u}{t}))^{-1}
\nonumber \\
\langle {\cal E}_{1}(u z_1) {\cal E}_{0}(u z_2){\cal E}_{-1}(\frac{u}{t}) 
{\cal E}_{0} (\frac{u}{t}) \rangle
&= \frac{1}{\zeta(\frac{u}{t})} (\zeta(u z_2)+1/\zeta(uz_2))
\nonumber \\
\langle {\cal E}_{2}(u z_1) {\cal E}_{-1}(u z_2){\cal E}_{-1}(\frac{u}{t}) 
{\cal E}_{0} (\frac{u}{t}) \rangle
&=\frac{\zeta(uz_1+2 uz_2)}{\zeta(\frac{u}{t})}
\nonumber \\
\langle {\cal E}_{0}(u z_1) {\cal E}_{0}(u z_2){\cal E}_{1}(\frac{u}{t}) 
{\cal E}_{-1} (\frac{u}{t}) \rangle
&= (\zeta(uz_1) \zeta(u z_2))^{-1}
\nonumber \\
\langle {\cal E}_{1}(u z_1) {\cal E}_{-1}(u z_2){\cal E}_{1}(\frac{u}{t}) 
{\cal E}_{-1} (\frac{u}{t}) \rangle
&= 1
\nonumber \\
\langle {\cal E}_{2}(u z_1) {\cal E}_{-2}(u z_2){\cal E}_{1}(\frac{u}{t}) 
{\cal E}_{-1} (\frac{u}{t}) \rangle
&= \frac{\zeta(2 u (z_1+z_2))}{\zeta(u(z_1+z_2)}
\nonumber \\
\langle {\cal E}_{0}(u z_1) {\cal E}_{1}(u z_2){\cal E}_{0}(\frac{u}{t}) 
{\cal E}_{-1} (\frac{u}{t}) \rangle
&= \frac{\zeta(\frac{u}{t})+1/\zeta(\frac{u}{t})}{\zeta(uz_1)}
\nonumber \\
\langle {\cal E}_{1}(u z_1) {\cal E}_{0}(u z_2){\cal E}_{0}(\frac{u}{t}) 
{\cal E}_{-1} (\frac{u}{t}) \rangle
&= \frac{1}{\zeta(u z_2) \zeta(\frac{u}{t})}+\frac{\zeta(\frac{u}{t})}{\zeta(uz_2)} + \frac{\zeta(u z_2)}{\zeta(\frac{u}{t})}
+ \zeta(uz_2) \zeta(\frac{u}{t})
\nonumber \\
\langle {\cal E}_{2}(u z_1) {\cal E}_{-1}(u z_2){\cal E}_{0}(\frac{u}{t}) 
{\cal E}_{-1} (\frac{u}{t}) \rangle
&= \zeta(uz_1+2u z_2) ( \zeta(\frac{u}{t})+1/\zeta(\frac{u}{t}))
\nonumber \\
\langle {\cal E}_{0}(u z_1) {\cal E}_{2}(u z_2){\cal E}_{-1}(\frac{u}{t}) 
{\cal E}_{-1} (\frac{u}{t}) \rangle
&= \frac{\zeta(\frac{2u}{t} + u z_2)}{\zeta(uz_1)}
\nonumber \\
\langle {\cal E}_{1}(u z_1) {\cal E}_{1}(u z_2){\cal E}_{-1}(\frac{u}{t}) 
{\cal E}_{-1} (\frac{u}{t}) \rangle
&=\zeta(\frac{u}{t}+uz_2) (\zeta(uz_2+\frac{u}{t}) +1/\zeta(uz_2+\frac{u}{t}))+1
\nonumber \\
\langle {\cal E}_{2}(u z_1) {\cal E}_{0}(u z_2){\cal E}_{-1}(\frac{u}{t}) 
{\cal E}_{-1} (\frac{u}{t}) \rangle
&=\frac{1}{\zeta(uz_2)}  \zeta(\frac{2u}{t}+uz_1)
+ \zeta(2 u z_2) \zeta(\frac{2u}{t}+uz_1+uz_2) \, .
\label{FifteenTerms}
\end{align}
The expansion contains the coefficient of  $z_1^2 z_2^4$:
\begin{align}
H(1,1,tz_1,tz_2,\frac{u}{t}) & \approx
%(
%(1081 t^8)/(576 u^4) + (25 t^6)/(1152 u^2) + (3577 %t^4)/13824  + \dots  ) z_1^2 z_2^2 
% \nonumber \\
%  & 
\dots+(\frac{23}{12} \frac{t^{10}}{u^4} - \frac{5}{2304} \frac{t^8}{u^2} + \frac{949}{7680} t^6 + \dots )
z_1^2 z_2^4+\dots
\end{align}
We can also  understand these numbers from diagrams and moduli space integrals and use them as  consistency checks. 
Instead, we summarize the results for half-diagrams that are slightly hidden in the disconnected oscillator results and discuss a few extra moduli space integrals. 

\subsection{The Values of Half Diagrams}
\label{ValuesHalfDiagrams}
We tally the results for half diagrams, obtained through moduli space integration or oscillator formalism. It is sufficient to evaluate connected parts of half diagrams. Disconnected diagrams have a value corresponding to the product of the values of the connected parts.
\subsubsection{The Values of the Left Diagrams}
We found for the left diagrams:
\begin{align}
L_{0,(2)} &= \frac{1}{2}\frac{t^2}{u^2} 
\qquad
L_{1,(2)} = \frac{1}{12}
\nonumber \\
L_{0,(1,1)_c}&= \frac{1}{2}\frac{t^2}{u^2} 
\qquad
L_{0,0,(1,1)_d} = \frac{t^4}{u^4}
\nonumber \\
L_{1,(1,1)_c} &= \frac{1}{24} \, .
\end{align}
%These are all the non-zero results up to an including genus one for the left hand diagrams. 
We have used a notation indicating the vertex genera and the edges (which are either connected or disconnected on the left hand side for $(1,1)$). The disconnected $(1,1)_d$ contribution is the square of $L_{0,(1)}=t^2/u^2$.  
%That settles the list of values for the left hand side diagrams. 
%Of course, these are in correspondence with the connected and disconnected Hurwitz numbers and moduli space integrals we computed before. 

\subsubsection{The  Right Diagrams without Edge}
We start with all right hand side components in which we have markings that are not attached to any edge. There are two possibilities. Either disconnected or connected. When disconnected, they are the product of one-point functions. We have for the unattached one-point functions:
\begin{align}
R_{\tau_1} &= \frac{1}{24} t^2  
\nonumber \\
R_{\tau_3} &= - \frac{1}{480}  t^2 u^2  \, .
\end{align}
For the connected two-point function (without edge and connected), we have:
\begin{align}
R_{\tau_1,\tau_3}^\circ &= - \frac{1}{160} t^4 u^2  \, .
\end{align}
The (purely) disconnected contribution corresponds to the product of $R_1$ and $R_3$. 
\begin{align}
R_{\tau_1,\tau_3}^d &
%= -\frac{1}{24\times 480} t^4  u^2
=- \frac{1}{11520} t^4  u^2
\, .
\end{align}
These results are exact in the genus expansion. 

\subsubsection{Right Diagrams with One Attached Marking}
We look at right diagrams where one marking is attached to one edge node. It can be the marking $\tau_1$ or $\tau_3$ and it can attach to a degree $2$ edge, a degree one edge or two degree one edges at once. We thus have six possibilities. 
Let us start with the degree $2$ edge. We take it to be at genus zero. The connected part for $\tau_1$ gives $-1/2$:
\begin{equation}
R_{\tau_1,0,(2)} = -\frac{1}{2} \frac{t^4}{u^2}
\end{equation}
The connected genus zero contribution for $\tau_3$ is
\begin{equation}
R_{\tau_3,0,(2)} = -\frac{1}{8} \frac{t^6}{u^2} \, .
\end{equation}
%There are also (known) genus one parts, if needed.  Needed ?
For one degree $1$ with one marker attached, the connected genus zero part equals:
\begin{equation}
R_{\tau_1,0,(1)} = - \frac{t^4}{u^2}
\end{equation}
and for $\tau_3$:
\begin{equation}
R_{\tau_3,0,(1)} = - \frac{t^6}{u^2} \, .
\end{equation}
%Again, genus one results are available. Do we need them ? 
%
Finally, we have the degree $(1,1)$ connected fork with one marker attached. We must go to genus one and have:
\begin{equation}
R_{\tau_1,1,(1,1)_c} = \frac{1}{12} t^2
\end{equation}
and for $\tau_3$ we must also choose genus one for the vertex and have:
\begin{equation}
R_{\tau_3,1,(1,1)_c} = \frac{1}{24} t^4 \, .
\end{equation}
%These are the six results we need for one attached marking. 
\subsubsection{Two Attached Markings}
Finally, we can have edges with two attached markings. We have three more possibilities. These are: a fork with an edge of degree $1$, a fork with an edge of degree $2$ and a crossroad with a connected $(1,1)$ edge  with two markings. %We need to compute these four possibilities. These correspond to (connected) degree $1$ and $2$ two-point functions. 
The relevant moduli space integrals are -- after realizing that the minimal and relevant genera are $g=(2,2,1)$ respectively --:
\begin{align}
R_{f,(1)} &= [z_1^2 z_2^4] \, t^2 z_1 z_2 \int_{2,3} \frac{\Lambda}{(1-\psi_e)(1-t z_1 \psi_1)(1-t z_2 \psi_2)} = t^6 \int_{2,3} \psi_1 \psi_2^3 (\lambda_2 -\lambda_1 \psi_e+ \psi_e^2 )
\nonumber \\
&=  t^6  (\frac{7}{1440} - \frac{1}{40}+ \frac{29}{1440}) = 0
\nonumber \\
R_{f,(2)} &= [z_1^2 z_2^4] \, 2 t^2  z_1 z_2 \int_{2,3} \frac{\Lambda}{(1-2\psi_e)(1-t z_1 \psi_1)(1-t z_2 \psi_2)} %=2  t^6  \int_{2,3} \psi_1 \psi_2^3 (\lambda_2 -2 \lambda_1 \psi_e+ 4\psi_e^2 )
\nonumber \\
&=  2 t^6  (\frac{7}{1440} - 2 \frac{1}{40}+ 4\frac{29}{1440}) 
= \frac{17}{240} t^6 
\nonumber \\
R_{cr,(1,1)_c} &=[z_1^2 z_2^4] \, t^2 z_1 z_2 \int_{1,4} \frac{\Lambda}{(1-\psi_e)(1-\psi_e')(1-t z_1 \psi_1)(1-t z_2 \psi_2)}
= \frac{1}{8} t^6  \, .
\end{align}
Shortly we will find that only the crossroads  contribute to our two-point function - the degree one fork is zero while the degree two fork will lead to too high a total genus.

\subsection{The Diagrams}
\label{TheDiagrams}

We have obtained ample data on all the degree $0,1$ and $2$ chiral zero, one and two-point functions $H$ and $H^\circ$. We combine them into the $27$ diagrams of Figures \ref{Diagrams1} and \ref{Diagrams2}, with the appropriate gluing symmetry factor (\ref{SymmetryFactor2}) or (\ref{SymmetryFactor11}).
A  subtler point is that for  $(1,1)$ edges that are not connected on the right, we need to distinguish between a marker attached to one or the other edge.
%\footnote{This is one example of a statement that becomes manifest when comparing to the oscillator approach. This is one among multiple reasons for our Fock space detour, which may be theoretically superfluous, but is practically valuable.} 
% \subsubsection{Left Hand Side Diagrams}
% We have three possible topologies. For the $(2)$ edge, there is a single topology. For the $(1,1)$ edge, there is the parallel bar diagram and the connected diagram. 
% The decoration by genus allows for genus zero everywhere but also genus one on the 2-edge or the fork $(1,1)$ edge. No other decoration contributes. 
%
We go through all diagrams systematically. 
\subsubsection{Full Diagrams with Disconnected Markings}
The  diagrams with  disconnected markings are  1, 5, 10, 15, 19 as well as 23, 24, 25, 26 and 27. 
We have two degree two diagrams. 
%
%  We should also flip the sign of $t$ in the second expression for $H$ (i.e. for the right hand diagram because the equivariant action at infinity is opposite to the one at zero but this will not make any difference in practice because they are even in the variable $t$). 
Let us apply our knowledge of all ingredients to the calculation of diagram $D_1$. We again find:
\begin{equation}
D_1 = -2 u^2 t^{-8} \times \frac{1}{2} \frac{t^2}{u^2} 
\times (-1) \frac{1}{11520} t^4 u^2 \times \frac{1}{2} \frac{t^2}{u^2}
= \frac{1}{23040}  \, .
\end{equation}
This is a total genus $1$ contribution and we showed the cancellation of powers of $u$ and $t$. 

Next, diagrams $5,10,15,19$ have a sum equal to:
\begin{align}
D_{5+10+15+19}
&= \frac{1}{2} \frac{u^4}{t^{10}} \times (\frac{t^4}{u^4}+ \frac{1}{2} \frac{t^2}{u^2} ) \times (\frac{t^4}{u^4}+ \frac{1}{2} \frac{t^2}{u^2} )
\times (-1) \frac{1}{11520} t^4 u^2
\nonumber \\
%&= - \frac{1}{23040} \frac{u^6}{t^6} 
%( t^8/u^8 + t^6/u^6+ \frac{1}{4} t^4/u^4)
%\nonumber \\
&= - \frac{1}{23040} (\frac{t^2}{u^2} + 1 + \dots)
\end{align}
The genus zero contribution (from diagram $D_5$) vanishes in the non-equivariant limit. The non-zero contributions  come from diagrams $D_{10}$ and $D_{15}$. 
We  compute the contribution from five other diagrams similarly. Firstly, diagram $D_{23}$:
\begin{equation}
D_{23} = -2 u^2 t^{-8} \times \frac{1}{2} \frac{t^2}{u^2} 
\times (-1) \frac{1}{160} t^4 u^2 \times \frac{1}{2} \frac{t^2}{u^2}
= \frac{1}{320}  \, ,
\end{equation}
and then the sum of four more:
\begin{align}
D_{24+25+26+27}
&= \frac{1}{2} \frac{u^4}{t^{10}} \times (\frac{t^4}{u^4}+ \frac{1}{2} \frac{t^2}{u^2} ) \times (\frac{t^4}{u^4}+ \frac{1}{2} \frac{t^2}{u^2} )
\times (-1) \frac{1}{160} t^4 u^2
\nonumber \\
&= - \frac{1}{320} \frac{u^6}{t^6} 
( t^8/u^8 + t^6/u^6+ \frac{1}{4} t^4/u^4)
\nonumber \\
&= - \frac{1}{320} (\frac{t^2}{u^2} + 1 + \dots)
\end{align}
The relevant contributions arise from diagrams $D_{25}$ and $D_{26}$. We observe that the ten diagrams with unconnected markings conspire to give zero net contribution.

\subsubsection{Full Diagrams with One Connected Marking}
The diagrams where one marking is connected to the rest are the diagrams $2,3,6,7, 11, 12$ and  $16,17,20,21$.  We first calculate the amplitude corresponding to diagrams $D_2$ and $D_3$ with the edge of degree $2$: 
\begin{align}
D_2 &= S_{(2)} \times L_{0,(2)} \times (R_3 \times R_{\tau_1,0,(2)})
\nonumber \\
&= -2 u^2 t^{-8} \times \frac{1}{2} \frac{t^2}{u^2}
\times (- \frac{1}{480} t^2 u^2 \times -\frac{1}{2} t^4 u^{-2})
\nonumber \\
&=-\frac{1}{960} \, ,
\end{align}
and
\begin{align}
D_3 &= S_{(2)} \times L_{1,(2)} \times (R_1 \times R_{\tau_3,0,(2)})
\nonumber \\
&= -2 u^2 t^{-8} \times \frac{1}{12} 
\times (\frac{1}{24} t^2  \times -\frac{1}{8} t^6 u^{-2})
\nonumber \\
&=\frac{1}{1152} \, .
\end{align}
Next up are all degree $(1,1)$ cases. We shall do them in the pairs $(6,11),(7,12),(16,20)$ and $(17,21)$.
We compute:
\begin{align}
D_{6+11} &=2 \times S_{(1,1)} \times 
L_{0,(1,1)} \times (R_3 \times R_{\tau_1,0,(1)} \times R_{0,(1)})
\nonumber \\
&= 2 \times \frac{1}{2} \frac{u^4}{t^{10}}
(t^4 u^{-4}+\frac{1}{2} t^2 u^{-2})
\times (- \frac{1}{480} t^2 u^2 \times (-t^4 u^{-2})
\times t^2 u^{-2})
\nonumber \\
&= 2 \times ( \frac{1}{960} u^{-2} t^2
+ \frac{1}{1920} + \dots)
\end{align}
We are in a special case in which we need to take into account that we can attach the markings on the edge of degree $1$ on the right to either one of two edges. This gives rise to the factor of two we put up front. 
% Since for the first term, we go to second order, we may have to do that for other terms as well, which we have not.  Danger for error here, and in following equations. 
%
% For diagram D6, we cannot have all genus $0$ on the bars. We must have genus $1$ inside, since otherwise we have zero. Thus, the diagram evaluates to:
% \begin{align}
% D_6 &= S_{1,1} L_{0,(1,1)} (R_3 R_{\tau_1,1,(1)} R_{0,(1)}
% \nonumber \\
% &= 0
% \end{align}
% where we used 
% \begin{align}
% t z_1 \int_{1,2} \frac{\Lambda}{(1-\psi_e)(1-t z_1\psi_1)}
% &=t^2 z_1^2 (\psi_e-\lambda_1) \psi_1 = 0
% \end{align}
% Thus, diagram D6 does not contribute.
The contribution arises from diagram $D_{11}$ only and equals:
\begin{align}
D_{11} &= 2 \times \frac{1}{1920} \, .
\end{align}
Next, we find -- for a choice of left vertices of genus zero --:
\begin{align}
D_{7+12} &= 2 \times S_{(1,1)} \times 
L_{0,(1,1)} \times (R_1 \times R_{\tau_3,0,(1)} \times R_{0,(1)})
\nonumber \\
&= 2 \times \frac{1}{2} \frac{u^4}{t^{10}}
(t^4 u^{-4}+\frac{1}{2} t^2 u^{-2})
\times ( \frac{1}{24} t^2  \times (-t^6 u^{-2})
\times t^2 u^{-2})
\nonumber \\
&= 2 \times (  -\frac{1}{48} u^{-4} t^2
- \frac{1}{96} u^{-2} + \dots )
\end{align}
We did not yet reach the desired order in our calculation. 
For diagram $D_7$ , we observe that we can only augment the inner genus of the $\tau_3$ connection to $2$ in order to get a contribution of the right order. We  compute the corresponding modular integral:
\begin{align}
I_7 &= t z_1 \int_{2,2} \frac{\Lambda}{(1-\psi_e)(1-\psi_1)}
=  t^4 z_1^4 \int_{2,2} (\lambda_2 - \lambda_1 \psi_e + \psi_e^2) \psi_1^3 = 0
\end{align}
and thus
\begin{align}
D_7 &= 0 \, .
\end{align}
For diagram $D_{12}$, we can put the left node at genus one.  Then it indeed contributes:
\begin{align}
D_{12} &= 2 \times S_{(1,1)} \times 
L_{1,(1,1)_c} \times (R_1 \times R_{\tau_3,0,(1)} \times R_{0,(1)})
\nonumber \\
&= 2 \times \frac{1}{2} \frac{u^4}{t^{10}} \times \frac{1}{24} \times (\frac{1}{24} t^2 (-t^6 u^{-2}) t^2 u^{-2})
\nonumber \\
&=  2 \times -\frac{1}{1152} 
\end{align}
The diagrams $D_{16}$ and $D_{20}$ can not contribute at the right order of total genus one. For instance, the node of diagram $D_{16}$ would need to be genus zero, but then the moduli space is a point and does not allow for a $\psi$ insertion corresponding to the descendant $\tau_1(\omega)$. 
% We continue:
% \begin{align}
% D_{16+20} &= S_{(1,1)} \times 
% L_{0,(1,1)} \times (R_3 \times R_{\tau_1,0,(1,1)_c} )
% \nonumber \\
% &= \frac{1}{2} \frac{u^4}{t^{10}}
% (t^4 u^{-4}+\frac{1}{2} t^2 u^{-2})
% \times (- \frac{1}{480} t^2 u^2 \times \frac{1}{12} t^2)
% \nonumber \\
% &=  \dots  u^{2} t^{-2}
% + \dots
% \end{align}
% Too high order. 
Next up are the diagrams:
\begin{align}
D_{17+21} &= S_{(1,1)} \times 
L_{0,(1,1)} \times (R_1 \times R_{\tau_3,1,(1,1)_c})
\nonumber \\
&= \frac{1}{2} \frac{u^4}{t^{10}}
(t^4 u^{-4}+\frac{1}{2} t^2 u^{-2})
\times (\frac{1}{24} t^2  \times \frac{1}{24} t^4)
\nonumber \\
&=  \frac{1}{1152} + \dots
\end{align}
We have a contribution from diagram $D_{17}$. 

\subsubsection{Full Diagrams with Two Connected Markings}
There are seven more diagrams, with the two markings connected to edges. We have diagrams $D_{4,8,9,13,14,18}$ and diagram $D_{22}$. The latter come in the pairs $(8,13),(9,14)$ and $(18,22)$. 
 Firstly, diagram $D_4$ will have the wrong genus.
% \begin{align}
% D_4 &= S_{(2)} \times L_{0,(2)} \times R_{f,(2)}
% \nonumber \\
% &= -2 u^2 t^{-8} \frac{1}{2} \frac{t^2}{u^2} \times \frac{17}{240} t^6 
% \nonumber \\
% &= - \frac{17}{240} \, .
% \end{align}
Secondly, the diagrams $(8,13)$ are zero because the fork $R_{f,(1)}$ is. 
Thirdly, we have the pair $(18,22)$ with crossroads at genus one:
\begin{align}
D_{18+22} &= S_{(1,1)} \times L_{0,(1,1)} \times R_{cr,g=1,(1,1)_c}
\nonumber \\
& = \frac{1}{2} u^4 t^{-10} (t^4 u^{-4} + \frac{1}{2} t^2 u^{-2}) \times \frac{1}{8} t^6
\nonumber \\
&= \frac{1}{16}  + \dots 
\end{align}
It is diagram $D_{18}$ that contributes here.
% However, we can also consider diagram D18 with crossroad genus one. We obtain:
% \begin{align}
% D_{18} &= S_{(1,1)} \times L_{0,(1,1)_d} \times R_{x,g=1,(1,1)_c}
% \nonumber \\
% &= \frac{1}{2} u^4 t^{-10}  \times t^4 u^{-4}
% \times t^6 \frac{1}{8}
% \nonumber \\
% & = \frac{1}{16}
% \end{align}
Finally, we have right-disconnected contributions in diagrams $D_9$ and $D_{14}$:
\begin{align}
D_{9+14'} &= 2 \times S_{(1,1)} \times L_{0,(1,1)} \times (R_{\tau_1,0,(1)} \times R_{\tau_3,0,(1)}) \nonumber \\
&=  2 \times u^4 t^{-10} (t^4 u^{-4} + \frac{1}{2} t^2 u^{-2}) \times (- t^4 u^{-2} (-) t^6 u^{-2})
\nonumber \\
&= 2 \times ( u^{-4} t^4 +  \frac{1}{2} u^{-2} t^2)
\end{align}
We  see that our choice of genera has been too naive. 
Diagram $D_{14}$ will rather contribute with a genus 1 on the left. We find:
\begin{align}
D_{14} &= 2 \times S_{(1,1)} \times L_{1,(1,1)} \times (R_{\tau_1,0,(1)} \times R_{\tau_3,0,(1)}) \nonumber \\
&=2 \times  \frac{1}{2} u^4 t^{-10} \times \frac{1}{24} \times (- t^4 u^{-2} (-) t^6 u^{-2})
\nonumber \\
&= 2 \times \frac{1}{48}
\end{align}
Again, we have two choices for which edge obtains marking $\tau_1$ (while the other than receives $\tau_3$). For diagram $D_9$, we cannot modify the  genera sensibly and it will not contribute. 

We summarize all non-zero diagram contributions:
\begin{align}
D_{1} &= \frac{1}{23040}
%\nonumber \\
&D_{10} %&
&= - \frac{1}{46080}
%\nonumber \\
&D_{15} %&
&= - \frac{1}{46080} 
% \nonumber \\
&D_{23} &= \frac{1}{320}
\nonumber \\
D_{25} &= - \frac{1}{640} 
%\nonumber \\
&  D_{26} %&
&= - \frac{1}{640} 
%\nonumber \\
& D_2 %&
&= - \frac{1}{960} 
%\nonumber \\
& D_3 %&
&= \frac{1}{1152}
\nonumber \\
D_{11} &= \frac{1}{960} 
%\nonumber \\
& D_{12} %&
&= -\frac{1}{576}  
%\nonumber \\
& D_{17} %&
&= \frac{1}{1152} 
%\nonumber \\
&
D_{18} %&
&= \frac{1}{16}
\nonumber \\
D_{14} &= \frac{1}{24} \, .
\end{align}
Their sum equals the predicted two-point function $5/48$. %The Gromov-Witten calculation matches the Hurwitz result. 

\subsection{Oscillators, Correspondences and Vanishings}
We could also have  used  the full degree $2$ two-point functions for the partitions $(2)$ and $(1,1)$ as displayed in equations  (\ref{22}) and (\ref{112}) directly. We then find the contributions $23/7680$  and $259/2560$ respectively for the same total.
It is easily checked that the sum of the diagrams $1,2,3,23$ indeed gives the  first coefficient.  Similarly, the other non-zero diagrams sum to the second coefficient. 
%
% We can divide this up by $(2)$ or $(1,1)$ partition origin:
% \begin{align}
% \frac{23 \text{c11} t^4}{24 u^4}+\frac{\frac{2203 \text{c11} t^2}{4608}+\frac{\text{c2} t^2}{192}}{u^2}+\left(\frac{259 \text{c11}}{2560}+\frac{23 \text{c2}}{7680}\right)+O\left(u^8\right)
% \end{align}
%with very intricate cancellations at orders $u^2$ and higher, between the two partitions.  
 %This is already a good check on our calculation. 
%
One can perform a careful comparison between the diagrams and terms in the oscillator expressions - we skip it.\footnote{For example, the first line in (\ref{FifteenTerms}) maps to the diagram $D_{10}$ while the second and third lines give rise to diagram $D_{25}$ and so on.}
% A more careful analysis shows the following correspondence between the contraction terms in the oscillator calculation and the various diagrams:
% \begin{align}
% a_1 & \leftrightarrow D_{10}
% \nonumber \\
% a_7 & \leftrightarrow D_{15}
% \nonumber \\
% a_2+a_3 & \leftrightarrow D_{25}
% \nonumber \\
% a_8+a_9 & \leftrightarrow D_{26}
% \nonumber \\
% a_4 & \leftrightarrow D_{12}
% \nonumber \\
% a_5+a_6 & \leftrightarrow D_{11} 
% \end{align}
%  Complete the picture ? 
%
The diagrams $D_{14}$ and $D_{18}$ are fully connected degree two two-point functions. They are responsible for the final result and all diagrams that are effectively lower-point functions or lower degree contributions sum to zero. This is as predicted in subsection \ref{Predictions}. 
% Also note that the one-point diagrams $D_{1,2,3,5,6,7,9,10,11,12,15,16,17,19,20,21}$ combine to a total of zero. The two-point function at degree zero (corresponding to diagrams $D_{23,25,26}$) also sums to zero. The two-point connected at degree one is zero as well. 
% The remaining diagrams are $D_{14}$ and $D_{18}$ that do indeed sum to the final result. 

We thus identify the world sheets in diagrams $D_{14}$ and $D_{18}$ as the crucial contributions. They are connected world sheets of genus one with the covering structure captured by their diagrams. We have a genus   one surface sent to one fixed point and  the two markers attached to the other fixed point on two separate covering spheres (in diagram $D_{14}$), or, two single covers connected at the one fixed point where the two  markers attach, along with a Riemann surface of genus one sent into that same fixed point (in diagram $D_{18}$).  

In passing, we note that our intermediate results are more powerful. Using the oscillator expressions we obtained and the symmetry factors, we can compute the exact equivariant two-point function:
\begin{align}
\langle \tau_1 \tau_3 \rangle_{\text{eq.}} &=  \frac{23 }{24 } \frac{t^4}{u^4} +\frac{2227}{4608 } \frac{t^2}{u^2} +\frac{5}{48} \, .
\end{align}
In this all genus expression, the negative powers of the equivariant parameter cancel as they must. 

\subsection{The Direct Non-Equivariant Calculation}
In order to identify the world sheets that contribute to the correlator, we delved into the equivariant calculation of the correlator. It is known that if one only wishes to compute the final result, one can take the $t \rightarrow 0$ limit earlier and work with simpler expressions.  Our correlator then corresponds to \cite{OkounkovPandharipande2}:
\begin{equation}
\langle \tau_1(\omega) \tau_3(\omega) \rangle = [z_1^2 z_2^4] \langle 0 | \alpha_1^2 {\cal E}_0(z_1) {\cal E}_0(z_2)
\alpha_{-1}^2 | 0 \rangle \, .
 \end{equation}
Again, this is calculated using the $\alpha$ and ${\cal E}$ commutators. 
We first move the first left oscillator to the right, then the second and compute the various commutators:
\begin{align}
G &= \frac{1}{4} \langle \alpha_{1}^2 {\cal E}_0(z_1) {\cal E}_0(z_2) \alpha_{-1}^2 \rangle
\nonumber \\
&= \frac{1}{4}  \langle \alpha_1
(\zeta(z_1) {\cal E}_1(z_1){\cal E}_0(z_2) \alpha_{-1}^2
+ \zeta(z_2) {\cal E}_0(z_1) {\cal E}_1(z_2)
\alpha_{-1}^2
+ 2 {\cal E}_0(z_1) {\cal E}_0 (z_2)
\alpha_{-1} 
\rangle
\nonumber \\
&=\frac{1}{4}  \langle 
(\zeta(z_1)^2 {\cal E}_2(z_1){\cal E}_0(z_2) 
\alpha_{-1}^2
 + 2 \zeta(z_1) \zeta(z_2)
{\cal E}_1(z_1){\cal E}_1(z_2) \alpha_{-1}^2 
+2 \zeta(z_1) {\cal E}_1(z_1) {\cal E}_0(z_2) \alpha_{-1})
\nonumber \\
&
+ \zeta(z_2)^2 {\cal E}_0(z_1) {\cal E}_2(z_2)
\alpha_{-1}^2
+ 2 \zeta(z_2) {\cal E}_0(z_1) {\cal E}_1(z_2) \alpha_{-1}
\nonumber \\ 
&
+ (2 \zeta(z_1) {\cal E}_1(z_1) {\cal E}_0 (z_2) \alpha_{-1}
+ 2 \zeta(z_2) {\cal E}_0 (z_1) {\cal E}_1(z_2))
\alpha_{-1}
+2 {\cal E}_0(z_1) {\cal E}_0(z_2)
\rangle
\end{align}
We still need to move the $\alpha_{-1}$ oscillators to the left. 
% Which diagrams does it take into account, one might ask. Any intermediate degree at once. The question is ill suited because we re-summed $\mu$.
We take into account that if we have a left operator with negative index, it will contribute zero. We find the end result:
% \begin{align}
% G &= \frac{1}{4}
% \Big( \zeta(z_1)^3/\zeta(z_2) + 2 \zeta(z_1)^3 \zeta(z_2)
%  +
% \zeta(z_1)^2 \zeta(z_2)^2
% \zeta(2 (z_1+z_2))/\zeta(z_1+z_2)
% \nonumber \\
% & + 4 \zeta(z_1) \zeta(z_2) + 2 \zeta(z_1) \zeta(z_2)^3
% + 2 \zeta(z_1)/\zeta(z_2) + 2 \zeta(z_1) \zeta(z_2)
%  + \zeta(z_2)^3/\zeta(z_1) + 2 \zeta(z_2)/\zeta(z_1)
% \nonumber \\
% &+ 2 \zeta(z_1)/\zeta(z_2) + 2 \zeta(z_1) \zeta(z_2)
%  + 2 \zeta(z_2)/\zeta(z_1) + 2 / \zeta(z_1) /\zeta(z_2) \Big) \, .
% \end{align}
\begin{align}
G &= \frac{1}{4}
\Big( \zeta(z_1)^3/\zeta(z_2) + 2 \zeta(z_1)^3 \zeta(z_2)
 +
\zeta(z_1)^2 \zeta(z_2)^2
\zeta(2 (z_1+z_2))/\zeta(z_1+z_2)
\nonumber \\
& + 8 \zeta(z_1) \zeta(z_2) + 2 \zeta(z_1) \zeta(z_2)^3
+ 4 \zeta(z_1)/\zeta(z_2) 
 + \zeta(z_2)^3/\zeta(z_1) + 4 \zeta(z_2)/\zeta(z_1)
\nonumber \\
& 
  + 2 / (\zeta(z_1) \zeta(z_2)) \Big) \, .
\end{align}
The answer is symmetric in $(z_1,z_2)$ as it must be.  The Taylor expansion does give rise to the term $5/48 \, z_1^2 z_2^4$ confirming the expected result. 

Our diagrams correspond to the  equivariant localization calculation.
The non-equivariant computation can also be given a diagrammatic interpretation in terms of oscillators and their Wick contractions. They allow for a geometric description as tropical Feynman diagrams \cite{Tropical}.
It would be interesting to attempt to interpret these tropical geometries directly in terms of string theory or two-dimensional topological gravity. Note that these diagrams would correspond to the calculation after an intermediate summation over the partitions $\mu$  has been performed and an exponential transposition interaction term has been canceled between the two fixed point localization factors.\footnote{See equations (\ref{Before}) and (\ref{After}) for a concrete example of this calculation in the relative Gromov-Witten context. The mechanism is generic.} For now, the tropical calculation \cite{Tropical} requires the input of the crucial relative one-point functions which in turn depends on the equivariant localization calculation.

\section{The Relative One-Point Function Localized}
\label{RelativeLocalization}
In sections \ref{BulkDuality} and \ref{TwoPoint}, we have concentrated on bulk Gromov-Witten correlation functions. The relative correlators are manifestly interesting as well. They require more technology to localize \cite{OkounkovPandharipande3}, but they provide an even more direct handle on the Gromov-Witten/Hurwitz correspondence. Indeed, the expression for bulk vertex operators in terms of ordinary cycles is fixed by the relative one-point functions \cite{OkounkovPandharipande1}. In this section, we therefore recall the equivariant localization of one-point functions and propose a diagrammatic picture for their calculation. 
\subsection{The Product Origin}
The non-equivariant relative one-point function $\langle \tau_k | \nu \rangle$ is calculated as the limit of its equivariant counterpart. It lives at a genus $g$ that satisfies $k=2g-2+|\nu|+l(\nu)$.
The generating series $G$ of equivariant relative invariants is:
\begin{equation}
G(z_i,w_j|\nu) =  \sum_{g,k_i,l_j} \prod_i z_i^{k_i+1} \prod_j w_j^{l_j+1} \langle \tau_{k_i} (0) \tau_{l_j}(\infty) | \nu \rangle_g \, .
\end{equation}
It is computed
once more from a factorized degeneration  \cite{OkounkovPandharipande3}:\footnote{We set $u=1$ in this section.}
\begin{equation}
G(z_i,w_j|\nu) = \sum_{|\mu|=d} \frac{1}{z(\mu)} t^{-l(\mu)-d-n} \prod_i \frac{\mu_i^{\mu_i}}{\mu_i!}
H(\mu,tz,t^{-1}) z(\mu) \tilde{G} (\mu,-t^{-1}|w_j|\nu) \, . \label{RelativeDegenerationFormula}
\end{equation}
The Hodge factor $H$ is familiar. The factor $\tilde{G}$ corresponds to a localization calculation that factors out a scaling transformation of the target $\mathbb{P}^1$ \cite{OkounkovPandharipande3}.
We  insert our operator at $0$ and consider a second factor without insertions. In that case, the second factor simplifies drastically and reduces to:
\begin{equation}
\tilde{G}(\mu|s|\nu) = \sum_{g,k} s^{k+1} \langle \mu ,k| | \nu \rangle_g^{\text{\textasciitilde}}
= \langle \mu | e^{s {\cal F}_2} | \nu \rangle \, .
\end{equation}
% The empty rubber tube is:
% \begin{equation}
% \sum_{k \ge -1} s^{k+1} \langle \mu,k||\nu\rangle^{tilde} = \sum_{k \ge -1} \frac{s^{k+1}}{(k+1)!} \langle \mu| {\cal F}_2^{k+1}| \nu \rangle= \langle \mu | e^{s {\cal F}_2} | \nu \rangle \, .
% \end{equation}
The summation index $k$ refers to a summation over descendants $\psi^k$ and it is computed using the dilaton equation (which is a recursion for reducing the degree of the descendant) \cite{OkounkovPandharipande3}.
Using this relation and the oscillator expression for Hodge integrals, one arrives at the oscillator form of the equivariant one-point function:
\begin{equation}
G(z|\nu) = \sum_\eta \langle A(tz,z) e^{\alpha_1} e^{\frac{{\cal F}_2}{t}} | \eta \rangle z(\eta) \langle \eta | e^{-\frac{{\cal F}_2}{t}} | \nu \rangle \, . \label{Before}
\end{equation}
Crucially, we can pull  the transposition insertion through the identity operator in the middle and annihilate these interactions. We then re-obtain equation (\ref{EquivariantOnePoint}):
\begin{equation}
G(z|\nu) = \sum_{k} z^{k+1} \langle \tau_k(0) | \nu \rangle = \langle A(tz,z) e^{\alpha_1} | \nu \rangle
\, . \label{After}
\end{equation}
We simplified the expression  at the cost of obscuring the localized world sheets that contribute to the amplitude. Again, one would ideally have a direct non-equivariant integration over the moduli space of Riemann surfaces to obtain the final result. Lacking such a tool, to understand the relative one-point functions,  we must identify more of the physical content of the equivariant theory. 

\subsection{The Localization and the Diagrams}
Again, we look at a detailed example. We concentrate on the relative one-point function $\langle \tau_3(\omega) | (2) \rangle$ at genus one.  Here, we compute 
the relevant term in the completed cycle $\overline{(4)}$  directly, using the localization of the relative Gromov-Witten invariant. 
\subsubsection{The Hurwitz Result}
We study the completed cycle $\overline{(4)}$ -- see equation (\ref{CompletedCycles}). It consists of degree $2,3$ and $4$ terms.  The degree $2$ cal\-cu\-lation gives:
\begin{equation}
\langle \tau_3(\omega) | (2) \rangle_{d=2}
= \frac{1}{2! 3!} \frac{5}{4} = \frac{5}{48}
\, .
\end{equation}
We want to reproduce this number from the Gromov-Witten theory.
\subsubsection{The Localization}
We calculate the ingredients in the relative degeneration formula (\ref{RelativeDegenerationFormula}).
At degree two, we can have two intermediate partitions namely $(1,1)$ and $(2)$. On the right hand side, we therefore need:
\begin{align}
\langle \alpha_2 e^{\frac{1}{t} {\cal F}_2} \alpha_{-2} \rangle
&=\langle \alpha_2 {\cal E}_{-2} (\frac{2 }{t}) \rangle
=  \frac{\zeta(\frac{4 }{t})}{\zeta(\frac{2}{t})}
 \approx 2 + \frac{1}{t^2} + \dots
\end{align}
as well as
\begin{align}
\langle \alpha_1 \alpha_1 e^{\frac{1}{t} {\cal F}_2} \alpha_{-2} \rangle
&=\langle \alpha_1 \alpha_1 {\cal E}_{-2} (\frac{2 }{t}) \rangle
=% \zeta(\frac{2 u}{t}) \zeta(\frac{2 u}{t})/\zeta(\frac{2u}{t})=
 \zeta(\frac{2 }{t})  \approx \frac{2}{t} + \frac{1}{3} \frac{1}{t^3} + \dots \, .
\end{align}
%We have that $z((1,1))=2=z((2))$. 
We need the degree two one-point functions $H$ as well, up to order $z_1^4$ - we determined them previously -- see equations (\ref{DegreeTwoOnePoint}) and (\ref{DegreeOneOneOnePoint}). 
The symmetry factors are:
\begin{equation}
\tilde{S}_{(2)} = 2 t^{-4} \, ,
\qquad \qquad
\tilde{S}_{(1,1)} = t^{-5} \, .
\end{equation}
Combining ingredients, we find the full result:
\begin{align}
[z^4] G_{d=2}(z| (2))
&= [z^4] \Big( 2 t^{-4} H((2),tz,\frac{1}{t}) \frac{1}{4} \langle \alpha_2 e^{-\frac{1}{t} {\cal F}_2} \alpha_{-2} \rangle
\nonumber \\
& + t^{-5} H((1,1),tz,\frac{1}{t}) \frac{1}{4} \langle \alpha_1^2 e^{-\frac{1}{t} {\cal F}_2} \alpha_{-2} \rangle \Big)
\nonumber \\
&= \frac{1}{2} t^{-4}  (-\frac{1}{8} {t^6} -\frac{1}{960} t^4+ \dots)   (2 + \frac{1}{ t^{2}} + \dots)
\nonumber \\
& + \frac{1}{4} t^{-5} (-2 {t^8} -\frac{1}{480} {t^6} + \dots) (-)(2 \frac{1}{t} + \frac{1}{3} \frac{1}{t^3} +\dots)
\nonumber \\
% & \approx -\frac{1}{8} {t^2} -\frac{1}{16} -\frac{1}{960} +  t^2 + \frac{1}{6} + \frac{1}{960} + \dots
% \nonumber \\
& \approx \frac{7}{8} t^2+ \frac{5}{48} + \dots
\label{Result}
\end{align}
We obtain the expected  final result at the non-equivariant order $t^0$. We note that both partitions contribute to the  amplitude. 

In order to identify the relevant world sheets, we develop a  diagrammar. 
The left hand side of our diagrams are familiar from the previous sections. They capture the moduli space integrals corresponding to the Hodge numbers.  The right hand side are diagrams for double Hurwitz numbers.  They obey different rules. We basically have a left and a right side (of the right hand diagram) representing the partitions $\eta$ and $\nu$ in equation (\ref{Before}). We moreover have a cut-and-join cubic interaction vertex corresponding to the transposition operator ${\cal F}_2$ 
\begin{equation}
{\cal F}_2 = \frac{1}{2} \sum_{i,j>0} \alpha_{-i-j} \alpha_i \alpha_j + \alpha_{-i} \alpha_{-j} \alpha_{i+j} \, 
\end{equation}
that cuts and joins strings represented by cycles. 
%that is given by:
The degeneration formula glues the left and right diagrams together to form a full relative Feynman diagram. 

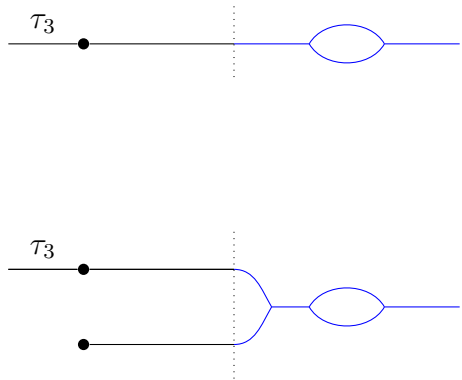
\begin{figure}[H]

\begin{tikzpicture}[
  scale=1,
  vertex/.style={circle, fill=black, inner sep=1.5pt}
]

\node[vertex] (d5) at (0,0) {};
\coordinate (d5r) at ($(d5) + (2,0)$) {};
\coordinate (d5p) at ($(d5) + (2,0.5)$) {};
\coordinate (d5pp) at ($(d5)+ (3,0)$) {};
\coordinate  (d5ppp) at ($(d5)+ (4,0)$) {};
\coordinate  (d5pppp) at ($(d5)+ (5,0)$) {};

\draw (d5) -- +(2,0);
\draw (d5) -- +(-1,0) node[fill=none,midway,above] {$\tau_3$};
\draw[color=blue] (d5r) -- (d5pp);
\draw[color=blue] (d5pp) to [out=60, in=120] (d5ppp) ;
\draw[color=blue] (d5pp) to [out=-60, in=-120] (d5ppp) ;
\draw[color=blue] (d5ppp) -- +(1,0);

\draw[dotted] (d5p) -- +(0,-1.0) ;

\node[vertex] (d6) at (0,-3) {};
\node[vertex] (d6b) at (0,-4) {};
\draw (d6) -- +(2,0);
\draw (d6) -- +(-1,0);
\draw (d6b) -- +(2,0);
\coordinate (d6r) at ($(d6) + (2,0)$) {};
\coordinate (d6br) at ($(d6b) + (2,0)$) {};
\coordinate (d6p) at ($(d6) + (2,0.5)$) {};
\coordinate (d61) at ($(d6) + (2.5,-0.5)$) {};
\coordinate (d6pp) at ($(d6)+ (3,-0.5)$) {};
\coordinate  (d6ppp) at ($(d6)+ (4,-0.5)$) {};
\coordinate  (d6pppp) at ($(d6)+ (5,-0.5)$) {};

\draw (d6) -- +(2,0);
\draw (d6) -- +(-1,0) node[fill=none,midway,above] {$\tau_3$};

\draw[color=blue] (d6r) to [out=0, in=120] (d61);
\draw[color=blue] (d6br) to [out=0, in=-120] (d61);
\draw[color=blue] (d61) -- (d6pp);
\draw[color=blue] (d6pp) to [out=60, in=120] (d6ppp) ;
\draw[color=blue] (d6pp) to [out=-60, in=-120] (d6ppp) ;
\draw[color=blue] (d6ppp) -- +(1,0);

\draw[dotted] (d6p) -- +(0,-2) ;

\end{tikzpicture}

\caption{The Factorised One-Point Function Diagrams}
\label{NonEquivariantOnePoint}
\end{figure}
\noindent
We conclude that we have the localization diagrams on the left and the cut and join interactions on the right. 
This again provides a surface, complemented with a string interaction diagram.

We illustrate the proposal in our example. 
See Figure \ref{NonEquivariantOnePoint}.
For $H((2),tz,u/t)$ we have a contributing genus zero vertex on the left with one mark and one edge of degree two attached. On the right, it combines with a string loop.  
%, and a non-contributing genus one with no string loop on the right. 
For the $(1,1)$ partition, we have two genus zero vertices on the left (of which one is marked) times a string loop on the right.
%We have a cancellation between  the genus zero respectively genus one disconnected contribution, where the former connects to trivial $(2)$ line.
The values of the diagrams sum to the constant term in the final result (\ref{Result}). The diagrams are a depiction of the relevant string world sheets.

\section{Conclusions}
\label{Conclusions}
In this paper, we showed that a standard open/closed topological quantum field theory
%\cite{Lauda:2006mn}
based on the partial permutation algebra %\cite{Troost:2025eqm} 
counts covering surfaces with boundaries up to degree $n$. Boundary conditions in terms of irreducible representations can be transformed into boundary holonomies. We demonstrated that the large $n$ limit of the theory is the Hurwitz theory that appears in the Gromov-Witten/Hurwitz correspondence \cite{OkounkovPandharipande1} and argued that boundary holonomies are equivalent to the specification of relative conditions in the Gromov-Witten theory.
An extension of the open theory that incorporates the multiplicities of irreducible representations associated to boundaries exists. 
Moreover, we provided a practical guide to the powerful results of \cite{OkounkovPandharipande1,OkounkovPandharipande2,OkounkovPandharipande3}. One goal was to better understand completed cycles dual to the bulk  vertex operators, which are descendants of the volume form on the target curve. In particular, we wanted to get a grip on the world sheet covering surfaces that are responsible for completion contributions in example amplitudes. To that end, we tracked the derivation of the correspondence \cite{OkounkovPandharipande1} to the equivariant calculation of bulk \cite{OkounkovPandharipande2} and relative Gromov-Witten correlators \cite{OkounkovPandharipande2}. We did identify the world sheets for example one and two-point functions as well as for a relative one-point function. For the crucial latter case, we proposed a diagrammatic way to represent all contributions. 

There are plenty of open research directions. The first may be to characterize the generic world sheets that are responsible for completed cycle contributions. Another is the matching of supersymmetry preserving $AdS_3$ one-point functions in the presence of a boundary with our topological boundary states.
One can formulate a gauge theory dual to the fully equivariant theory. There is also the intriguing possibility to interpret the branch point insertions in terms of the twist deformation in an $AdS_3$ bulk string theory. One would like a geometric W-algebra description of the relative amplitudes. The integrable hierarchy for the open/closed theory, including the weighing of the number of boundaries, is also worth pursuing. In conclusion, there is more work to be done.

\subsubsection*{Acknowledgments}
It is a pleasure to thank my colleagues for creating a stimulating research environment and Lior Benizri in particular for discussions on these and unrelated topics.

\appendix

\section{The Moduli Space Integrals}
\label{ModuliSpaceIntegrals}
All integrals we encounter are over the compactified moduli spaces of Riemann surface ${\overline{M}}_{g,n}$ which we abbreviate to the indices $(g,n)$. The moduli spaces include stable nodal curves. We also provide definitions of integrals over moduli spaces of unstable curves that are compatible with the integrable structure of our model.

\subsection{The Integrals over Moduli Spaces of Unstable Curves}
Firstly, we define integrals over the moduli spaces of unstable curves through the equalities \cite{OkounkovPandharipande1}:
% \begin{align}
% \int_{{0,1}} \frac{1}{1-x_1 \psi} &= \frac{1}{x_1^2}
% \nonumber \\
% \int_{{0,2}} \frac{1}{(1-x_1 \psi)(1-x_2\psi)}
% &= \frac{1}{x_1+x_2} \, ,
% \end{align}
% or in the form in which we use them most frequently:
\begin{align}
\int_{{0,1}} \frac{x_1 }{1-x_1 \psi} &= \frac{1}{x_1}
\nonumber \\
 \int_{{0,2}} \frac{x_1 x_2}{(1-x_1 \psi)(1-x_2\psi)}
&= \frac{x_1 x_2}{x_1+x_2} \, . \label{IntegralsUnstable}
\end{align}
These equalities render the combinatorics of describing the full partition function easier.

\subsection{The Integrals over Compactified Moduli Spaces}
We make a summary of the compactified moduli space integrals over stable curves we use. At genus zero we mostly meet the unstable integrals which are easily computed from formula (\ref{IntegralsUnstable}).
%We also recall that at genus zero, the moduli space $\overline{M}_{0,3}$ is a point. 
% 
At higher genus, we make use of the Hodge class $\Lambda$:
\begin{equation}
\Lambda = 1 - \lambda_1 + \lambda_2 + \dots + (-1)^g \lambda_g \, 
\end{equation}
in the definition of the linear Hodge integrals $H$. 
At genus one we exploited the  concrete results:
\begin{align}
\int_{1,1} \lambda_1 &= \frac{1}{24}
&
\int_{1,1} \psi_1 &= \frac{1}{24}
\nonumber  \\
 \int_{1,1} \frac{\Lambda}{1-\psi} &= 
 %\int_{1,1} (\psi-\lambda_1) =
 0 
% \nonumber \\  
&
 \int_{1,1} \frac{2 \Lambda}{1-2 \psi}  &=% 2 \int_{1,1} ((-\lambda_1) + 2 \psi)=
 \frac{1}{12}
\nonumber \\
\int_{1,2} \frac{\Lambda}{(1-\psi_1)(1-\psi_2)} &=%\int_{1,2} (-\lambda_1 (\psi_1+\psi_2)+\psi_1^2+\psi_2^2 + \psi_1 \psi_2) \dots=
\frac{1}{24}
%\nonumber \\
&
\int_{1,2} \psi_1^2 &= \frac{1}{24}
\, .
\end{align}
At genus two we found the following integrals useful:
\begin{align}
\int_{2,1} \lambda_2 \psi_1^2 &= \frac{7}{5760}
%\nonumber \\
\qquad \qquad 
\int_{2,1} (-\lambda_1) \psi^3_1 %&
= -\frac{1}{480}
\nonumber \\
\int_{2,2} (-\lambda_1) \psi_1 \psi_2^3 &= - \frac{1}{160}
%\nonumber \\
\qquad \qquad \qquad
%\int_{1,3} \frac{\Lambda}{(1-\psi_1)(1-\psi_2)(1-\psi_3)}  &=_{restr|}
\int_{1,3} \psi_2^3 = \frac{1}{24}
\nonumber \\
\int_{1,4} \psi_1 \psi_2^3  &= \frac{1}{8}
\nonumber \\
%\int_{1,3} \frac{\Lambda}{(1-\psi_1)(1-\psi_2)(1-\psi_3)}  &=_{restr|}
\int_{1,3} (\psi_1^2+\psi_2^2+\psi_1 \psi_2 -\lambda_1(\psi_1+\psi_2))\psi_3
&= \frac{1}{12} 
 \nonumber \\
% z_2 \int_{2,2} \frac{\Lambda}{1-\psi_e} z_2^3 \psi_2^3
%&= z_2^4 
\int_{2,2} (\psi_1^2-\psi_1 \lambda_1+\lambda_2)  \psi_2^3
&=% z_2^4 
(\frac{29}{5760}-\frac{1}{160}+\frac{7}{5760}) =0
\nonumber \\
%z_1^2 z_2^4 
\int_{2,3} \psi_1 \psi_2^3 \frac{\Lambda}{1-\psi_3}
&=
%z_1^2 z_2^4 
%\int_{2,3} \psi_1 \psi_2^3 (\psi_e^2-\psi_e \lambda_1 + \lambda_2)
%=
%z_1^2 z_2^4 
(\frac{29}{1440}-\frac{1}{40}+\frac{7}{1440}) = 0
\, .
\end{align}
%Most of these integrals are mentioned in the text.
Other integrals can be obtained using analytic techniques or software  \cite{LinearHodge,AdmcyclesPaper,AdmcyclesAuthors,Witten:1989ig}.

\bibliographystyle{JHEP}

\end{document}